\documentclass[manuscript,screen]{acmart}
\renewcommand\footnotetextcopyrightpermission[1]{}
\acmJournal{TOSEM}
\usepackage[many]{tcolorbox}
\usepackage{booktabs}
\usepackage{pifont}
\usepackage{array}
\usepackage{subcaption}
\usepackage{xltabular}
\usepackage{threeparttablex}
\usepackage{ragged2e}

\usepackage{tikz}
\usepackage{graphicx}
\usetikzlibrary{arrows.meta, positioning, calc}

\AtBeginDocument{%
  }

\setcopyright{acmlicensed}
\copyrightyear{2018}
\acmYear{2018}
\begin{document}

\title{Understanding Maintenance and Support in a Community-Driven Scientific Workflow Ecosystem: A Cross-Space Study of Galaxy}

\author{Khairul Alam}
\email{kha060@usask.ca}
\orcid{1234-5678-9012}
\correspondingauthor
\authornotemark[1]
\affiliation{%
  \institution{University of Saskatchewan}
  \city{Saskatoon}
  \state{Saskatchewan}
  \country{Canada}
}

\author{Kowsik Roy}
\affiliation{%
  \institution{BRAC University}
  \city{Dhaka}
  \country{Bangladesh}}
\email{kowsik.roy@g.bracu.ac.bd}

\author{Md Shamimur Rahman}
\orcid{0009-0001-5355-4600}
\affiliation{%
  \institution{University of Saskatchewan}
  \city{Saskatoon}
  \state{Saskatchewan}
  \country{Canada}
}
\email{shamimur.rahman@usask.ca}

\author{Banani Roy}
\affiliation{%
  \institution{University of Saskatchewan}
  \city{Saskatoon}
  \state{Saskatchewan}
  \country{Canada}
}
\email{banani.roy@usask.ca}

\renewcommand{\shortauthors}{Alam et al.}

\begin{abstract}
Galaxy is a widely used, community-driven scientific workflow system whose sustainability depends on continuous maintenance across its software, tools, workflows, infrastructure, documentation, and user-support ecosystem. However, maintenance knowledge in Galaxy is distributed across development and community-support spaces, making it difficult to understand what is maintained, how maintenance artifacts are resolved, and how user-facing concerns connect to repository-level development. We conduct a large-scale empirical study of Galaxy using 11,762 GitHub issues, 52,203 pull requests, and 6,235 Community Forum discussions. We characterize maintenance and support concerns, examine factors associated with resolution outcomes and resolution time, and investigate explicit and candidate connections among maintenance artifacts across these spaces.

Using BERTopic modeling, we identify nine issue topics, 14 pull-request topics, and 14 forum topics, revealing a maintenance landscape spanning workflow execution, data management, tools and dependencies, infrastructure, testing, scientific resources, documentation, and user support. Resolution analyses show that coordination, diagnostic, contributor, automation, and engagement characteristics exhibit different associations with whether artifacts are resolved and how quickly resolution occurs. We further find limited explicit traceability between development and support spaces: 97.77\% of 16,426 resolved explicit relationships occur within GitHub, while only 294 connect GitHub artifacts with Community Forum discussions, despite additional semantic and technical relatedness across these spaces. Together, these findings characterize Galaxy maintenance as a distributed ecosystem-level process and identify opportunities to improve diagnostic reporting, lifecycle-aware triage, cross-space traceability, and the reuse of community-support knowledge.
\end{abstract}

\begin{CCSXML}
<ccs2012>
   <concept>
       <concept_id>10011007.10011074.10011134.10003559</concept_id>
       <concept_desc>Software and its engineering~Open source model</concept_desc>
       <concept_significance>300</concept_significance>
       </concept>
   <concept>
       <concept_id>10011007</concept_id>
       <concept_desc>Software and its engineering</concept_desc>
       <concept_significance>500</concept_significance>
       </concept>
   <concept>
       <concept_id>10011007.10011074</concept_id>
       <concept_desc>Software and its engineering~Software creation and management</concept_desc>
       <concept_significance>500</concept_significance>
       </concept>
   <concept>
       <concept_id>10011007.10011074.10011134</concept_id>
       <concept_desc>Software and its engineering~Collaboration in software development</concept_desc>
       <concept_significance>300</concept_significance>
       </concept>
 </ccs2012>
\end{CCSXML}

\ccsdesc[300]{Software and its engineering~Open source model}
\ccsdesc[500]{Software and its engineering}
\ccsdesc[500]{Software and its engineering~Software creation and management}
\ccsdesc[300]{Software and its engineering~Collaboration in software development}

\keywords{Scientific workflow systems, Galaxy, Software maintenance, Maintenance and support, Resolution outcomes, Cross-space traceability, Community-driven software, Empirical software engineering}

\maketitle

\section{Introduction}
\label{sec:introduction}

Modern scientific research is increasingly data-intensive, creating a growing need for computational platforms that can support the design, execution, sharing, and reproduction of complex analyses. In domains such as genomics, transcriptomics, proteomics, and biomedical data science, researchers frequently combine multiple software tools, parameters, datasets, reference resources, and execution environments into multi-step computational analyses. When these analyses are implemented through ad hoc scripts, informal descriptions, or local computing environments, reproducing and transferring them across users and infrastructures can become difficult~\cite{gruning2018practical,DBLP:journals/bib/Leipzig17}. \textit{Scientific Workflow Systems} (SWSs) address these challenges by providing structured mechanisms for defining computational steps, managing dependencies and data flow, executing and monitoring analyses, and supporting their reuse and reproducibility across computing environments~\cite{DBLP:journals/fgcs/DeelmanGST09,DBLP:journals/fgcs/SilvaFPJSD17,DBLP:journals/fgcs/SuterCABBCCDTFGJKKLMOPP26}. These capabilities also align with broader efforts to make computational research findable, accessible, interoperable, and reusable~\cite{DBLP:journals/datasci/LamprechtGKMAPA20,barker2022introducing}.

A diverse ecosystem of SWSs has consequently emerged to support different scientific domains, execution models, workflow representations, and computing infrastructures~\cite{deelman2009workflows, DBLP:journals/fgcs/SuterCABBCCDTFGJKKLMOPP26}. Systems such as Pegasus~\cite{DBLP:journals/fgcs/DeelmanVJRCMMCS15}, Nextflow~\cite{di2017nextflow}, Snakemake~\cite{DBLP:journals/bioinformatics/KosterR18}, and Galaxy~\cite{goecks2010galaxy} provide different abstractions for constructing and executing scientific workflows. Among these, \textit{Galaxy} has developed into a large, community-driven ecosystem for accessible, reproducible, and collaborative computational science~\cite{giardine2005galaxy,goecks2010galaxy,DBLP:journals/nar/AfganNGBGSOMLSF22,galaxy2024galaxy,galaxy2026galaxy}. Galaxy enables researchers to construct and execute analyses through a web-based environment while supporting reusable workflows, extensive collections of scientific tools, shared datasets and histories, public and institutional deployments, training resources, and community-developed extensions~\cite{goecks2010galaxy,DBLP:journals/nar/AfganNGBGSOMLSF22,galaxy2024galaxy,galaxy2026galaxy}. Its operation therefore depends not only on the Galaxy core software, but also on a broad network of tools, workflows, dependencies, reference resources, execution infrastructure, documentation, and community contributions~\cite{blankenberg2014dissemination}.

Sustaining such an ecosystem introduces maintenance challenges that extend well beyond conventional source-code defect correction. Scientific tools are continually added and updated; dependencies and container environments evolve; reference genomes and scientific databases change; workflow interfaces and execution mechanisms are revised; and distributed computing infrastructure must remain compatible with changing software and resource requirements. Workflows themselves can become difficult to reuse when tools disappear, versions diverge, dependencies change, or workflow specifications no longer match the target execution environment~\cite{DBLP:conf/apsec/AlamRS23}. Prior research has documented challenges related to scientific workflow execution, reproducibility, portability, and reuse~\cite{gruning2018practical,DBLP:journals/fgcs/SilvaFPJSD17,DBLP:journals/ese/AlamRRM25}. These challenges suggest that the long-term utility of an SWS depends as much on continuous maintenance and support as on its ability to execute workflows in the first place.

Maintenance in Galaxy is inherently \emph{socio-technical}, involving developers, tool authors, infrastructure maintainers, administrators, contributors, trainers, and scientific users across multiple development and support spaces. GitHub issues~\cite{github-issue-tracker} capture defects, enhancement requests, infrastructure problems, and other maintenance needs, while pull requests~\cite{pr-tracker-action} document implementation, testing, dependency updates, infrastructure changes, and documentation work. The Galaxy Community Help Forum~\cite{galaxycommunityhelp} provides a complementary user-facing space where researchers seek help with execution failures, configuration, data handling, reference resources, and domain-specific analyses. Prior empirical software-engineering research has shown that such repository and discussion artifacts provide valuable evidence about collaboration, maintenance, contribution evaluation, and user support~\cite{tsay2014influence,DBLP:journals/ese/HataNBKT22,DBLP:journals/ese/KalliamvakouGBS16,DBLP:conf/icse/GousiosPD14,DBLP:conf/indiaSE/DhasadeVC20,DBLP:conf/icse-chase/HellmanCUCG22}. However, repository-centered analyses capture only part of the maintenance process: user-facing problems may be resolved in community-support spaces without becoming formal issues, while repository changes may address recurring problems without being explicitly linked back to their original discussions. Examining these spaces together is therefore necessary to understand how maintenance work is distributed and how problem and solution knowledge is connected across the Galaxy ecosystem.

Despite growing research on scientific workflow execution, reproducibility, reuse, and repository-centered software maintenance, an integrated empirical understanding of \emph{maintenance and support in Galaxy as an ecosystem} remains limited. In particular, three aspects remain insufficiently understood. First, maintenance and support concerns across developer-facing and community-facing spaces have not been systematically characterized. Second, little is known about how the characteristics of issues, pull requests, and community discussions are associated with both resolution outcomes and resolution time. Third, although user-reported problems, repository issues, and implementation changes may reflect related technical needs, the extent to which these artifacts are explicitly connected or otherwise related across development and support spaces remains unclear. Addressing these gaps requires moving beyond a repository-centric view of maintenance. We argue that maintenance in a community-driven SWS should be studied as a distributed process in which user-facing problems, formalized maintenance tasks, implementation changes, and reusable support knowledge are produced across multiple spaces. Such a perspective can reveal not only \emph{what} the community maintains, but also \emph{how} maintenance work is resolved and \emph{whether} the knowledge generated through these processes remains connected across the ecosystem.

To address these gaps, we conduct a large-scale empirical study of Galaxy using \emph{11,762 GitHub issues}, \emph{52,203 pull requests}, and \emph{6,235 Galaxy Community Help Forum discussions}. We combine topic modeling and qualitative interpretation with statistical and time-to-event analyses, explicit cross-artifact traceability analysis, semantic matching, and cross-space technical-signal analysis. Together, these analyses examine \emph{what} maintenance and support concerns arise across Galaxy, \emph{how} their resolution varies, and \emph{how} problem and solution knowledge is connected across development and support spaces. Based on these objectives, we formulate the following research questions:

\begin{itemize}
    \item \textbf{RQ1:} What types of maintenance and support concerns emerge across Galaxy GitHub issues, pull requests, and Community Forum discussions?
    \item \textbf{RQ2:} What factors are associated with resolution outcomes and
    resolution time across Galaxy GitHub issues, pull requests, and Community
    Forum discussions?
    \item \textbf{RQ3:} How are maintenance and support artifacts connected across repository-centered development spaces and community-centered support spaces?
\end{itemize}

Our results provide three complementary views of Galaxy maintenance and support. First, using BERTopic modeling \cite{DBLP:journals/corr/abs-2203-05794}, we identify nine issue topics, 14 pull-request topics, and 14 forum topics, showing that maintenance spans workflow execution, data and history management, tools and dependencies, distributed infrastructure, testing, deployment, scientific resources, documentation, and user-facing analytical support. Second, we find that resolution is multidimensional: coordination, diagnostic, contributor, automation, and engagement characteristics exhibit different associations with resolution outcomes and resolution speed across the three artifact types. Third, we find limited explicit traceability between development and community-support spaces. Among 16,426 resolved explicit relationships, 97.77\% occur between GitHub artifacts, whereas only 294 directly bridge GitHub and the Galaxy Community Help Forum. Nevertheless, candidate semantic connections and recurring technical concerns indicate that related maintenance knowledge is distributed across these spaces even when explicit links are absent. Together, these findings provide an integrated empirical view of Galaxy maintenance and support across whole lifecycles.

Based on these findings, the study makes four major contributions:

\begin{itemize}

    \item \textbf{An ecosystem-level characterization of Galaxy maintenance and support.} We provide a multi-source analysis of maintenance and support concerns across GitHub issues, pull requests, and Community Forum discussions, connecting developer-centered maintenance activities with user-facing support.

    \item \textbf{A multidimensional analysis of maintenance resolution.} We examine both resolution outcomes and resolution time across issues, pull requests, and forum discussions, identifying coordination, diagnostic, contributor, automation, and engagement characteristics associated with their resolution lifecycles.

    \item \textbf{An empirical analysis of cross-space traceability.} We quantify explicit relationships among development and support artifacts and complement them with candidate semantic connections and shared technical signals, revealing where related maintenance knowledge exists without explicit links.

    \item \textbf{Empirically grounded recommendations for Galaxy and implications for community-driven scientific software ecosystems.} We translate our combined findings into practical recommendations for Galaxy and broader implications for improving diagnostic reporting, lifecycle-aware triage, selective automation, cross-space traceability, and the reuse of community-support knowledge.

\end{itemize}

Overall, this study provides an ecosystem-level view of Galaxy maintenance as a distributed socio-technical process rather than a collection of isolated software changes. By examining maintenance concerns, resolution behavior, and cross-space connections together, we provide a more comprehensive account of how a mature scientific workflow ecosystem is sustained and identify opportunities to make maintenance knowledge more visible, traceable, and reusable.

\noindent \textbf{Replication Package: }To support transparency and reproducibility, the replication package for this study is publicly available and can be accessed via \cite{alam_2026_22882299}.

\section{Background and Related Work}
\label{sec:background}
\subsection{Scientific Workflows and Scientific Workflow Systems}
Scientific workflows represent multi-step computational analyses as connected tasks with explicit data and control dependencies~\cite{deelman2009workflows,DBLP:journals/concurrency/LudascherABHJJLTZ06,talia2013workflow}. They are widely used to coordinate analyses involving multiple tools, parameters, datasets, and intermediate outputs, including applications such as RNA-seq analysis~\cite{DBLP:journals/nar/GruningFYWEEHBV17} and metagenomics~\cite{batut2018asaim}. By making analysis steps and their dependencies explicit, workflows can support the repeatability, inspection, sharing, and reuse of computational experiments~\cite{DBLP:journals/fgcs/BoulakiaBCCFGHL17}.

An SWS is a specialized software platform designed to automate, manage, and execute complex sequences of computational or data processing tasks, often referred to as scientific workflows, in scientific research whose execution order is driven by a computerized representation of the workflow logic \cite{DBLP:journals/tsc/LinLFCPLFH09, DBLP:journals/ese/AlamRRM25}. SWSs provide the infrastructure for defining, executing, monitoring, and managing scientific workflows. Their responsibilities often extend beyond workflow orchestration to include data management, provenance tracking, resource allocation, dependency management, and execution across heterogeneous computing environments~\cite{DBLP:journals/fgcs/DeelmanGST09,DBLP:journals/fgcs/SilvaFPJSD17}. Because scientific workflows depend on evolving tools, software packages, data formats, reference resources, and execution environments, their continued operation requires ongoing adaptation to changes in the surrounding software and computational ecosystem.

In Galaxy, these dependencies are distributed across several interconnected components. Beyond the core Galaxy platform, the ecosystem includes scientific tool wrappers and the ToolShed~\cite{blankenberg2014dissemination}, workflow resources, software dependencies, reference data, distributed execution infrastructure, documentation, and training resources. Galaxy also treats datasets and histories as first-class analysis objects and supports workflow execution across heterogeneous computing environments~\cite{goecks2010galaxy,DBLP:journals/nar/AfganNGBGSOMLSF22,galaxy2024galaxy,galaxy2026galaxy}. Consequently, maintenance may involve changes not only to the core platform but also to tools, workflows, dependencies, infrastructure, scientific resources, and supporting documentation. This broader ecosystem provides the technical context for the maintenance and support artifacts examined in this study.

\subsection{Galaxy as a Community-Driven Scientific Workflow Ecosystem}
Galaxy is an open-source, web-based platform designed to make computational analyses accessible, reproducible, and transparent \cite{goecks2010galaxy, DBLP:journals/nar/AfganNGBGSOMLSF22}. It allows researchers to execute computational tools through a graphical interface, organize datasets within histories, combine tools into reusable workflows, preserve provenance, share analyses, and execute workflows across different computational infrastructures  without requiring extensive programming or systems-administration expertise \cite{DBLP:journals/nar/AfganNGBGSOMLSF22, galaxy2024galaxy,galaxy2026galaxy}.

The Galaxy ecosystem extends beyond the core platform and includes the Galaxy framework, thousands of integrated scientific tools, workflow repositories, public and institutional Galaxy deployments, training resources, documentation, APIs, and infrastructure for distributed execution~\cite{DBLP:journals/nar/AfganNGBGSOMLSF22,galaxy2024galaxy,galaxy2026galaxy}. Community initiatives such as the Intergalactic Workflow Commission~\cite{galaxyIWC2026} curate and test reusable workflows, while the Galaxy Training Network~\cite{galaxyTraining2026} develops training materials for scientific and technical users. Development itself is distributed across an international community involving software developers, tool developers, workflow authors, system administrators, trainers, contributors, and researchers~\cite{galaxy2026galaxy,DBLP:journals/nar/AfganNGBGSOMLSF22}.

This distributed structure creates multiple forms of maintenance. Core developers evolve the Galaxy framework and user interface; tool developers maintain wrappers and dependencies; infrastructure maintainers address deployment, job execution, storage, and authentication concerns; workflow developers adapt workflows to changing tools and data requirements; and contributors maintain tests, documentation, training resources, and configuration infrastructure. Changes in one part of the ecosystem may also propagate to others. For example, a tool update may require corresponding changes to workflows and tests, while infrastructure changes may affect workflow execution across individual Galaxy deployments.

Galaxy also provides several channels through which maintenance and support activities occur. GitHub issues are used to report defects, propose enhancements, discuss technical problems, and coordinate maintenance tasks. Pull requests capture implementation and integration activities, including bug fixes, feature development, refactoring, dependency updates, testing, documentation changes, and infrastructure modifications. In parallel, the Galaxy Community Help Forum~\cite{galaxycommunityhelp} provides a community-facing space where users seek assistance with workflow execution, tools, datasets, histories, storage, accounts, configuration, reference resources, and scientific analyses. Together, these artifacts provide complementary perspectives on the Galaxy ecosystem: issues expose reported problems and maintenance needs, pull requests document implementation and integration activities, and forum discussions reveal operational and analytical problems encountered by users in practice.

\subsection{Development and Maintenance of Scientific Workflows and SWSs}

Prior research on scientific workflows has largely focused on reproducibility, portability, execution, reuse, and the technical challenges of workflow development~\cite{DBLP:journals/fgcs/DeelmanGST09,DBLP:journals/fgcs/SilvaFPJSD17,DBLP:journals/fgcs/SuterCABBCCDTFGJKKLMOPP26,gruning2018practical}. This literature shows that workflow sustainability depends not only on workflow logic but also on software versions, dependencies, input data, parameters, execution environments, provenance, and the surrounding infrastructure. For Galaxy workflows, prior empirical work has further shown that reuse can be disrupted by tool upgrades, unavailable tools, incomplete workflows, design problems, and workflow-loading failures~\cite{DBLP:conf/apsec/AlamRS23}.

More recent empirical studies have examined how such dependencies manifest in SWS development and maintenance. Alam et al.~\cite{DBLP:journals/ese/AlamRRM25} analyzed Stack Overflow and GitHub discussions across multiple SWSs and identified recurring concerns involving workflow execution, data operations, dependencies, documentation, task management, scheduling, and system evolution. Alam and Roy~\cite{DBLP:conf/saner/AlamR26} subsequently analyzed more than 21,000 GitHub issues across SWSs and showed that issue-management practices, including labeling and assignment, are associated with resolution behavior. More focused work on the nf-core ecosystem examined GitHub issues and pull requests and identified maintenance activities involving pipeline development, integration, testing, continuous integration, bug fixing, containerized execution, and version updates~\cite{DBLP:conf/msr/AlamR26}.

Together, these studies provide important evidence about workflow-development challenges and repository-centered maintenance, but they offer a limited view of how maintenance and support operate across the broader ecosystem of a mature SWS. This limitation is particularly relevant to Galaxy, which combines a workflow system, graphical analysis environment, extensive tool ecosystem, public computational services, distributed execution infrastructure, documentation, training resources, and a heterogeneous user and developer community. Maintenance therefore extends beyond workflows and source code to include tool integration, dependencies, infrastructure, scientific resources, and user-facing support. Our study extends this perspective by jointly examining GitHub issues, pull requests, and Galaxy Community Help Forum discussions to characterize maintenance and support concerns, their resolution, and the relationships among maintenance knowledge distributed across development and community-support spaces.

\subsection{Issue Tracking, Pull-Based Development, Community Support, and Cross-Space Knowledge}
\label{sec:background_channels}

Software maintenance is increasingly distributed across multiple collaborative artifacts and communication spaces. Issue trackers provide a primary mechanism for reporting defects, requesting enhancements, documenting diagnostic information, and coordinating work among users and developers~\cite{DBLP:conf/icse/AnvikHM06,DBLP:journals/tse/ZimmermannPBJSW10,DBLP:conf/icse/ZhangGV13}. Prior research has shown that the content and quality of issue reports influence developers' ability to understand and diagnose reported problems. Information such as reproduction steps, stack traces, test cases, and contextual details can improve the usefulness of bug reports, while missing information often results in additional interaction between reporters and developers~\cite{DBLP:journals/tse/ZimmermannPBJSW10,DBLP:conf/cscw/BreuPSZ10}. Issue characteristics and management practices can consequently influence how maintenance tasks are triaged, coordinated, and resolved.

Pull requests complement issue tracking by capturing the implementation, review, testing, and integration of proposed software changes. Whereas issues commonly formalize problems, requests, and maintenance needs, pull requests provide evidence of the concrete changes through which those needs may be addressed. Studies of pull-based development show that pull-request outcomes and processing time are associated with both technical and social factors, including contribution characteristics, contributor experience, discussion, testing, review activity, and prior interactions~\cite{DBLP:conf/icse/GousiosPD14,DBLP:conf/icse/GousiosZSD15,tsay2014influence,DBLP:journals/tse/ZhangYGR23}. Issues and pull requests therefore capture complementary stages of maintenance, ranging from problem reporting and coordination to implementation, evaluation, and integration.

Maintenance knowledge, however, is not confined to software repositories. Developers and users also rely on Q\&A sites, forums, mailing lists, and other discussion platforms to report problems, seek guidance, exchange workarounds, and share technical knowledge~\cite{DBLP:journals/ese/BaruaTH14,DBLP:journals/tse/StoreyZFSG17,DBLP:conf/msr/ZagalskyTGSP16,DBLP:journals/ese/HataNBKT22}. Studies of developer Q\&A communities show that such platforms can accumulate reusable knowledge around programming problems, conceptual questions, troubleshooting practices, and software usage~\cite{DBLP:conf/icse/TreudeBS11}. Research on open-source user forums similarly demonstrates that end users play an active role in initiating discussions and contributing responses, providing a user-centered perspective on problems encountered in practice~\cite{DBLP:conf/icse-chase/HellmanCUCG22}. Discussion platforms may therefore capture questions, emerging problems, contextual information, and support interactions that are not yet---and may never become---formal issues or pull requests~\cite{DBLP:journals/ese/HataNBKT22}.

The resulting maintenance knowledge is consequently distributed across heterogeneous development and support spaces. Studies comparing software-development communication channels show that different platforms serve complementary purposes and that relevant knowledge may be fragmented across repositories, Q\&A systems, and communication platforms~\cite{DBLP:conf/socialcom/VasilescuFS13,DBLP:conf/msr/ZagalskyTGSP16,DBLP:journals/tse/StoreyZFSG17}. This fragmentation also creates a traceability challenge: artifacts discussing the same software problem or change may exist in different spaces without an explicit connection between them. Prior research has demonstrated that linking communication artifacts to software artifacts is non-trivial because discussions are often informal and explicit references are incomplete; information-retrieval and textual-similarity techniques have therefore been explored to recover potentially related artifacts~\cite{DBLP:conf/icse/BacchelliLR10}. Preserving or recovering such relationships can make it easier to connect problem context, implementation activity, and reusable technical knowledge across otherwise separate information sources.

This cross-space perspective is particularly important for Galaxy. User-facing problems are often shaped not only by source-code defects but also by the broader analytical and computational context in which Galaxy operates. Community Forum discussions may concern workflow invocation, failed jobs, tool behavior, dataset uploads, histories and collections, reference genomes, visualization outputs, storage limitations, authentication, server access, or the behavior of particular public Galaxy deployments. Some of these concerns may subsequently become formal GitHub issues or motivate pull requests, whereas others may remain within the support space because their resolution depends on a user's data, workflow configuration, server, or computational environment. Restricting analysis to GitHub issues and pull requests would therefore capture repository-centered maintenance while overlooking part of the user-facing operational and analytical context in which maintenance needs emerge.

We consequently examine GitHub issues, pull requests, and Galaxy Community Help Forum discussions as complementary maintenance and support artifacts. This combined perspective allows us to characterize how maintenance concerns manifest across problem reporting, implementation, and user support; examine how their resolution differs across these spaces; and investigate whether related maintenance knowledge remains explicitly connected or can be identified through cross-space relatedness. Together, these perspectives motivate an integrated analysis of repository-centered development and community-centered support to understand how maintenance concerns, resolution processes, and related knowledge are distributed across the Galaxy ecosystem.

\subsection{Topic Modeling for Maintenance and Support Analysis}

Software repositories and community-support platforms contain large volumes of unstructured textual data, including issue descriptions, pull-request discussions, error reports, and user questions. Topic modeling has been widely used in software-engineering research to uncover recurring themes in such data, particularly in developer communication, issue and bug reports, and repository artifacts~\cite{DBLP:journals/ese/BaruaTH14,DBLP:conf/msr/Treude019,DBLP:journals/ese/SilvaGG21}. Prior studies have used topic models to characterize developer concerns, software problems, and maintenance activities across GitHub, Stack Overflow, and other software-development communities~\cite{DBLP:journals/ese/HanSWDX20,DBLP:conf/msr/YangWSHKLXL23,DBLP:journals/ese/AlamRRM25}.

Latent Dirichlet Allocation (LDA) is one of the most widely used traditional topic-modeling techniques~\cite{DBLP:journals/jmlr/BleiNJ03}. It represents documents as mixtures of latent topics and topics as distributions over words, making it useful for discovering recurring lexical patterns in large document collections. However, prior software-engineering studies have shown that LDA-based results can be sensitive to parameter settings, corpus characteristics, and random initialization, and that topic quality may vary when applied to short, heterogeneous, and technically specialized software artifacts~\cite{DBLP:journals/infsof/AgrawalFM18,DBLP:journals/ese/AlamRRM25}. These challenges are particularly relevant to maintenance and support data, where semantically related discussions may use different terminology and where documents may contain short descriptions, technical identifiers, logs, code fragments, and informal language \cite{DBLP:conf/msr/AlamR26, DBLP:journals/ese/AlamRRM25}.

More recent embedding-based topic-modeling approaches seek to capture semantic similarity beyond direct word co-occurrence. BERTopic~\cite{DBLP:journals/corr/abs-2203-05794} combines transformer-based document embeddings with dimensionality reduction and density-based clustering, and represents discovered topics using class-based TF--IDF. By relying on contextual embeddings, BERTopic can group semantically related documents even when they share relatively few surface-level terms, while its clustering-based formulation does not require the number of topics to be fixed in advance. These properties make it suitable for heterogeneous software-maintenance corpora containing short and domain-specific textual artifacts.

Topic modeling is particularly useful for studying maintenance across multiple development and support spaces because different artifact types capture distinct perspectives on the same ecosystem. GitHub issues primarily represent reported problems and maintenance needs, pull requests capture implementation and integration activities, and community-support discussions expose operational and analytical difficulties encountered by users. In this study, we therefore apply BERTopic independently to Galaxy GitHub issues, pull requests, and Community Forum discussions, preserving their distinct communicative roles while enabling comparison of recurring concerns across channels. This analysis provides an empirical basis for characterizing the breadth of Galaxy maintenance and identifying concerns that recur across, or are specific to, development and support spaces.

\section{Study Design}
\label{sec:study-design}

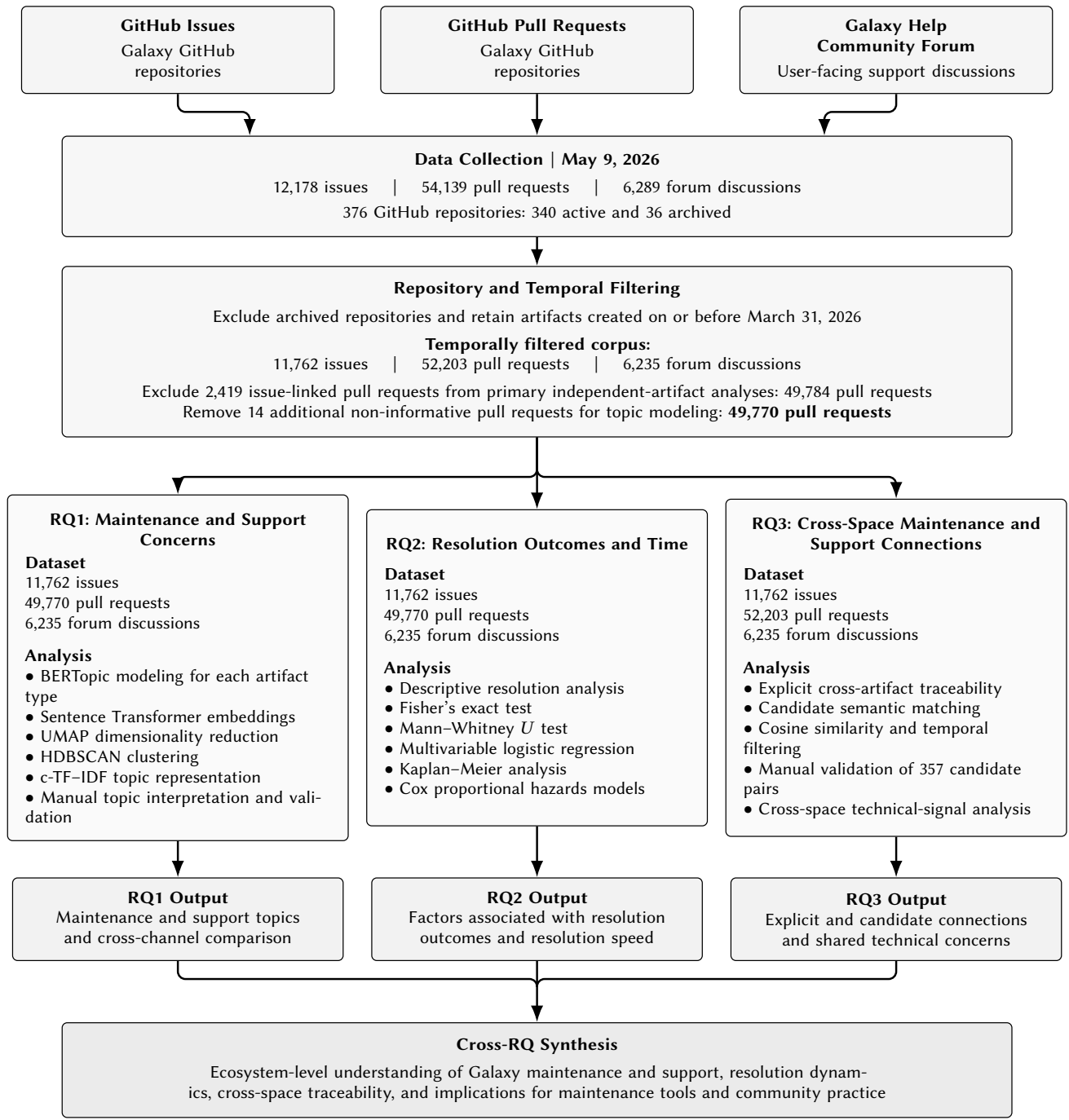
\begin{figure*}[t]
\centering
\resizebox{\textwidth}{!}{%
\begin{tikzpicture}[
    font=\sffamily\footnotesize,
    >=Latex,
    source/.style={
        draw,
        rounded corners=2pt,
        align=center,
        text width=4.0cm,
        minimum height=1.2cm,
        fill=black!4,
        inner sep=5pt
    },
    process/.style={
        draw,
        rounded corners=2pt,
        align=center,
        text width=12.8cm,
        minimum height=1.2cm,
        fill=black!3,
        inner sep=6pt
    },
    rq/.style={
        draw,
        rounded corners=2pt,
        align=left,
        text width=4.25cm,
        minimum height=4.4cm,
        fill=black!2,
        inner sep=7pt
    },
    output/.style={
        draw,
        rounded corners=2pt,
        align=center,
        text width=4.25cm,
        minimum height=1.15cm,
        fill=black!5,
        inner sep=5pt
    },
    synthesis/.style={
        draw,
        rounded corners=2pt,
        align=center,
        text width=12.8cm,
        minimum height=1.15cm,
        fill=black!8,
        inner sep=6pt
    },
    arrow/.style={
        ->,
        thick,
        rounded corners=3pt
    },
    connector/.style={
        thick,
        rounded corners=3pt
    }
]

% =========================================================
% Data Sources
% =========================================================

\node[source] (issues) at (-5,0) {
    \textbf{GitHub Issues}\\[2pt]
    Galaxy GitHub\\
    repositories
};

\node[source] (pullrequests) at (0,0) {
    \textbf{GitHub Pull Requests}\\[2pt]
    Galaxy GitHub\\
    repositories
};

\node[source] (forum) at (5,0) {
    \textbf{Galaxy Help}\\
    \textbf{Community Forum}\\[2pt]
    User-facing support discussions
};

% =========================================================
% Data Collection
% =========================================================

\node[process] (collection) at (0,-1.9) {
    \textbf{Data Collection $\vert$ May 9, 2026}\\[3pt]
    12,178 issues
    \quad $\vert$ \quad
    54,139 pull requests
    \quad $\vert$ \quad
    6,289 forum discussions\\[2pt]
    376 GitHub repositories:
    340 active and 36 archived
};

\draw[arrow]
    (issues.south) -- ++(0,-0.25)
    -| ([xshift=-4cm]collection.north);

\draw[arrow]
    (pullrequests.south) -- (collection.north);

\draw[arrow]
    (forum.south) -- ++(0,-0.25)
    -| ([xshift=4cm]collection.north);

% =========================================================
% Filtering and Dataset Construction
% =========================================================

\node[process] (filtering) at (0,-4.2) {
    \textbf{Repository and Temporal Filtering}\\[3pt]
    Exclude archived repositories and retain artifacts created
    on or before March 31, 2026\\[3pt]

    \textbf{Temporally filtered corpus:}\\
    11,762 issues
    \quad $\vert$ \quad
    52,203 pull requests
    \quad $\vert$ \quad
    6,235 forum discussions\\[3pt]

    Exclude 2,419 issue-linked pull requests from
    primary independent-artifact analyses:
    49,784 pull requests\\
    Remove 14 additional non-informative pull requests for topic modeling:
    \textbf{49,770 pull requests}
};

\draw[arrow]
    (collection.south) -- (filtering.north);

% =========================================================
% RQ1
% =========================================================

\node[rq] (rq1) at (-5.0,-8.6) {
    \centering
    \textbf{RQ1: Maintenance and Support Concerns}\\
    \raggedright
    \vspace{1pt}

    \textbf{Dataset}\\
    11,762 issues\\
    49,770 pull requests\\
    6,235 forum discussions

    \vspace{5pt}
    \textbf{Analysis}\\
    $\bullet$ BERTopic modeling for each artifact type\\
    $\bullet$ Sentence Transformer embeddings\\
    $\bullet$ UMAP dimensionality reduction\\
    $\bullet$ HDBSCAN clustering\\
    $\bullet$ c-TF--IDF topic representation\\
    $\bullet$ Manual topic interpretation and validation
};

% =========================================================
% RQ2
% =========================================================

\node[rq] (rq2) at (0,-8.6) {
    \centering
    \textbf{RQ2: Resolution Outcomes and Time}\\
    \raggedright
    \vspace{4pt}

    \textbf{Dataset}\\
    11,762 issues\\
    49,770 pull requests\\
    6,235 forum discussions

    \vspace{5pt}
    \textbf{Analysis}\\
    $\bullet$ Descriptive resolution analysis\\
    $\bullet$ Fisher's exact test\\
    $\bullet$ Mann--Whitney $U$ test\\
    $\bullet$ Multivariable logistic regression\\
    $\bullet$ Kaplan--Meier analysis\\
    $\bullet$ Cox proportional hazards models
};

% =========================================================
% RQ3
% =========================================================

\node[rq] (rq3) at (5,-8.6) {
    \centering
    \textbf{RQ3: Cross-Space Maintenance and Support Connections}\\
    \raggedright
    \vspace{4pt}

    \textbf{Dataset}\\
    11,762 issues\\
    52,203 pull requests\\
    6,235 forum discussions

    \vspace{5pt}
    \textbf{Analysis}\\
    $\bullet$ Explicit cross-artifact traceability\\
    $\bullet$ Candidate semantic matching\\
    $\bullet$ Cosine similarity and temporal filtering\\
    $\bullet$ Manual validation of 357 candidate pairs\\
    $\bullet$ Cross-space technical-signal analysis
};

% =========================================================
% Angular Arrows to Research Questions
% =========================================================

\coordinate (rqsplit) at ($(filtering.south)+(0,-0.55)$);
\coordinate (rq1turn) at (rq1.north |- rqsplit);
\coordinate (rq3turn) at (rq3.north |- rqsplit);

% Left arrow made symmetric with right arrow
\draw[arrow]
    (filtering.south) -- (rqsplit) -| (rq1.north);

% Center arrow
\draw[arrow]
    (filtering.south) -- (rq2.north);

% Right arrow
\draw[arrow]
    (filtering.south) -- (rqsplit) -| (rq3.north);

% =========================================================
% Outputs
% =========================================================

\node[output] (out1) at (-5,-12.1) {
    \textbf{RQ1 Output}\\
    Maintenance and support topics\\
    and cross-channel comparison
};

\node[output] (out2) at (0,-12.1) {
    \textbf{RQ2 Output}\\
    Factors associated with resolution\\
    outcomes and resolution speed
};

\node[output] (out3) at (5,-12.1) {
    \textbf{RQ3 Output}\\
    Explicit and candidate connections\\
    and shared technical concerns
};

\draw[arrow] (rq1.south) -- (out1.north);
\draw[arrow] (rq2.south) -- (out2.north);
\draw[arrow] (rq3.south) -- (out3.north);

% =========================================================
% Cross-RQ Synthesis
% =========================================================

\node[synthesis] (synthesis) at (0,-14.2) {
    \textbf{Cross-RQ Synthesis}\\[3pt]
    Ecosystem-level understanding of Galaxy maintenance and support,
    resolution dynamics, cross-space traceability, and implications
    for maintenance tools and community practice
};

% =========================================================
% Merge outputs correctly, then one arrow to synthesis
% =========================================================

\coordinate (mergecenter) at ($(synthesis.north)+(0,0.55)$);
\coordinate (mergeleft) at (out1.south |- mergecenter);
\coordinate (mergeright) at (out3.south |- mergecenter);

% Plain connector lines from outputs
\draw[connector] (out1.south) -- ++(0,-0.25) -| (mergecenter);
\draw[connector] (out2.south) -- (mergecenter);
\draw[connector] (out3.south) -- ++(0,-0.25) -| (mergecenter);

% Single correct direction arrow to synthesis
\draw[arrow] (mergecenter) -- (synthesis.north);

\end{tikzpicture}%
}

\caption{Overview of the study design. We collect maintenance and support
artifacts from Galaxy GitHub repositories and the Galaxy Help Community
Forum, apply repository and temporal filtering, and analyze the resulting
artifacts through three complementary research-question-specific pipelines.
}

\label{fig:study_design}
\end{figure*}

We design a multi-artifact empirical study to investigate maintenance and support in the Galaxy ecosystem from complementary development and community-support perspectives. Our analysis integrates \emph{11,762 GitHub issues}, \emph{52,203 pull requests}, and \emph{6,235 Galaxy Community Help Forum discussions}. These artifacts capture different stages and perspectives of maintenance: issues document reported problems and requested changes, pull requests record implementation and integration activities, and forum discussions reflect user-facing operational and analytical support needs. We use these data to characterize recurring maintenance and support concerns (\textbf{RQ1}), examine factors associated with resolution outcomes and resolution time (\textbf{RQ2}), and investigate explicit and candidate connections among maintenance and support artifacts across development and support spaces (\textbf{RQ3}). Figure~\ref{fig:study_design} summarizes the overall study design.

\subsection{Data Collection}
We collected maintenance and support artifacts from complementary data sources within the Galaxy ecosystem: GitHub issues and pull requests from the Galaxy Project, UseGalaxy.eu, and UseGalaxy.org.au repositories\footnote{\url{https://github.com/galaxyproject}, \url{https://github.com/usegalaxy-eu}, and \url{https://github.com/usegalaxy-au}.} and discussions from the Galaxy Help Community Forum\footnote{\url{https://help.galaxyproject.org/}}. The initial data collection was conducted on May 9, 2026. At that time, we retrieved 12,178 GitHub issues, 54,139 pull requests, and 6,289 forum discussions. The GitHub artifacts were distributed across 376 repositories, comprising 340 active and 36 archived repositories. Because archived repositories are no longer actively maintained, we excluded them from subsequent analyses and focused on the 340 active repositories.

We collected GitHub issues and pull requests using the \textit{GitHub REST API}. For the Galaxy Help Community Forum, we developed a custom Python-based data-collection script to systematically retrieve the available discussion records and their associated metadata.

To reduce the influence of artifacts created immediately before data collection, we introduced an observation buffer between the artifact-creation cutoff and the collection date. Although the data were collected on May 9, 2026, we retained only artifacts created on or before March 31, 2026. This cutoff provided approximately five weeks for recently created artifacts to receive responses, undergo maintenance activity, or reach an observable resolution state. The buffer reduces immediate observation truncation but does not eliminate right-censoring; unresolved artifacts are therefore handled explicitly as censored observations in the time-to-event analyses described in Section~\ref{sec:resolution-analysis}.

After applying the repository-status and temporal filters, the dataset contained 11,762 issues from active Galaxy repositories with issue activity, 52,203 pull requests from active repositories with pull-request activity, and 6,235 forum discussions created on or before March 31, 2026. These artifacts formed the temporally filtered corpus from which the RQ-specific analysis datasets were subsequently constructed.

We then examined explicit overlap between GitHub issues and pull requests. A pull request may be directly associated with an existing issue and therefore represent implementation activity corresponding to the same underlying maintenance task. For analyses intended to treat issues and pull requests as distinct maintenance artifacts, such explicit associations may introduce overlapping representations of the same maintenance activity. Following prior work~\cite{DBLP:conf/msr/AlamR26, DBLP:journals/ese/AlamRRM25}, we used the \textit{GitHub Search API} to identify pull requests explicitly linked to existing issues. This procedure identified 2,419 linked pull requests, which were excluded from the primary pull-request dataset for analyses requiring independent artifact observations.

After this filtering step, the primary analysis dataset comprised \emph{11,762 issues}, \emph{49,784 pull requests}, and \emph{6,235 Community Forum discussions}. We use this dataset for \textbf{RQ1} and \textbf{RQ2} analyses in which issues, pull requests, and forum discussions are treated as distinct units of observation. For \textbf{RQ3}, however, we retained the complete set of 52,203 temporally filtered pull requests, including the 2,419 pull requests explicitly linked to issues. These associations constitute direct evidence of connections among maintenance artifacts and are therefore essential for examining cross-artifact traceability.

\subsection{Data preprocessing}

We adopted separate preprocessing strategies for the textual content used for topic modeling and the metadata, structural attributes, and technical signals used for the resolution and cross-space connection analyses. This distinction is important because the analyses impose different data requirements. Topic modeling benefits from removing repetitive markup, boilerplate, and syntactic noise that may obscure the semantic structure of maintenance discussions. In contrast, \textbf{RQ2} and \textbf{RQ3} require preserving diagnostic and coordination indicators such as code fragments, URLs, error descriptions, configuration information, and other forms of technical evidence.

For topic modeling, we constructed a single textual document for each artifact. For GitHub issues and pull requests, the document was formed by concatenating the artifact title and body. For Community Forum discussions, we combined the discussion title with the rendered discussion content. We removed HTML markup, URLs, code fragments, command-line snippets, stack traces, configuration blocks, and other structured technical fragments. Although such content can be useful for diagnosing individual problems, retaining it directly in the topic-modeling corpus can cause topic representations to be disproportionately influenced by frequently repeated file paths, commands, environment variables, configuration syntax, and other implementation-specific tokens rather than by the underlying maintenance or support concern.

We converted the remaining natural-language text to lowercase, removed punctuation and non-alphabetic symbols, and filtered standard English stopwords using an established stopword list~\cite{hardeniya2016natural}. We also removed corpus-specific high-frequency terms, including \emph{galaxy}, \emph{usegalaxy}, and \emph{galaxyproject}, because these terms characterize the overall study context but contribute little to distinguishing among maintenance and support topics. Finally, we lemmatized the text using the \texttt{spaCy} \texttt{en\_core\_web\_sm} model~\cite{vasiliev2020natural} to normalize inflected word forms and reduce lexical sparsity.

Removing technical fragments from the topic-modeling text did not mean discarding their diagnostic information from the study. Before textual cleaning, we separately derived and retained structural and technical indicators from the original artifacts. These included the presence of code blocks, URLs, error-related expressions, version information, command-line content, system-information descriptions, container-related terms, cloud-related terms, high-performance-computing (HPC) terms, and reproduction-related information. These indicators were subsequently used in \textbf{RQ2} and \textbf{RQ3} to examine their associations with resolution and their prevalence across maintenance and support spaces.

Following preprocessing, we assessed whether each artifact retained sufficient textual information for topic modeling. All selected GitHub issues and Community Forum discussions contained meaningful textual content after preprocessing. In contrast, 14 pull requests were identified as non-informative. These pull requests had empty body fields and titles that either consisted solely of numeric values or contained vague, non-descriptive phrases, such as \emph{pull request}, \emph{bump}, and \emph{you can do it}. Because these records did not provide sufficient semantic information from which a maintenance topic could be meaningfully inferred, we excluded them from the topic-modeling corpus. This filtering reduced the PR corpus from 49,784 to \emph{49,770 pull requests}.

The final corpus used for topic modeling therefore comprised \emph{67,767 artifacts}: \emph{11,762 GitHub issues}, \emph{49,770 pull requests}, and \emph{6,235 Galaxy Community Help Forum discussions}. The resulting textual representations were used exclusively for topic discovery and interpretation.

For the resolution and cross-space connection analyses, we retained the original artifact metadata and relevant structural attributes rather than relying only on the cleaned textual representations. Depending on artifact type, these attributes included labels, assignees, comments, milestones, requested reviewers, linked issues, draft status, author information, reply counts, accepted-answer status, timestamps, and diagnostic elements such as code blocks and URLs.

\subsection{Identifying Maintenance and Support Topics}
\label{sec:topic-modeling}

To answer \textbf{RQ1}, we applied BERTopic~\cite{DBLP:journals/corr/abs-2203-05794} independently to the GitHub issue, pull-request, and Community Forum corpora. We modeled the three artifact types separately because they serve different roles in the Galaxy ecosystem: issues primarily capture reported problems and maintenance coordination, pull requests document implementation and review activities, and forum discussions reflect user-facing troubleshooting and support. Separate modeling allows each corpus to develop its own topic structure without forcing heterogeneous artifacts into a shared clustering structure.

BERTopic combines transformer-based document representations with dimensionality reduction, density-based clustering, and class-based term weighting to identify semantically related groups of documents~\cite{DBLP:journals/corr/abs-2203-05794}. This approach is appropriate for our dataset because GitHub and community-support artifacts often contain short descriptions, domain-specific terminology, and lexically different expressions of related maintenance concerns. We generated document embeddings using the \texttt{all-mpnet-base-v2} Sentence Transformer model~\cite{DBLP:conf/emnlp/ReimersG19,sentence_transformers_huggingface,huggingface_models}. The model builds on the MPNet architecture~\cite{DBLP:conf/nips/Song0QLL20} and maps textual documents into 768-dimensional dense representations suitable for semantic comparison and clustering. We selected \texttt{all-mpnet-base-v2} because the corpus contains short and technically specialized descriptions in which semantically related maintenance concerns may share relatively few surface-level terms.

We reduced the high-dimensional embeddings using Uniform Manifold Approximation and Projection (\textit{UMAP})~\cite{DBLP:journals/corr/abs-1802-03426}. UMAP reduces dimensionality while preserving neighborhood relationships among semantically similar documents. We then applied \textit{HDBSCAN}~\cite{DBLP:journals/jossw/McInnesHA17} to the UMAP-reduced representations. HDBSCAN does not require the number of clusters to be specified in advance, can identify clusters of varying densities, and can distinguish documents that do not strongly belong to any cluster. Together, the embedding--UMAP--HDBSCAN pipeline enabled us to derive topic structures from the semantic characteristics of each corpus rather than imposing a predefined taxonomy.

We tuned the BERTopic configuration independently for issues, pull requests, and forum discussions because the three corpora differ substantially in size, textual structure, and thematic diversity. We explored UMAP \texttt{n\_neighbors} values between 15 and 60 and \texttt{n\_components} values between 5 and 40. We used cosine distance during UMAP dimensionality reduction to remain consistent with the semantic geometry of the Sentence Transformer embeddings. For HDBSCAN, we varied \texttt{min\_cluster\_size} between 30 and 300 and used Euclidean distance on the UMAP-reduced representations. Corpus-specific tuning allowed us to balance topic separation, granularity, and interpretability.

We assessed candidate configurations using topic coherence together with the interpretability and distinctiveness of the resulting topic structures. Topic coherence provides a quantitative indication of whether highly ranked terms within a topic form a semantically meaningful concept~\cite{DBLP:conf/wsdm/RoderBH15}. Because a configuration with a high coherence score does not necessarily provide the level of granularity most useful for an empirical maintenance study, we considered coherence alongside substantive interpretation of the discovered clusters. This follows established software-engineering topic-modeling practice, where quantitative topic-quality measures are complemented by contextual inspection of representative terms and documents~\cite{DBLP:journals/ese/AlamRRM25,DBLP:conf/icsm/aounLKO21,DBLP:conf/msr/Abdellatif0BAS20,DBLP:conf/msr/AlamR26}.

For topic representation, we configured \texttt{CountVectorizer} to consider unigram and bigram features. Including bigrams preserves domain-relevant multiword expressions frequently appearing in Galaxy maintenance discussions, such as \emph{workflow execution}, \emph{tool installation}, \emph{reference genome}, and \emph{dependency update}. BERTopic then used these features to construct c-TF--IDF representations and rank the terms most characteristic of each discovered topic.

After corpus-specific parameter tuning, the final BERTopic models identified \textbf{nine topics from GitHub issues}, \textbf{14 topics from pull requests}, and \textbf{14 topics from Galaxy Community Help Forum discussions}.

\paragraph{\textbf{Manual topic validation.}}
We validated the topic interpretations through structured expert inspection. For each topic, the first author examined the representative terms and manually reviewed at least 25 representative artifacts. The first author assigned an initial topic label based on recurring technical concerns, artifact context, and representative examples. Another author with experience in scientific workflow research subsequently reviewed the proposed labels and interpretations. Ambiguous, overlapping, or borderline cases were discussed and the topic labels refined through repeated inspection of representative terms and artifacts until consensus was reached. This procedure was applied consistently across the issue, pull-request, and forum corpora.

\subsection{Analyzing Factors Associated with Resolution Outcomes and Time}
\label{sec:resolution-analysis}

To answer \textbf{RQ2}, we examined which characteristics of Galaxy maintenance and support artifacts were associated with both resolution outcomes and resolution time. We analyzed GitHub issues, pull requests, and Community Forum discussions separately because resolution has a different operational meaning for each artifact type. Examining both dimensions allows us to distinguish characteristics associated with whether an observable resolution occurred from those associated with the pace at which artifacts progressed toward resolution.

\paragraph{\textbf{Resolution outcomes.}}
For GitHub issues, we defined resolution as issue closure and measured resolution time as the number of days between issue creation and closure. Issues that remained open at the end of the observation period were retained in the time-to-event analysis as right-censored observations.

For pull requests, we distinguished among merged pull requests, pull requests closed without merging, and pull requests that remained open. We examined merging as the primary integration outcome and separately analyzed closure without merge among pull requests that had reached a final decision. For merged pull requests, we measured the time from creation to merge. For lifecycle analysis, a final decision corresponded to either merging or closure without merge, while pull requests remaining open at the end of observation were treated as right-censored.

For Community Forum discussions, we defined observed resolution as the presence of an accepted answer. For discussions with an accepted answer, the \texttt{accepted\_answer} record identifies the corresponding post and its creation timestamp. We therefore measured the elapsed time from discussion creation to the creation of the post that was ultimately accepted. Because the timestamp of the acceptance action itself was not available, this measure represents time to the creation of the ultimately accepted answer rather than time to the acceptance action.

\paragraph{\textbf{Artifact-level factors.}}
For GitHub issues, we examined coordination and maintenance characteristics including comments, labels, assignees, milestones, and author association. We also derived indicators from the original issue title and body to capture diagnostic and contextual information, including code blocks, URLs, error or failure descriptions, version information, command-line information, system information, reproduction information, and references to containers, cloud infrastructure, HPC environments, Galaxy tools, workflows, histories, and datasets. We additionally distinguished selected issue-label categories, including bug, enhancement, linting, and good-first-issue labels.

For pull requests, we considered draft status, labels, assignees, requested reviewers and teams, milestones, auto-merge configuration, target branch, author association, and automated or bot-authored contributions. We also derived characteristics of the pull request description, including body and text length, code blocks, URLs, checklists, and references to fixes, testing, documentation, dependency management, security, automation, Galaxy tools and wrappers, workflows, and configuration or deployment activities. Selected bug, enhancement, and dependency labels were also examined.

For forum discussions, we examined engagement and discussion-structure characteristics including the number of posts, replies, views, likes, tags, and duration of discussion activity. We also extracted structural indicators from the original discussion content, including code blocks, links, images, and uploaded material. Diagnostic and contextual indicators captured references to errors, Galaxy tools, workflows, histories and datasets, jobs and queues, data transfer and storage, reference genomes, training materials, versions, reproduction information, containers, cloud infrastructure, HPC environments, and account or authentication problems. We additionally considered forum category and commonly occurring Galaxy forum tags.

\paragraph{\textbf{Bivariate analysis.}}
We first characterized resolution outcomes using frequencies, proportions, and distributions of resolution times. For binary artifact characteristics, we compared resolution rates using Fisher's exact test~\cite{freeman2007analysis} and reported odds ratios to quantify the magnitude and direction of associations. For count and continuous characteristics, we used the Mann--Whitney $U$ test because these variables and resolution-time distributions were generally non-normal. We reported Cliff's $\delta$ as a non-parametric effect-size measure. We also compared resolution times among resolved artifacts using the Mann--Whitney $U$ test and Cliff's $\delta$. To account for multiple statistical comparisons, we adjusted $p$-values using the Benjamini--Hochberg procedure~\cite{benjamini1995controlling}.

\paragraph{\textbf{Multivariable analysis.}}
We fitted multivariable logistic regression models to examine associations among multiple artifact characteristics and resolution outcomes simultaneously. We estimated nested models to distinguish characteristics available from artifact content and context from characteristics reflecting subsequent coordination or engagement. The first model included author, textual, diagnostic, and contextual characteristics. The second model additionally incorporated lifecycle or coordination characteristics, including comments, labels, assignees, and milestones for issues; reviewer and assignment characteristics for pull requests; and engagement characteristics for forum discussions. These lifecycle variables are interpreted as retrospective characteristics of artifact evolution rather than as information necessarily available at artifact creation.

For the GitHub analyses, we used repository-clustered standard errors to account for potential dependence among artifacts originating from the same repository. Forum models additionally controlled for discussion category. Because post and reply counts capture closely related dimensions of forum interaction, we used post count in the primary engagement model and repeated the analysis with reply count as a sensitivity analysis. We report adjusted odds ratios and 95\% confidence intervals from the multivariable models.

\paragraph{\textbf{Lifecycle analysis.}}
We complemented the binary-outcome analyses with time-to-event methods. We used Kaplan--Meier estimators~\cite{d2021methods} to compare time-to-resolution patterns for selected artifact characteristics, with differences assessed using log-rank tests. We then fitted Cox proportional hazards models~\cite{kalbfleisch2023fifty} to examine multivariable associations with the rate at which artifacts progressed toward resolution.

For issues, the event was issue closure; for pull requests, the event was a final decision through either merging or closure without merge; and for forum discussions, the event was creation of the post that was ultimately accepted. Artifacts that had not experienced the corresponding event by the end of observation were right-censored. Hazard ratios greater than one indicate comparatively faster progression toward the specified event, whereas values below one indicate slower progression. We assessed the proportional-hazards assumption as part of the model diagnostics.

\subsection{Analyzing Cross-Space Connections Among Maintenance and Support Artifacts}
\label{sec:rq3_approach}

To answer \textbf{RQ3}, we examined how maintenance and support artifacts are connected across Galaxy GitHub issues, pull requests, and Community Forum discussions. We considered three complementary forms of evidence: explicit cross-artifact references, candidate semantic connections among otherwise unlinked artifacts, and recurring technical signals across the three artifact types.

\paragraph{\textbf{Explicit cross-artifact traceability.}}
We first identified explicit references among issues, pull requests, and forum discussions. For GitHub artifacts, we examined the available title and body text; for forum discussions, we examined both the rendered discussion content and embedded HTML so that hyperlinks contained in anchor elements were preserved. We extracted GitHub issue and pull-request URLs, Community Forum URLs, qualified references such as \texttt{galaxyproject/galaxy\#123}, same-repository references such as \texttt{\#123}, and closing-keyword expressions such as \emph{fixes}, \emph{closes}, and \emph{resolves}.

Each extracted reference was resolved against the collected Galaxy corpus using repository and artifact identifiers. Resolved references were classified according to their source and target artifact types, including \textit{issue-to-issue}, \textit{issue-to-pull request}, \textit{pull request-to-issue}, \textit{pull request-to-pull request}, \textit{forum-to-GitHub}, \textit{GitHub-to-forum}, and \textit{forum-to-forum} connections. Multiple references between the same source and target artifacts were consolidated into a single source--target relationship while preserving the observed evidence types. Extracted references that could not be resolved to an artifact in the collected corpus were retained separately and were not included in the resolved-link counts. Consequently, this analysis represents explicit traceability observable in the analyzed artifact fields rather than all possible relationships in the Galaxy ecosystem.

\paragraph{\textbf{Candidate semantic connections.}}
Explicit references capture only relationships that contributors record directly. We therefore complemented the traceability analysis with semantic matching to identify potentially related artifacts lacking an explicit link. Prior information-retrieval and traceability-recovery studies have similarly used textual similarity to recover missing relationships among software artifacts~\cite{DBLP:conf/sigsoft/WuZKC11,DBLP:journals/access/AlsharaSSS23,DBLP:journals/re/LudersPM23}. We examined cross-artifact families including forum--issue, forum--pull request, issue--pull request, and pull request--issue pairs.

For each artifact, we concatenated its title and natural-language textual content. We removed URLs and commit-like hexadecimal hashes before constructing semantic representations to reduce the likelihood that similarity scores merely reproduced explicit references. The resulting text was represented using TF--IDF with unigram and bigram features, English stopword removal, sublinear term frequency, and $L_2$ normalization. Terms appearing in fewer than two documents or in more than 95\% of documents were excluded. We capped the TF--IDF vocabulary at 50,000 features to limit dimensionality and computational cost while retaining informative unigram and bigram features. Cosine similarity was then computed between artifacts within each candidate family.

Before semantic matching, we removed pairs already connected through explicit traceability. We used a cosine-similarity threshold of 0.30 as a \emph{candidate-generation} threshold rather than as evidence of a confirmed relationship. This choice was informed by prior empirical evidence showing that explicitly linked software issues can exhibit relatively modest TF--IDF cosine similarity~\cite{DBLP:journals/re/LudersPM23}. To reduce weak candidates, we retained at most the five highest-scoring matches for each source artifact, consistent with traceability-recovery studies that present a small ranked set of candidate links for subsequent inspection~\cite{DBLP:conf/saner/YasaOAKDUT25}.

We further required the target artifact to occur between 30 days before and 180 days after the source artifact. Combining textual similarity with temporal proximity follows established traceability-recovery approaches~\cite{DBLP:conf/sigsoft/WuZKC11,DBLP:journals/jss/RuanCPZ19}. The wider forward window accommodates maintenance activity that may follow an initial artifact after some delay, whereas the shorter backward window captures near-contemporaneous or previously initiated work. Because the temporal window is asymmetric, the relative frequency of the two temporal orders is descriptive of the configured candidate-generation procedure and is not interpreted as an unbiased estimate of directional flow. We refer to the resulting pairs as \emph{candidate semantic connections}. These connections indicate textual and temporal relatedness only; they do not establish that one artifact caused, motivated, or resolved another.

\paragraph{\textbf{Manual validation of candidate connections.}}
Because textual similarity and temporal proximity do not necessarily indicate a meaningful relationship between two artifacts, we manually validated a stratified random sample of the 4,882 candidate semantic connections. The sample size was determined using a 95\% confidence level and a 5\% margin of error. To prevent the substantially larger issue--pull-request family from dominating the validation set, we sampled an equal number of candidate pairs from each cross-artifact family. The resulting validation sample comprised 357 pairs.

Two authors independently reviewed and classified all sampled pairs. Both authors have experience in SWSs, software maintenance, and empirical software engineering, providing relevant domain and methodological expertise for interpreting relationships among Galaxy maintenance and support artifacts. Each pair was classified into one of four levels of relatedness: \emph{exact or strongly related}, when both artifacts concerned the same underlying problem, change, or maintenance need; \emph{broadly related}, when they concerned the same component or technical concern but a direct relationship could not be established; \emph{weakly related}, when only limited contextual or topical overlap was evident; and \emph{unrelated}, when no meaningful relationship could be identified.

We measured inter-rater agreement using Cohen's $\kappa$ \cite{cohen1960coefficient}, obtaining a value of $\kappa = 0.86$, indicating high agreement between the two authors \cite{landis1977measurement}. Disagreements were subsequently discussed by the two authors, who revisited the corresponding artifact content and reached consensus on the final classification. The manual validation was used to assess the extent to which semantic matching surfaced substantively related artifacts beyond those connected through explicit references. Because textual and temporal relatedness alone cannot establish traceability or causality, even manually supported matches are treated as \emph{candidate semantic connections} rather than confirmed problem--solution links.

\paragraph{\textbf{Cross-space technical signals.}}
Finally, we examined whether similar technical concerns occur across the three artifact types. Using the original artifact text, we derived regular-expression indicators for execution errors, workflow execution and invocation, history and data management, tool-wrapper integration, testing and validation, dependency management, container runtimes, distributed execution, infrastructure and deployment, installation and configuration, authentication and account management, training and documentation, reference-genome and annotation resources, and upload, storage, and transfer concerns. We compared the prevalence of these signals across issues, pull requests, and forum discussions. This analysis provides complementary evidence of shared technical concerns across development and support spaces; it does not establish pairwise traceability between individual artifacts.

\section{Results}

This section presents the findings for the three \textbf{RQs}. We first
characterize the maintenance and support concerns that emerge across GitHub
issues, pull requests, and Community Forum discussions (\textbf{RQ1}). We then examine
the factors associated with resolution outcomes and resolution time across the
three artifact types (\textbf{RQ2}). Finally, we investigate explicit and candidate
connections among maintenance and support artifacts across development and
community-support spaces (\textbf{RQ3}).

\subsection{RQ1. Maintenance and Support Topics}
\label{sec-RQ1}
Following the topic-modeling and validation procedure described in Section~\ref{sec:study-design}, we analyzed GitHub issues, pull requests, and Community Forum discussions independently and compared the resulting topic structures across the three channels. Our analysis identified nine topics from GitHub issues, 14 from pull requests, and 14 from Community Forum discussions (Tables~\ref{tab:galaxy_issue_topics}--\ref{tab:galaxy_forum_topics}). Together, these topics reveal that maintenance and support in Galaxy extend well beyond source-code defect correction, encompassing workflow execution, data and history management, scientific tools and reference resources, software dependencies, testing, computing infrastructure, deployment, documentation, training, and user-facing analytical support.

\subsubsection{Motivation}

Maintenance and support in a community-driven SWS span multiple technical and community-facing spaces. In Galaxy, GitHub issues capture reported problems and maintenance needs, pull requests reflect implementation and integration activities, and Community Forum discussions reveal user-facing operational and analytical difficulties. Examining these channels together allows us to characterize the breadth of maintenance and support concerns and identify themes that recur across, or are distinctive to, different parts of the ecosystem. \textbf{RQ1} therefore investigates the major maintenance and support concerns that emerge across the Galaxy ecosystem.

\subsubsection{Approach}

As described in Section~\ref{sec:study-design}, we analyzed Galaxy GitHub issues, pull requests, and community forum discussions separately to preserve the distinct maintenance and support concerns expressed in each channel. After preprocessing the textual content, we applied BERTopic modeling independently to each artifact type to identify recurring topics. We then interpreted the resulting topics through their representative terms and representative artifacts. Following established practices in prior studies~\cite{DBLP:journals/jcst/YangLXWS16, DBLP:conf/msr/ScocciaMA21, DBLP:journals/ese/AlamRRM25, DBLP:journals/corr/abs-2601-09612, DBLP:journals/corr/abs-2205-03181, DBLP:conf/icsm/OpenjaAK20}, the first author initially assigned candidate topic labels after examining the representative terms and manually reviewing at least 25 representative artifacts from each topic. This interpretation was informed by the first author's more than five years of experience working with the Galaxy SWS and over a decade of professional software-development experience. The proposed labels and descriptions were then independently reviewed by another author with more than nine years of experience in scientific workflow research. Disagreements were resolved through iterative discussion and repeated inspection of the representative terms and artifacts until consensus was reached. This validation process ensured that the final topic labels were grounded in both the BERTopic output and the domain-specific context of Galaxy maintenance and support. Finally, we compared the identified topics across issues, pull requests, and forum discussions to distinguish channel-specific concerns from maintenance themes that recur across the broader Galaxy ecosystem.

\subsubsection{Results of RQ1}

Our analysis identified nine topics from GitHub issues, 14 from pull requests, and 14 from Community Forum discussions (Tables~\ref{tab:galaxy_issue_topics}--\ref{tab:galaxy_forum_topics}). Together, these topics show that maintenance and support in Galaxy span substantially more than source-code defect correction, covering workflow operation, scientific tools and reference resources, data and history management, computing infrastructure, software dependencies, testing, deployment, documentation, and user support.

\begin{table*}[!t]
\centering
\begin{threeparttable}

\caption{Maintenance and support topics identified from issue discussions.}
\label{tab:galaxy_issue_topics}

\scriptsize
\renewcommand{\arraystretch}{1.08}
\setlength{\tabcolsep}{3pt}

\begin{tabularx}{\textwidth}{
>{\raggedright\arraybackslash}p{0.025\textwidth}
>{\raggedright\arraybackslash}p{0.14\textwidth}
>{\raggedright\arraybackslash}p{0.15\textwidth}
>{\raggedright\arraybackslash}X
}
\toprule
\textbf{SL} &
\textbf{Topic Name} &
\textbf{Representation} &
\textbf{Concise Description with Example} \\
\midrule

1 &
\textbf{Workflow Execution and Data Management}
&
tool, version, test, error, workflow, datum, dataset, history, upload
&
Captures issues related to running and managing workflows in Galaxy, including workflow invocation and execution, handling workflow inputs and outputs, and managing datasets, histories, and dataset collections. Examples include \textit{Failed to schedule WorkflowInvocation..}, which reports a failure when scheduling a workflow for execution, and \textit{Cannot change storage location for dataset collections}, which concerns difficulties in changing the storage location of dataset collections.
\\\cmidrule{3-4}

2 &
\textbf{Genomic Data Sources and Assembly Resources}
&
genome, datum, assembly, tutorial, read, add, sequence, ncbi, dataset, ucsc 
&
Covers issues related to making genomic reference data and assembly-specific resources available and usable in Galaxy, including reference databases, lineage datasets, genome assemblies, and prebuilt indices required by tools and workflows. Examples include \textit{Please add DADA2-formatted reference databases to test/main}, which requests reference datasets required by DADA2 tools, and \textit{Busco: The option to download lineage data has been removed, but there are missing files for many lineages in CVMFS data}, which reports failures caused by missing BUSCO lineage files in Galaxy's CVMFS reference data.
\\\cmidrule{3-4}

3 &
\textbf{Training Material and Educational Content Maintenance}
&
tutorial, content, slide, hub, topic, update, create, video, visitor, site
&
Covers issues related to creating, updating, organizing, and improving Galaxy training and educational resources, including tutorials, slides, learning pathways, videos, and GTN content. Examples include \textit{Help needed for Updating, Testing, Reviewing and Recording of Tutorial ``Data Manipulation Olympics''}, which concerns reviewing and updating an existing tutorial, and \textit{Associate Learning pathways with a topic and display them on the topic page}, which proposes linking relevant learning pathways to GTN topic pages so that learners can more easily find structured training routes.
\\\cmidrule{3-4}

4 &
\textbf{Distributed Job Execution and Infrastructure}
&
job, pulsar, run, error, docker, ansible, package list, version, try, playbook
&
Discusses issues related to executing Galaxy jobs across distributed or remote computing infrastructure, including Pulsar, Slurm, Docker, cloud nodes, job runners, data transfer, and deployment configuration. For example, an issue \textit{Incorrect html.files\_path with docker/pulsar job} reports an incorrect job-directory structure when running AlphaFold remotely across Pulsar deployments using Azure nodes and Slurm/Docker runners, preventing Galaxy from locating generated output files. Another issue, \textit{Error on data transfer from Pulsar to Galaxy}, reports intermittent failures when jobs are executed remotely through Pulsar and results are transferred back to Galaxy in a RabbitMQ-based configuration.
\\\cmidrule{3-4}

5 &
\textbf{Workflow Invocation and Configuration}
&
input, step, invocation, subworkflow, behavior, workflow editor, tool, observe bug, galaxy version
&
Discusses issues related to invoking and configuring Galaxy workflows, including workflow inputs, step dependencies, subworkflows, invocation behavior, and workflow-editor configuration. For example, an issue \textit{Partial workflow invocation} reports that when a multi-step workflow is invoked through BioBlend, an upstream step executes successfully while a dependent downstream step is never invoked or added to the history. Another issue, \textit{Workflow Editor: Insert steps from starred workflow broken}, reports that users are unable to insert steps from an existing workflow into the current workflow in Firefox because the expected confirmation dialog does not appear.
\\\cmidrule{3-4}

6 &
\textbf{History and Collection Management}
&
history, dataset, collection, observe bug, server, galaxy version, expect, delete, steps reproduce, reproduce behavior
&
This topic gathers issues related to managing and interacting with Galaxy histories, datasets, and collections, including filtering, visibility, organization, deletion, import/export, and collection handling. For example, an issue \textit{History filtering on state doesn't return collections} reports that filtering a history by dataset state returns individual datasets but fails to return matching dataset collections. Another issue, \textit{Dataset not visible in a user's history (compressed/uncompressed, same hid)}, reports that large datasets listed in the storage manager are not visible in the corresponding history, preventing users from accessing, deleting, or purging them.
\\\cmidrule{3-4}

7 &
\textbf{Software Dependency and Package Management}
&
conda, dependency, glibcxx, installation, import, test, module, numpy, fix, wheel
&
Discusses issues related to installing, resolving, and maintaining software dependencies and packages required by Galaxy tools and related components, including Conda environments, package-version conflicts, missing or incompatible dependencies, library compatibility, and installation failures. For example, an issue \textit{Potential dependency conflicts between bioblend and requests} identifies potentially incompatible dependency requirements between BioBlend, Requests, and Requests-Toolbelt that could result in future build failures. Another issue, \textit{sha256 checksum error downloading trinity 2.2.0 dependency}, reports that installation of the Trinity package fails because the checksum of the downloaded dependency does not match the expected value.
\\\cmidrule{3-4}

8 &
\textbf{Tool and Workflow Testing and Validation}
&
tool, run, planemo, lint, python, test fail, version, test, directory, install
&
This topic covers issues related to testing and validating Galaxy tools and workflows, particularly through Planemo, including linting, test discovery and execution, tool serving, validation rules, and compatibility across Planemo versions. For example, an issue \textit{planemo lint: empty citation tag is likely valid} reports that Planemo lint accepts an empty citation tag as valid, while the same tool subsequently fails to load with \texttt{planemo serve}. Another issue, \textit{planemo test is not recursive anymore}, reports that after upgrading Planemo, running \texttt{planemo test} from the project root no longer discovers and executes tests located within the tools directory.
\\\cmidrule{3-4}

9 &
\textbf{Authentication and Runtime Issues}
&
sentry issue, issue, token, oidc, email, log, azure, attributeerror, username, credential
&
This topic includes Galaxy issue reports related to authentication, credential handling, user identity, and runtime exceptions across Galaxy and associated components, including login mechanisms, OIDC, usernames, credentials, and Sentry-reported application errors. For example, an issue \textit{Correct credentials not being used} reports that CloudLaunch fails to respect the credentials selected by a user and instead requires one credential set to be configured as the default before cloud resources can be retrieved. Another issue, \textit{AttributeError: 'NoneType' object has no attribute 'id'}, reports a Sentry-detected runtime exception in Galaxy's workflow API when the application attempts to access a missing user object while retrieving stored workflows.
\\

\bottomrule
\end{tabularx}

\begin{tablenotes}[flushleft]
\scriptsize
\item \textit{Note.} Topic names were assigned based on BERTopic representations and inspection of representative issue discussions.
\end{tablenotes}

\end{threeparttable}
\end{table*}

\paragraph{\textbf{GitHub issues topics.}}
GitHub issues primarily expose reported problems, limitations, and maintenance needs across the Galaxy ecosystem. Several topics concern the execution and management of scientific workflows and their associated data. \textit{Workflow Execution and Data Management} captures broader problems involving workflow operation, datasets, histories, and data handling, whereas \textit{Workflow Invocation and Configuration} focuses more specifically on workflow inputs, steps, subworkflows, invocation behavior, and Workflow Editor configuration. \textit{History and Collection Management} further isolates problems involving the organization, visibility, deletion, and handling of histories, datasets, and collections.

A second group of issue topics reflects the technical environment required to operate and maintain Galaxy. \textit{Distributed Job Execution and Infrastructure} concerns remote and distributed execution environments, including Pulsar, job runners, containers, and data transfer. \textit{Software Dependency and Package Management} captures installation, package, and dependency compatibility problems, while \textit{Tool and Workflow Testing and Validation} reflects testing, linting, and validation concerns associated with Galaxy tools and workflows. \textit{Authentication and Runtime Issues} additionally captures authentication, credential, identity, and runtime-exception problems.

Maintenance concerns also extend beyond the core software platform. \textit{Genomic Data Sources and Assembly Resources} reflects the maintenance and availability of scientific reference resources, while \textit{Training Material and Educational Content Maintenance} captures the creation, organization, and updating of Galaxy training materials. Overall, the issue topics demonstrate that reported maintenance needs span not only Galaxy's software functionality but also its workflows, data resources, execution infrastructure, dependencies, testing mechanisms, and educational resources.

% =========================================================
% Table: Part 1
% =========================================================

\begin{table*}[!t]
\centering
\begin{threeparttable}

\caption{Maintenance and support topics identified from pull request discussions.}
\label{tab:pr_topics2}

\scriptsize
\setlength{\tabcolsep}{3pt}
\renewcommand{\arraystretch}{1.08}

\begin{tabularx}{\textwidth}{
>{\centering\arraybackslash}p{0.025\textwidth}
>{\RaggedRight\arraybackslash}p{0.14\textwidth}
>{\RaggedRight\arraybackslash}p{0.15\textwidth}
>{\RaggedRight\arraybackslash}X
}
\toprule
\textbf{SL} &
\textbf{Topic Name} &
\textbf{Representation} &
\textbf{Concise Description with Example} \\
\midrule

1 &
\textbf{Feature Evolution and Maintenance}
&
change, tool, version, workflow, include, component, make, automate test, contribution, agree
&
Discusses broad development and maintenance changes across the Galaxy ecosystem, including new feature implementation, tool and workflow updates, software-component refactoring, version changes, automated testing, and contributor-oriented maintenance. Examples include \textit{Add Galaxy Notification System}, which introduces a notification system spanning database models, REST APIs, user preferences, automated tests, and front-end interfaces for user-specific and server-wide messages; \textit{Migrate Workflow Components from BTable to GTable}, which refactors several workflow-related Galaxy UI components to use the newer GTable component while retaining test coverage; and \textit{Drop wrong help text on bwa-mem / bwa-mem2 output sorting}, which corrects misleading help text in existing tool wrappers and coordinates the change with version-checking and deployment considerations.
\\\cmidrule{3-4}

2 &
\textbf{Tool Integration and Updates}
&
document tool, appropriate tools-iuc, update exist, exist tool, collection explain, educational commercial, use educational, permit unrestricted, read, appropriate
&
Discusses pull requests related to integrating new bioinformatics tools into Galaxy and updating existing Galaxy tool wrappers, including tool versions, parameters, dependencies, inputs and outputs, tests, and associated configuration. For example, a pull request \textit{New tool addition: chromap} adds the Chromap tool to the Galaxy tools-iuc repository as a new Galaxy tool integration. Another pull request, \textit{Update ncbi datasets downloader v.14.6.0 -> 14.13.2}, updates the existing NCBI Datasets Downloader wrapper to a newer software version and adjusts the associated tool implementation and tests.
\\\cmidrule{3-4}

3 &
\textbf{Core Functionality and Interface Changes}
&
automate, change, coverage, past contribution, mit, core codebase, exist test, option apply, test refactor, component
&
Covers pull requests that modify the behavior and user-facing interfaces of Galaxy's core application, including histories, dataset collections, tool panels, workflows, jobs, object stores, and related frontend or backend functionality. For example, a pull request \textit{Mark dataset collections as deleted when purging a history} changes Galaxy's history-purging behavior so that dataset collections are correctly marked as deleted and purged, with accompanying tests for the affected functionality. Another pull request, \textit{Group Tool Versions in IT Panel}, modifies the Galaxy tool panel to group different versions of the same tool, improving how available tool versions are presented and accessed through the interface.
\\\cmidrule{3-4}

4 &
\textbf{Bioinformatics Analysis and Training Resources}
&
tutorial, add, annotation, sequence analysis, metagenomics, read, rna, new tool, plot, protein
&
Discusses pull requests that expand or refine Galaxy resources for domain-specific bioinformatics analyses and their associated training content, including genome annotation, sequence analysis, RNA analysis, metagenomics, assembly, and other scientific workflows. For example, a pull request \textit{Add new tutorial for synthetic biology: basic assembly analysis} introduces a new Galaxy Training Network tutorial for performing basic synthetic-biology assembly analysis. Another pull request, \textit{Genome annotation statistics tool: jcvi\_gff\_stats}, adds a Galaxy tool for calculating statistics from GFF genome annotations, such as gene, exon, and intron counts and lengths, with the tool intended for use in a related GTN genome-annotation tutorial.
\\\cmidrule{3-4}

5 &
\textbf{Community Activities}
&
add news, blog post, freiburg, community, create, training, eurosciencegateway, list, talk, bgruene
&
Discusses pull requests related to communicating Galaxy community activities, including news items, blog posts, workshops, events, project announcements, and other outreach content published through Galaxy community channels. For example, a pull request \textit{Add blog post: Freiburg Galaxy Team at CE-IUSSI Meeting 2025} adds a Galaxy Hub blog post highlighting the Freiburg Galaxy Team's participation in the CE-IUSSI meeting. Another pull request, \textit{Create event page for Galaxy Workshop on Digital Humanities in Freiburg}, creates a Galaxy Hub event page to publicize a community workshop in Freiburg.
\\\cmidrule{3-4}

6 &
\textbf{Compute Resource Allocation and Job Scheduling}
&
slurm, offline, run, increase, limit, send, qld, ram, queue, pulsarqldgpu
&
Focuses on the allocation of computational resources and scheduling of Galaxy jobs across execution environments, including CPU and memory requests, GPU resources, job queues, Slurm and HTCondor configuration, Pulsar, and job destinations. For example, a pull request \textit{Support HTCondor in CLUSTER\_SLOTS\_STATEMENT.sh} enables Pulsar to derive the number of allocated CPU slots from HTCondor environment variables and propagate this information through \texttt{GALAXY\_SLOTS}. Another pull request, \textit{[tpv\_db\_optimizer] Automatically update resource requests based on usegalaxy historical data}, introduces automated adjustment of tool resource requests using historical Galaxy job-usage data, including observed memory requirements.
\\\cmidrule{3-4}

7 &
\textbf{Code Quality and Contribution Compliance}
&
test, agree, contribution, uncheck, condition, galaxycommitterslistsgalaxyprojectorg, allow committer, change, issue, automate
&
Focuses on code-quality and contribution practices within Galaxy, including linting and coding conventions, automated test requirements, refactoring under existing test coverage, dependency-related maintenance, and contributor licensing requirements. For example, \textit{Stricter docstrings parsing} introduces stricter docstring validation using \texttt{flake8-docstrings} and includes fixes to affected tests. Another pull request, \textit{Use absolute imports}, standardizes Python imports to avoid MyPy errors and improve code consistency while following Galaxy's testing and contribution requirements.
\\\cmidrule{3-4}
8 &
\textbf{Workflow Editing and Configuration}
&
step, input, automate, add, parameter, workflow editor, select, test change, codebase mit, past contribution
&
Covers changes to the creation and configuration of Galaxy workflows, including workflow steps, inputs, parameters, subworkflows, and Workflow Editor functionality. For example, \textit{Add download option to workflow editor menu} adds an option for downloading workflows directly from the Workflow Editor. Another pull request, \textit{Fix workflow invocation report pdf generate}, corrects the generation and download of reports associated with workflow invocations.
\\\cmidrule{3-4}

9 &
\textbf{Dependency Update and CI Automation}
&
fix, dependabot create, rebase, upgrade close, version reopen, ci, changelog, manually, default future, language
&
Captures automated and routine updates to software dependencies and continuous-integration infrastructure across the Galaxy ecosystem, including package-version upgrades, GitHub Actions, Dependabot-generated changes, compatibility checks, rebasing, and related CI maintenance. For example, \textit{Update Python dependencies} uses Galaxy's automated dependency-update process to refresh Python dependencies in the core Galaxy repository. Another pull request, \textit{Bump actions/checkout from 4 to 6}, is generated by Dependabot to upgrade the GitHub Actions checkout dependency, incorporating release and compatibility information for the newer version.
\\\cmidrule{3-4}

10 &
\textbf{Deployment and Service Infrastructure}
&
jenkins, failed, fail install, update error, jenkin install, execute test, file, runt, traceback recent, install chipseeker
&
Addresses operational and infrastructure changes that support the deployment and maintenance of Galaxy services, including cloud migration, virtual-machine and storage configuration, service deployment, training infrastructure, and administrative updates. For example, \textit{Migrate new cloud} moves Galaxy compute infrastructure to a new cloud environment by updating credentials, Jenkins configuration, virtual machines, and HTCondor-connected compute resources. Another pull request, \textit{Move vgcn-mounts.yml.j2 template from mounts repository to userdata.yaml.j2}, simplifies Galaxy infrastructure deployment by moving mount configuration into the VM user-data template, thereby removing the need for Jenkins to generate the mount configuration during deployment.
\\\cmidrule{3-4}

11 &
\textbf{Containerized Tool Execution and Resolution}
&
container, docker image, singularity, job, default, resolver, version, mount, automate test, include
&
Encompasses changes related to running Galaxy tools in containerized environments, including Docker, Singularity/Apptainer, container-image resolution, volume mounting, runtime configuration, and container dependencies. For example, \textit{Fixes to get Docker Galaxy Stable running as an --engine again} addresses Docker API and CLI changes affecting environment handling and volume mounting when running Galaxy through Planemo's Docker engine. Another pull request, \textit{Share docker group between pulsar and dind}, updates Pulsar's container environment by sharing the Docker group with Docker-in-Docker and adding the required Apptainer \texttt{starter-suid} binary for container execution.
\\\cmidrule{3-4}

12 &
\textbf{Installation and Upgrade Validation}
&
install successful, inspect ci, output expect, expect change, change deploy, test use, sequence, description installation, replace header
&
Addresses changes that improve and verify Galaxy installation and upgrade processes, including installation testing, compatibility checks, configuration fixes, component initialization, and validation of upgraded environments. For example, \textit{fix python version for bioblend} corrects the Python environment used to test BioBlend within Galaxy flavor testing after tests fail because of incompatible Python versions. Another pull request, \textit{Fix the upgrade process}, improves Galaxy Helm upgrades by handling cached templates and database-readiness problems that can cause web-handler startup and database-upgrade failures.
\\\cmidrule{3-4}

13 &
\textbf{Automated Tool Version and Metadata Maintenance}
&
comment query, criticism bot, relate tool, update create, available conda, version project, automate, create, date new, follow tool
&
Captures routine and automated maintenance of existing Galaxy tools, including software-version synchronization, dependency updates, tool metadata enrichment, and related wrapper adjustments. For example, \textit{Updating tools/cutadapt from version 4.6 to 4.7} is automatically generated by Planemo to update the Cutadapt wrapper after detecting a newer dependency version available through Conda. Another pull request, \textit{add bio.tools id to xml}, enriches the Trycycler Galaxy wrapper with a bio.tools identifier, improving its metadata and linkage to external tool registries.
\\\cmidrule{3-4}

14 &
\textbf{Documentation and Web Resource Maintenance}
&
url, unknown error, link fix, image, update, add, matrix, internal, open, dead
&
Encompasses maintenance of Galaxy documentation and web-based resources, including correcting broken or outdated links, updating URLs, maintaining images and media, and improving access to supporting online content. For example, \textit{Fix links} corrects links in Galaxy Training Network tutorials that incorrectly redirected users back to the tutorial instead of the intended GitHub repository. Another pull request, \textit{Auto Compress Images}, automatically optimizes images used in Galaxy training materials to reduce file size while preserving the associated educational content.
\\

\bottomrule
\end{tabularx}

\begin{tablenotes}[flushleft]
\scriptsize
\item \textit{Note.} Topic names were assigned based on BERTopic representations and inspection of representative pull request discussions.
\end{tablenotes}

\end{threeparttable}
\end{table*}

\paragraph{\textbf{Pull requests topics.}}
Pull requests emphasize implementation, integration, evolution, and preventive maintenance activities across the Galaxy ecosystem. The broad \textit{Feature Evolution and Maintenance} topic captures cross-cutting development work involving new functionality, software-component changes, workflow and tool updates, and associated testing. More focused topics, including \textit{Core Functionality and Interface Changes} and \textit{Workflow Editing and Configuration}, reflect changes to particular areas of Galaxy's core application and workflow-development functionality.

Several topics concern the continued evolution of Galaxy's scientific capabilities. \textit{Tool Integration and Updates} captures the addition and updating of Galaxy tool wrappers, while \textit{Bioinformatics Analysis and Training Resources} reflects changes to domain-specific analytical and educational resources. \textit{Automated Tool Version and Metadata Maintenance} captures routine and automated updates to tool versions, dependencies, and metadata.

Infrastructure and execution constitute another substantial area of pull-request activity. \textit{Compute Resource Allocation and Job Scheduling} concerns resource requests, queues, schedulers, and job destinations; \textit{Deployment and Service Infrastructure} captures changes to cloud, virtual-machine, storage, and service-deployment infrastructure; and \textit{Containerized Tool Execution and Resolution} reflects maintenance of Docker, Singularity/Apptainer, and container-resolution mechanisms. \textit{Installation and Upgrade Validation} further captures changes intended to improve or verify installation and upgrade processes.

The pull-request topics also reveal substantial preventive and supporting maintenance. \textit{Code Quality and Contribution Compliance} concerns coding conventions, testing requirements, refactoring, and contribution practices, while \textit{Dependency Update and CI Automation} captures routine dependency and continuous-integration maintenance. \textit{Documentation and Web Resource Maintenance} reflects maintenance of documentation, links, images, and web resources, and \textit{Community Activities} captures updates to news, events, workshops, and other community-facing content. Thus, the pull-request topics show that Galaxy maintenance involves not only changes to the core application but also continuing work on tools, workflows, infrastructure, automation, documentation, and community resources.

\begin{ThreePartTable}

\begin{TableNotes}[flushleft]
\scriptsize
\item \textit{Note.} Topic names were assigned based on BERTopic
representations and inspection of representative community forum discussions.
\end{TableNotes}

\scriptsize
\renewcommand{\arraystretch}{1.08}
\setlength{\tabcolsep}{3pt}

\begin{longtable}{
>{\centering\arraybackslash}p{0.025\textwidth}
>{\RaggedRight\arraybackslash}p{0.14\textwidth}
>{\RaggedRight\arraybackslash}p{0.15\textwidth}
>{\RaggedRight\arraybackslash}p{0.62\textwidth}
}

\caption{Maintenance and support topics identified from Galaxy Community Help Forum discussions.}
\label{tab:galaxy_forum_topics}\\

\toprule
\textbf{SL} &
\textbf{Topic Name} &
\textbf{Representation} &
\textbf{Concise Description with Example} \\
\midrule
\endfirsthead

\multicolumn{4}{c}{
\tablename\ \thetable{} -- \textit{continued from previous page}
} \\

\toprule
\textbf{SL} &
\textbf{Topic Name} &
\textbf{Representation} &
\textbf{Concise Description with Example} \\
\midrule
\endhead

\midrule
\multicolumn{4}{r}{\textit{Continued on next page}} \\
\endfoot

\bottomrule
\insertTableNotes
\endlastfoot

1 &
\textbf{Analysis Tool and Dataset Troubleshooting} &
file, tool, error, datum, read, genome, dataset, workflow &
Covers user support discussions involving errors and unexpected results when
using Galaxy analysis tools and processing datasets, including file handling,
sequence reads, workflow operations, and tool-specific analysis problems.
For example, \textit{collection operations column join, result has missing
datasets and unnamed columns} reports that joining datasets within a collection
produces missing samples and unnamed output columns, requiring troubleshooting
of the collection-processing operation. Another discussion,
\textit{Troubleshooting with Deseq2}, seeks help interpreting unexpected DESeq
results, particularly problems with the resulting p-values and their
distribution. \\
\cmidrule{3-4}

2 &
\textbf{Job Queuing and Execution Delays} &
queue, long, start, rna star, run job, process, tool, submit &
Covers user concerns about Galaxy jobs remaining queued or taking unusually
long to start or complete, often because of resource availability, cluster
workload, or tool-specific computational demands. For example,
\textit{Job waiting to run, but others are working?} reports that a small
dataset-joining job remains queued while other jobs in the same history execute
normally, prompting clarification that different tools may be routed to
different compute resources. Another discussion,
\textit{Long run time for RNA Star jobs}, describes RNA-STAR jobs remaining in
the waiting state for almost 24 hours and asks whether such delays are expected
or whether jobs can be submitted to faster resources. \\
\cmidrule{3-4}

3 &
\textbf{Data Upload and FTP Transfer Issues} &
server, directory, datum, problem, ftp upload, transfer, time, large, use filezilla &
Discusses difficulties with uploading or transferring datasets to Galaxy
servers, particularly large files and FTP-based transfers, including connection
failures, server-specific upload behavior, directory access, and FileZilla
configuration. For example, \textit{Can't connect to ftp server on Galaxy}
reports that multiple FTP clients, including FileZilla, CyberDuck, and CoreFTP,
are unable to connect for transferring a 64 GB dataset. Another discussion,
\textit{Failled Data upload}, describes repeated failures when uploading data
to a Galaxy server, including files that had previously uploaded successfully,
with the server returning a ``please make sure the file is available'' error. \\
\cmidrule{3-4}

4 &
\textbf{Installation and Dependency Troubleshooting} &
venvlibpythonsitepackage requirementstxt, error, dependency, try install,
sh runsh, conda, version, venv, init &
Concerns about installation and software-environment problems, including
Python packages, Conda dependencies, virtual environments, version
incompatibilities, and dependencies required for local or tool execution.
For example, \textit{Help installing DRMAA Python for use with Slurm?}
reports that a locally installed Galaxy instance cannot start its Slurm job
runner because the Python DRMAA package is unavailable. Another discussion,
\textit{HicDetectLoop bug cleanlab version issue}, reports that HiCExplorer
fails because an incompatible Cleanlab dependency prevents the tool from
importing the required Python class. \\
\cmidrule{3-4}

5 &
\textbf{Storage Management and Data Purging} &
purge, space, dataset, gb, account, export, disk quota, delete history,
delete file &
Focuses on challenges in managing storage usage, including account quotas,
deleting and purging histories or datasets, reclaiming storage space, and
recovering accidentally purged data. For example,
\textit{History purged but not deleted from the account storage} reports that
purging several histories does not immediately reduce the user's reported
storage usage, requiring the Storage Dashboard to recalculate quota usage.
Another discussion, \textit{How to retrieve purged dataset}, asks whether an
accidentally purged dataset can be recovered. \\
\cmidrule{3-4}

6 &
\textbf{Differential Expression Troubleshooting} &
error, factor, edgeR, rowname, RNA-seq, enUSUTF, follow, DESeq error,
count file, featureCounts &
Highlights difficulties encountered during RNA-seq differential-expression
analysis, particularly with edgeR, DESeq2, FeatureCounts, count matrices,
factor definitions, and sample or gene identifiers. For example,
\textit{Unable to run edgeR tool} describes a case where edgeR does not
proceed when supplied with a single count matrix and a contrast file,
prompting investigation of the job status and input configuration.
Another discussion, \textit{Error in DESeq2 output}, reports a DESeq2 failure
when using HTSeq-count files, where incompatible factor levels across the input
samples produce an \texttt{Ops.factor} error. \\
\cmidrule{3-4}

7 &
\textbf{Reference Genome Availability and Configuration} &
reference genome, list, map, NCBI, HISAT, RNA-seq, version, annotation, DM &
Concerns the availability, selection, and configuration of reference genomes
and annotations required by Galaxy tools, including requests for missing
genome builds, use of custom references, and tool-specific genome indexes.
For example,
\textit{Request to upload/add new version of reference genome of Aedes albopictus
(Genome assembly AalbF5)} asks for the latest Aedes albopictus assembly to be
made available for RNA-seq analysis, with guidance provided on using the genome
and annotation as custom reference data. Another discussion,
\textit{reference genome for STARsolo}, requests the Sus scrofa
GCF\_000003025.6\_Sscrofa11.1 genome and annotation for STARsolo, with the
response explaining how to use a custom reference genome when a built-in index
is unavailable. \\
\cmidrule{3-4}

8 &
\textbf{Genome and Sequence Analysis Errors} &
annotation, genome, galaxyreplmainfilesdatasetdat, fatal error, FASTA,
sequence, delete false, historycontenttype, updatetime, dataset &
Includes user-reported failures encountered during genome and sequence analyses
in Galaxy, particularly errors associated with FASTA/FASTQ data, genome
annotations, reference sequences, and tool-specific input requirements.
For example, \textit{Bismark mapping with Fatal Error exit} reports that
Bismark mapping fails for several samples because the uploaded sequencing
files are not correctly recognized as FASTQ data; re-uploading the files
resolves the problem. Another discussion, \textit{MiModd variant calling},
reports a fatal error during variant calling because sequence names in the
custom reference genome contain characters that are not permitted by MiModD,
requiring the FASTA identifiers to be corrected before analysis. \\

9 &
\textbf{Workflow Invocation and Execution Problems} &
run workflow, invocation, step, import, history, tool, API, scheduling fail,
share &
Highlights problems encountered when invoking and executing workflows,
including invocation scheduling failures, stalled workflow steps, and cases
where individual tools function correctly but fail when executed together as
a workflow. For example,
\textit{Invocation scheduling failed - Galaxy administrator may have additional
details in logs} reports a workflow that repeatedly fails to schedule jobs at
a particular step because the workflow expects a collection input but receives
a single dataset. Another discussion, \textit{Cannot run my workflow},
describes a workflow extracted from a history that remains at the
``Workflow Invocation State'' and step-scheduling stage without progressing,
even though each individual tool step runs successfully on its own. \\
\cmidrule{3-4}

10 &
\textbf{Account Access Management} &
authentication, proxy, admin, local, log, email address, help, activation,
new account, galaxyyml &
Discusses problems involving account access and administration, including
account activation, authentication, login failures, verification emails,
local user accounts, and administrator configuration. For example,
\textit{Verification email to verify account is not being send} reports that
a newly registered user cannot activate their Galaxy account because repeated
requests for a verification email produce no message. Another discussion,
\textit{Local account vs. Galaxy account vs. admin account and login not working},
describes confusion when configuring users and administrator privileges on a
local Galaxy installation, clarifying that local accounts are managed
independently and that administrator access is configured through the local
Galaxy settings. \\
\cmidrule{3-4}

11 &
\textbf{LEfSe Analysis and Visualization Errors} &
LEfSe analysis, error, traceback, cladogram, input file, recent file, size,
plot, format datum, feature &
Discusses LEfSe analysis issues, including input-data formatting, LDA
effect-size computation, traceback and tool errors, and generation or
interpretation of plots and cladograms. For example,
\textit{Need Help with Galaxy LEfSe LDA Step Error} reports that the analysis
cannot proceed beyond the LDA Effect Size step even when using an input file
that had worked successfully in the past. Another discussion,
\textit{Cannot run my workflow for LEfSe}, reports that the data-formatting
and LDA steps complete successfully, but the LEfSe plotting step produces no
visible figure despite appearing to finish successfully. \\
\cmidrule{3-4}

12 &
\textbf{Microbial Sequence and Taxonomic Analysis} &
datum, sequence, analysis, taxonomic, bacterial genome, microbial analysis,
want, amplicon, generate, UniPept &
Relates to the analysis and taxonomic characterization of microbial sequence
data, including metagenomic classification, amplicon analysis, sequence
processing, database selection, and interpretation of taxonomic results.
For example, \textit{Galaxy Kraken2 Perform Quality Filtering?} asks whether
Kraken2 performs read-quality filtering before metagenomic classification or
whether sequencing reads should be preprocessed separately. Another discussion,
\textit{Different totals depending on database used for Krona pie chart},
reports different taxonomic totals when the same Kraken2-classified sequences
are analysed using SILVA, Greengenes, and RDP databases, requiring
clarification of how reference-database composition influences the resulting
taxonomic summaries. \\
\cmidrule{3-4}

13 &
\textbf{Kraken--Bracken Database Compatibility and Reporting} &
Bracken, pie chart, database Kraken, kmer, add, error, GTDB, work &
Centers on problems with Kraken/Kraken2 and Bracken database compatibility,
taxonomic reporting, and downstream abundance estimation, including missing
reference databases, K-mer distributions, incompatible report inputs, and
database-specific errors. For example, \textit{Kraken-report Database} asks
how to generate a Kraken2 report compatible with Bracken after using the
PlusPF database, with guidance to enable Kraken2's aggregate-count report
output for downstream analysis. Another discussion,
\textit{Bracken database error}, reports that Bracken fails when processing
Kraken2 results generated with the PlusPF reference because the required K-mer
distribution file cannot be located, indicating a server-side database
configuration problem. \\
\cmidrule{3-4}

14 &
\textbf{Paired-End Read and FASTQ Collection Handling} &
read, datum, forward reverse, sequence, trim, FASTQ file, pair end, list,
merge, set &
Concerns the preparation and processing of paired-end sequencing data,
including organizing forward and reverse reads, splitting interleaved FASTQ
files, trimming reads, managing paired-end collections, and handling reads
that lose their mate during quality control. For example,
\textit{Manipulating interleaved/interlaced fastq data} describes difficulty
using paired-end sequencing reads stored together in a single interleaved
FASTQ file with Trimmomatic and explains how to split the reads or import them
as an appropriate paired collection. Another discussion,
\textit{Convert Single End Read to Paired End??}, asks whether high-quality
unpaired reads remaining after trimming can be converted into artificial read
pairs, with guidance recommending that genuine paired-end and single-end reads
be retained and analysed according to their original sequencing structure. \\

\end{longtable}

\end{ThreePartTable}

\paragraph{\textbf{Community forum topics.}}
Community Forum discussions reveal the user-facing operational and analytical difficulties that arise while working with Galaxy. Several topics involve general troubleshooting of platform functionality and scientific analyses. \textit{Analysis Tool and Dataset Troubleshooting} captures broad problems involving tools, datasets, workflows, and unexpected analytical results, whereas \textit{Workflow Invocation and Execution Problems} focuses specifically on workflow scheduling, step execution, and input compatibility. \textit{Job Queuing and Execution Delays} exposes problems associated with waiting times and computational-resource availability.

Other forum topics concern data and environment management. \textit{Data Upload and FTP Transfer Issues} captures difficulties transferring large or remote datasets, while \textit{Storage Management and Data Purging} reflects questions about quotas, deletion, purging, and storage recovery. \textit{Installation and Dependency Troubleshooting} concerns local installation, Python and Conda environments, and software compatibility, and \textit{Account Access Management} reflects authentication, account activation, and administrative-access problems. Users also seek support for scientific resources through \textit{Reference Genome Availability and Configuration}.

In contrast to the broader platform-oriented topics visible in GitHub, the forum contains several highly domain-specific analytical concerns. \textit{Differential Expression Troubleshooting}, \textit{LEfSe Analysis and Visualization Errors}, \textit{Genome and Sequence Analysis Errors}, \textit{Microbial Sequence and Taxonomic Analysis}, \textit{Kraken--Bracken Database Compatibility and Reporting}, and \textit{Paired-End Read and FASTQ Collection Handling} reflect difficulties encountered while configuring, executing, and interpreting particular scientific analyses. These topics show that community support involves not only resolving platform-level operational problems but also helping users correctly prepare data, select resources, configure analytical tools, and interpret workflow behavior.

\paragraph{\textbf{Cross-channel comparison.}} Comparison across the three channels reveals both recurring maintenance concerns and distinct channel-specific perspectives. Workflow execution, data management, tools and dependencies, computing infrastructure, and scientific resources recur across issues, pull requests, and forum discussions, but they manifest differently in each space. GitHub issues primarily formalize reported defects, limitations, and maintenance needs; pull requests emphasize implementation, integration, testing, automation, and preventive maintenance activities; and forum discussions expose user-facing operational and analytical difficulties.

For example, workflow execution appears as invocation and configuration problems in issues, workflow-editing and execution-related changes in pull requests, and workflow scheduling or execution difficulties in forum discussions. Infrastructure concerns similarly appear as distributed-job and execution problems in issues, resource-allocation and deployment activities in pull requests, and job-queuing or installation difficulties in the forum. Scientific resources also span the channels through reference-data and assembly issues, bioinformatics and tool-related pull requests, and reference-genome or domain-specific support discussions.

The channels nevertheless contain distinctive concerns. Pull requests foreground code quality, CI automation, containerized execution, documentation, and community-resource maintenance, reflecting implementation and preventive maintenance activities that are less visible in the other channels. Forum discussions, in contrast, foreground domain-specific analytical support, including differential expression, LEfSe, Kraken--Bracken, microbial analysis, and paired-end data handling. Overall, the three channels provide complementary perspectives on Galaxy maintenance and support: issues foreground reported maintenance needs, pull requests capture implementation and integration activities, and forum discussions reveal the operational and analytical difficulties encountered by users.

\begin{tcolorbox}[
    title=\textbf{RQ1 Summary},
    colback=gray!5,
    colframe=black!60,
    boxrule=0.5pt,
    arc=1mm,
    left=4pt,
    right=4pt,
    top=4pt,
    bottom=4pt
]
\small
Maintenance and support concerns recur across the three channels, particularly around workflow execution and configuration, tool and dependency management, scientific data and reference resources, and computing infrastructure. However, each channel provides a distinct perspective: GitHub issues primarily document reported defects, limitations, and maintenance needs; pull requests capture implementation, integration, testing, automation, and preventive maintenance activities; and Community Forum discussions reveal operational and scientific-analysis difficulties encountered by users. Together, these channels show that Galaxy maintenance is distributed across the core platform, workflows, scientific tools and data resources, computing infrastructure, documentation and training, and community support.

\end{tcolorbox}

\subsection{RQ2: Factors Associated with Resolution Outcomes and Time}
\label{sec:rq2}

\textbf{RQ1} characterized the maintenance and support concerns that emerge across Galaxy's development and support spaces. \textbf{RQ2} extends this analysis by examining how these artifacts progress toward resolution. Specifically, we investigate both whether an observable resolution is reached and how quickly resolution occurs across GitHub issues, pull requests, and Community Forum discussions. Because these artifact types follow different resolution processes, we analyze them separately while considering coordination, diagnostic, contributor, automation, and engagement characteristics associated with their respective resolution lifecycles.

\subsubsection{Motivation}
Identifying the maintenance and support concerns that occur in Galaxy does not reveal whether those artifacts ultimately reach an observable resolution or how long that process takes. Artifacts concerning similar maintenance topics may differ substantially in both resolution outcome and resolution time. This distinction is important in a community-driven scientific software ecosystem, where unresolved issues, delayed pull-request decisions, and prolonged support discussions may contribute to maintenance burden and affect the evolution and use of the platform. \textbf{RQ2} therefore examines which characteristics of issues, pull requests, and Community Forum discussions are associated with observable resolution outcomes and with the pace at which those artifacts progress toward resolution.

\subsubsection{Approach}
Because GitHub issues, pull requests, and Community Forum discussions follow different resolution processes, we analyze each artifact type separately. For issues, we examine closure and time-to-closure. For pull requests, we distinguish among merging, closure without merge, and remaining open, and examine time to a final decision. For forum discussions, we examine accepted-answer status and the time from discussion creation to the creation of the post that was eventually accepted.

We first characterize resolution outcomes and observed resolution-time distributions, followed by bivariate comparisons and effect-size analysis. We then use multivariable logistic regression to examine characteristics associated with resolution outcomes and Kaplan--Meier and Cox proportional-hazards analyses to examine resolution trajectories while retaining unresolved artifacts as right-censored observations. Because several coordination and engagement characteristics, such as comments, milestones, reviews, posts, replies, likes, and activity span, may emerge or accumulate during an artifact's lifecycle, the corresponding estimates are interpreted as retrospective associations with the resolution process rather than as causal effects or prospective predictors.

\subsubsection{Results of RQ2}

\paragraph{\textbf{Factors Associated with GitHub Issue Resolution}}

\begin{figure}[t]
    \centering
    \includegraphics[width=0.7\textwidth, height=0.3\textheight]
    {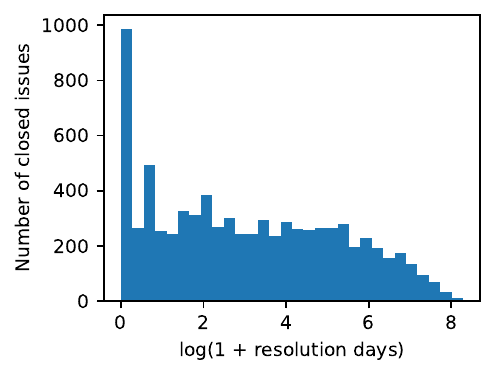}
    \caption{Log-transformed distribution of time-to-closure for closed issues.}
    \label{fig:issue_time_to_close}
\end{figure}

Among the 11,762 GitHub issues, 7,754 (65.92\%) were closed, while 4,008 remained open. Among closed issues, the median time-to-closure was 15.29 days, substantially lower than the mean of 164.15 days, indicating a strongly right-skewed distribution (Fig.~\ref{fig:issue_time_to_close}). Although 25\% of closed issues were resolved within 1.89 days, the slowest 10\% required more than 496.87 days and the slowest 5\% more than 930.66 days.

\textbf{Bivariate associations with issue closure.}

Bivariate analysis showed that several coordination characteristics were associated with eventual closure (Table~\ref{tab:issue_bivariate_closure}). Issues receiving comments had a higher closure rate than those without comments (74.50\% vs.\ 47.05\%; $\mathrm{OR}=3.29$). Higher closure rates were also observed for issues associated with milestones (85.45\% vs.\ 64.78\%; $\mathrm{OR}=3.18$) and assignees (76.85\% vs.\ 61.36\%; $\mathrm{OR}=2.09$). Diagnostic characteristics were likewise positively associated with closure, including bug labels ($\mathrm{OR}=1.40$), explicit error mentions ($\mathrm{OR}=1.34$), and code blocks ($\mathrm{OR}=1.23$).

\begin{table}[t]
\caption{Selected bivariate associations with issue closure.}
\label{tab:issue_bivariate_closure}
\centering
\small
\setlength{\tabcolsep}{8pt}
\renewcommand{\arraystretch}{1.10}

\begin{tabular}{lrrr}
\toprule
\textbf{Factor} &
\textbf{Absent (\%)} &
\textbf{Present (\%)} &
\textbf{OR} \\
\midrule
Comments       & 47.05 & 74.50 & 3.29 \\
Milestone      & 64.78 & 85.45 & 3.18 \\
Assignee       & 61.36 & 76.85 & 2.09 \\
Bug label      & 65.60 & 72.77 & 1.40 \\
Mentions error & 63.70 & 70.15 & 1.34 \\
Code block     & 65.02 & 69.65 & 1.23 \\
\bottomrule
\end{tabular}

\begin{minipage}{0.80\textwidth}
\vspace{1mm}
\footnotesize
\textit{Note.} Absent and Present denote issue closure rates when the
corresponding factor is absent or present, respectively. OR denotes the
odds ratio for issue closure. All reported associations remained
statistically significant after Benjamini--Hochberg correction.
\end{minipage}
\end{table}

Descriptive comparisons among closed issues showed that characteristics associated with eventual closure did not necessarily correspond to shorter resolution times (Table~\ref{tab:issue_time_to_close}). Issues explicitly mentioning errors had a median time-to-closure of 9.51 days compared with 22.91 days for issues without such mentions. Issues containing code blocks similarly closed in a median of 9.24 days compared with 18.25 days. In contrast, issues receiving comments required a median of 19.20 days compared with 8.19 days for issues without comments, while labelled issues required 23.61 days compared with 10.99 days. Issues containing checklists showed a particularly long median time-to-closure of 104.83 days compared with 14.14 days. Because these comparisons include only closed issues, we use the subsequent survival analyses to characterize resolution trajectories while retaining unresolved issues as censored observations.

Taken together, the bivariate results distinguish between \emph{eventual closure} and \emph{resolution speed}. Coordination characteristics such as comments, milestones, and assignees were associated with higher closure rates at the bivariate level, whereas concrete diagnostic information, particularly explicit error descriptions and code examples, was associated with shorter resolution times among closed issues. Because several coordination characteristics can accumulate during an issue's lifecycle, these patterns are interpreted as descriptive maintenance-process associations rather than causal effects; their independent associations are examined subsequently using the multivariable models.

\begin{table}[t]
\caption{Selected binary factors associated with time-to-close for issues.}
\label{tab:issue_time_to_close}
\centering
\small
\renewcommand{\arraystretch}{1.10}
\setlength{\tabcolsep}{10pt}

\begin{tabular}{lrr}
\toprule
\textbf{Factor} &
\multicolumn{1}{c}{\textbf{Absent}} &
\multicolumn{1}{c}{\textbf{Present}} \\
&
\multicolumn{1}{c}{\textbf{Median days}} &
\multicolumn{1}{c}{\textbf{Median days}} \\
\midrule
Mentions error      & 22.91 & 9.51   \\
Contains code block & 18.25 & 9.24   \\
Has comments        & 8.19  & 19.20  \\
Has label           & 10.99 & 23.61  \\
Contains checklist  & 14.14 & 104.83 \\
\bottomrule
\end{tabular}

\vspace{1mm}
\begin{minipage}{0.72\textwidth}
\footnotesize
\textit{Note.} Time-to-close is calculated for closed issues and
represents the median number of days from issue creation to closure.
\end{minipage}
\end{table}

\textbf{Topic-level variation in issue resolution.}

Issue resolution also varied across the maintenance topics identified in RQ1 (Table~\ref{tab:issue_topic_resolution}). \emph{Software Dependency and Package Management} had the highest closure rate (70.56\%), followed by \emph{Workflow Invocation and Configuration} (69.54\%) and \emph{Genomic Data Sources and Assembly Resources} (68.14\%). \emph{Tool and Workflow Testing and Validation} had the lowest closure rate (61.54\%). Resolution speed showed a different pattern. \emph{Authentication and Runtime Issues} had the shortest median time-to-closure (7.13 days), whereas \emph{Software Dependency and Package Management}, despite having the highest closure rate, had the longest median time-to-closure (44.94 days). Thus, topics with comparatively high closure rates were not necessarily resolved most quickly.

\begin{table}[t]
\caption{Topic-level variation in issue resolution.}
\label{tab:issue_topic_resolution}
\centering
\small
\renewcommand{\arraystretch}{1.10}
\setlength{\tabcolsep}{6pt}

\begin{tabular}{
p{0.45\textwidth}
rr
}
\toprule
\textbf{Issue Topic} &
\multicolumn{1}{c}{\textbf{Closure (\%)}} &
\multicolumn{1}{c}{\textbf{Median Days}} \\
\midrule

Workflow Execution and Data Management
& 66.28 & 14.99 \\

Genomic Data Sources and Assembly Resources
& 68.14 & 20.70 \\

Training Material and Educational Content Maintenance
& 64.55 & 20.98 \\

Distributed Job Execution and Infrastructure
& 64.56 & 22.18 \\

Workflow Invocation and Configuration
& 69.54 & 8.75 \\

History and Collection Management
& 65.56 & 11.94 \\

Software Dependency and Package Management
& \textbf{70.56} & 44.94 \\

Tool and Workflow Testing and Validation
& \textbf{61.54} & 29.45 \\

Authentication and Runtime Issues
& 64.71 & \textbf{7.13} \\

\bottomrule
\end{tabular}

\vspace{1mm}
\begin{minipage}{0.7\textwidth}
\footnotesize
\textit{Note.} Median Days denotes the median time from issue creation
to closure among closed issues within each topic.
\end{minipage}
\end{table}

\textbf{Multivariable model of issue closure.}

We fitted a multivariable logistic regression using all 11,762 issues to examine which characteristics remained associated with closure after considering the other included factors and issue type, with repository-clustered standard errors used to account for within-repository dependence. The model converged successfully and had a McFadden pseudo-$R^{2}$ of 0.082.

As shown in Table~\ref{tab:issue_logit}, milestone assignment exhibited the strongest positive adjusted association with closure ($\mathrm{OR}=3.09$, 95\% CI: 2.43--3.93), while each additional assignee was associated with higher odds of closure ($\mathrm{OR}=1.95$, 95\% CI: 1.77--2.14). Bug labels ($\mathrm{OR}=1.64$), explicit error mentions ($\mathrm{OR}=1.34$), code blocks ($\mathrm{OR}=1.31$), and URLs ($\mathrm{OR}=1.17$) also remained positively associated with closure. In contrast, enhancement labels ($\mathrm{OR}=0.69$), member/owner authorship ($\mathrm{OR}=0.74$), greater text length ($\mathrm{OR}=0.86$), and a larger number of labels ($\mathrm{OR}=0.88$) were associated with lower odds of closure.

Notably, the strong bivariate association between comments and closure was no longer present after adjustment ($\mathrm{OR}=1.00$, $p_{\mathrm{adj}}=.626$). This difference suggests that comment presence co-occurs with other characteristics of the issue lifecycle rather than showing an independent adjusted association with closure.

\begin{table}[t]
\caption{Selected adjusted associations with issue closure.}
\label{tab:issue_logit}
\centering
\small
\renewcommand{\arraystretch}{1.10}
\setlength{\tabcolsep}{8pt}

\begin{tabular}{lccc}
\toprule
\textbf{Factor} &
\textbf{OR} &
\textbf{95\% CI} &
\textbf{Adj. $p$} \\
\midrule
Milestone            & 3.09 & 2.43--3.93 & $<.001$ \\
Number of assignees  & 1.95 & 1.77--2.14 & $<.001$ \\
Bug label            & 1.64 & 1.33--2.02 & $<.001$ \\
Mentions error       & 1.34 & 1.20--1.50 & $<.001$ \\
Code block           & 1.31 & 1.16--1.48 & $<.001$ \\
Has URL              & 1.17 & 1.07--1.29 & $<.001$ \\
Number of labels     & 0.88 & 0.84--0.93 & $<.001$ \\
Text length          & 0.86 & 0.81--0.91 & $<.001$ \\
Member/owner author  & 0.74 & 0.66--0.84 & $<.001$ \\
Enhancement label    & 0.69 & 0.56--0.85 & $<.001$ \\
\bottomrule
\end{tabular}

\vspace{1mm}
\begin{minipage}{0.78\textwidth}
\footnotesize
\textit{Note.} OR denotes the adjusted odds ratio for issue closure.
The model controls for issue type, with repository-clustered standard errors
used to account for within-repository dependence. Adjusted $p$-values use the
Benjamini--Hochberg correction.
\end{minipage}
\end{table}

\textbf{Cox proportional hazards model.}

To examine issue-resolution trajectories while retaining unresolved issues as
right-censored observations, we fitted a multivariable Cox proportional hazards
model. As shown in Table~\ref{tab:issue_cox}, milestone presence showed the
strongest association with faster closure ($HR=1.65$, 95\% CI:
$1.52$--$1.80$), followed by having an assignee ($HR=1.35$) and a bug
label ($HR=1.31$). Diagnostic specificity also remained associated with
faster closure: issues mentioning errors ($HR=1.21$), containing code blocks
($HR=1.16$), or providing reproduction information ($HR=1.10$) progressed
more rapidly toward closure. In contrast, issues authored by project members
or owners ($HR=0.82$), enhancement-labelled issues ($HR=0.84$), issues
carrying labels more generally ($HR=0.87$), and issues providing workflow
context ($HR=0.94$) exhibited slower closure trajectories. Comments,
version mentions, command information, and system information did not show
significant adjusted associations. These results reinforce the distinction
between lifecycle activity and resolution speed: structured coordination and
actionable diagnostic evidence are associated with closure trajectories, but
not all forms of discussion or contextual detail correspond to faster
resolution.

\begin{table}[t]
\caption{Selected Cox proportional hazards results for issue closure.}
\label{tab:issue_cox}
\centering
\small
\renewcommand{\arraystretch}{1.10}
\setlength{\tabcolsep}{8pt}

\begin{tabular}{lccc}
\toprule
\textbf{Factor} &
\textbf{HR} &
\textbf{95\% CI} &
\textbf{Adj. $p$} \\
\midrule
Milestone           & 1.65 & 1.52--1.80 & $<.001$ \\
Assignee            & 1.35 & 1.25--1.45 & $<.001$ \\
Bug label           & 1.31 & 1.20--1.44 & $<.001$ \\
Mentions error      & 1.21 & 1.14--1.27 & $<.001$ \\
Code block          & 1.16 & 1.10--1.23 & $<.001$ \\
Number of assignees & 1.13 & 1.07--1.20 & $<.001$ \\
Mentions reproduce  & 1.10 & 1.02--1.19 & .022 \\
Has URL             & 1.07 & 1.02--1.12 & .006 \\
Workflow context    & 0.94 & 0.89--0.98 & .006 \\
Number of labels    & 0.95 & 0.92--0.98 & $<.001$ \\
Has label           & 0.87 & 0.82--0.93 & $<.001$ \\
Enhancement label   & 0.84 & 0.76--0.93 & .002 \\
Has body            & 0.83 & 0.75--0.92 & .001 \\
Member/owner author & 0.82 & 0.78--0.87 & $<.001$ \\
\bottomrule
\end{tabular}

\vspace{1mm}
\begin{minipage}{0.82\textwidth}
\footnotesize
\textit{Note.} HR denotes the hazard ratio for issue closure.
An $HR>1$ indicates a higher instantaneous rate of closure, whereas
an $HR<1$ indicates a lower instantaneous rate of closure.
Adjusted $p$-values use the Benjamini--Hochberg correction.
Unresolved issues were retained as right-censored observations.
\end{minipage}
\end{table}

\textbf{Kaplan--Meier analysis.}

\begin{figure*}[t]
    \centering

    \begin{subfigure}[t]{0.32\textwidth}
        \centering
        \includegraphics[width=\linewidth]
        {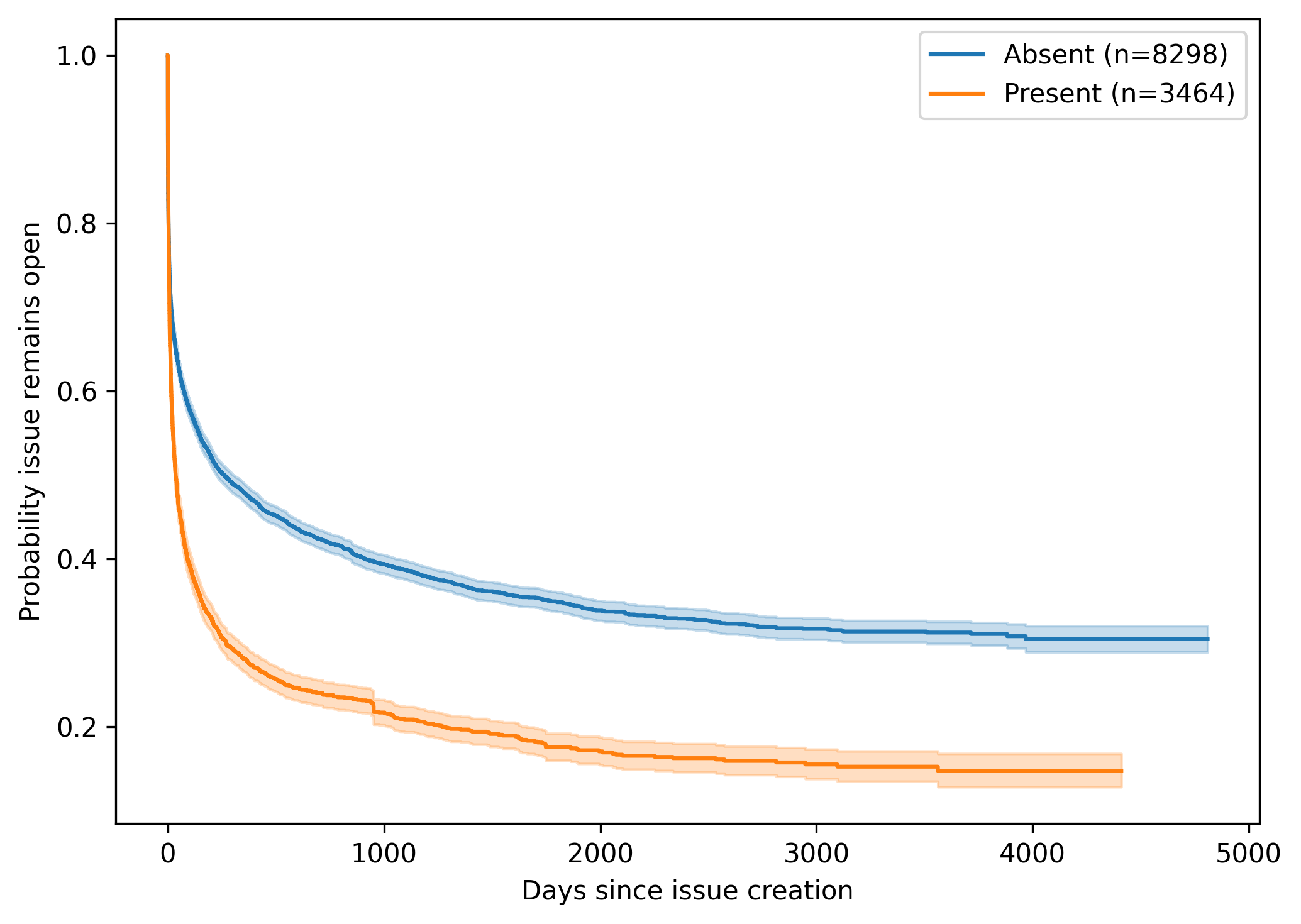}
        \caption{Assignee presence}
        \label{fig:km_assignee}
    \end{subfigure}
    \hfill
    \begin{subfigure}[t]{0.32\textwidth}
        \centering
        \includegraphics[width=\linewidth]
        {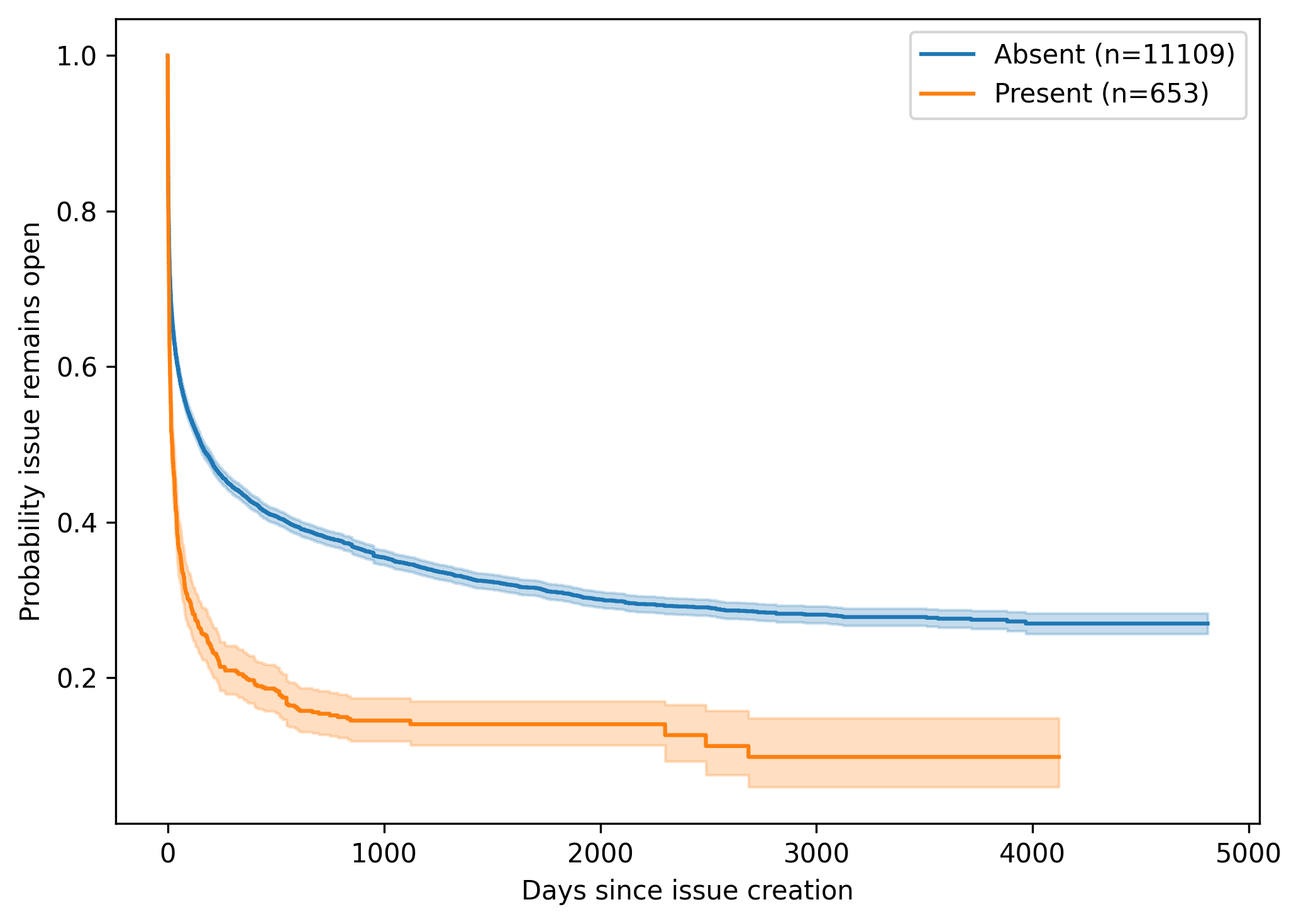}
        \caption{Milestone presence}
        \label{fig:km_milestone}
    \end{subfigure}
    \hfill
    \begin{subfigure}[t]{0.32\textwidth}
        \centering
        \includegraphics[width=\linewidth]
        {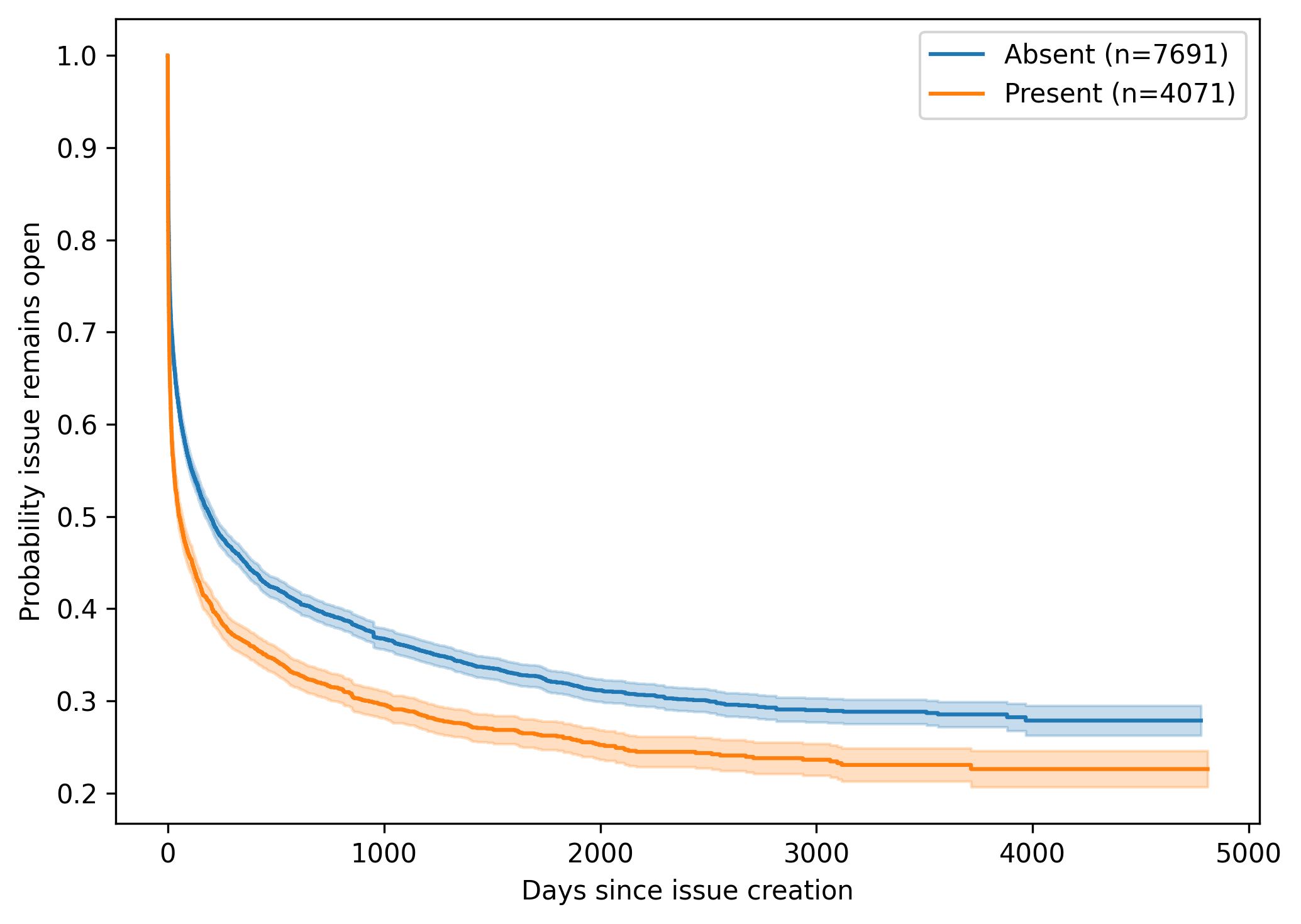}
        \caption{Error mention}
        \label{fig:km_error}
    \end{subfigure}

    \caption{Kaplan--Meier estimates of time-to-closure for selected issue characteristics. The survival probability denotes
    the probability that an issue remains open. Representative curves are
    shown for factors exhibiting clear and statistically robust differences;
    all selected binary factors were examined in the analysis.}
    \label{fig:issue_km}
\end{figure*}

We examined Kaplan--Meier curves for all selected binary issue-level factors
to compare their unadjusted time-to-closure trajectories.
Fig.~\ref{fig:issue_km} presents three representative factors that showed
clear and statistically robust differences and that also remained important
in the Cox model: assignee presence, milestone presence, and explicit error
mentions. The y-axis represents the estimated probability that an issue
remains open; consequently, a more rapidly declining curve indicates faster
progression toward closure. Issues with an assignee or milestone exhibited
faster closure trajectories than those without these coordination signals,
while issues explicitly mentioning errors showed a similar pattern. The
remaining Kaplan--Meier comparisons were examined as part of the analysis
and shared in the replication package \cite{alam_2026_22882299}. These unadjusted patterns complement
the multivariable Cox analysis, with unresolved issues retained as
right-censored observations.

\begin{figure}[t]
    \centering
    \includegraphics[width=0.7\textwidth, height=0.3\textheight]
    {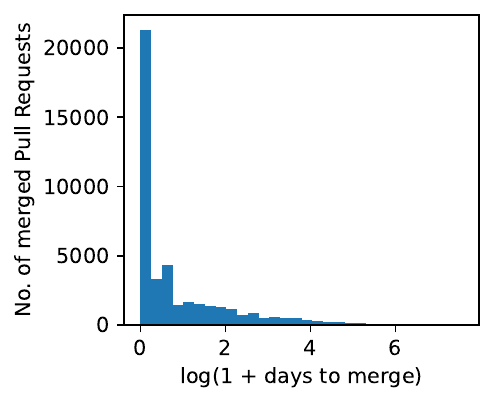}
    \caption{Log-transformed distribution of time-to-merge for merged pull requests.}
    \label{fig:pull_request_merge_time}
\end{figure}

\paragraph{\textbf{Pull-request resolution.}}
Among the 49,770 pull requests analyzed, 42,957 (86.31\%) were merged, 5,419 (10.89\%) were closed without merging, and 1,394 (2.80\%) remained open. Among merged pull requests, the median time-to-merge was 0.30 days, substantially lower than the mean of 9.51 days, indicating a strongly right-skewed distribution (Fig.~\ref{fig:pull_request_merge_time}). Among pull requests that reached a final decision through either merging or closure without merge, the median time-to-decision was 0.41 days, compared with a mean of 20.85 days. Thus, although most pull requests were integrated rapidly, a non-negligible proportion were closed without integration, and a smaller subset remained unresolved at the end of the observation period.

\textbf{Automation and pull-request resolution.}

Pull-request outcomes varied substantially by author type. Among 44,785 human-authored pull requests, 88.74\% were merged, compared with 68.61\% of 790 Dependabot pull requests and 63.72\% of 4,195 pull requests generated by other bots. In contrast, median time-to-merge differed only modestly across the three groups, ranging from 0.29 days for human-authored pull requests to 0.33 days for Dependabot and 0.35 days for other bot-authored pull requests.

\textbf{Requested review and pull-request resolution.} 

Pull requests with a requested reviewer showed only a small difference in merge rate but a more noticeable difference in merge time. Among 45,226 pull requests without a requested reviewer, 86.26\% were merged, compared with 86.77\% of the 4,544 pull requests with a requested reviewer. However, pull requests with a requested reviewer had a longer median time-to-merge (0.64 days) than those without one (0.28 days). Thus, requested review was associated with only a marginally higher merge rate but with a longer integration process among merged pull requests.

\textbf{Topic-level variation in pull-request resolution.}

Pull-request resolution differed substantially across the maintenance topics identified in RQ1 (Table~\ref{tab:pr_topic_resolution}). Merge rates ranged from 76.44\% to 93.87\%. \emph{Deployment and Service Infrastructure} had both the highest merge rate (93.87\%) and the shortest median time-to-merge (0.05 days). In contrast, \emph{Installation and Upgrade Validation} had the lowest merge rate (76.44\%) and the longest median time-to-merge (1.26 days). \emph{Tool Integration and Updates} also showed a comparatively lower merge rate (79.23\%) and a longer median time-to-merge (0.99 days). These results indicate that both integration likelihood and integration time differ across forms of maintenance activity.

\begin{table}[t]
\caption{Topic-level variation in pull-request resolution.}
\label{tab:pr_topic_resolution}
\centering

\small
\renewcommand{\arraystretch}{1.08}
\setlength{\tabcolsep}{4pt}

\begin{tabularx}{0.75\textwidth}{
@{}
>{\RaggedRight\arraybackslash}X
>{\centering\arraybackslash}p{0.14\textwidth}
>{\centering\arraybackslash}p{0.16\textwidth}
@{}
}
\toprule
\textbf{Pull Request Topic} &
\textbf{Merge (\%)} &
\textbf{Median Days} \\
\midrule

Feature Evolution and Maintenance
& 86.33 & 0.35 \\

Tool Integration and Updates
& 79.23 & 0.99 \\

Core Functionality and Interface Changes
& 86.88 & 0.86 \\

Bioinformatics Analysis and Training Resources
& 87.60 & 0.32 \\

Community Activities
& 88.60 & 0.13 \\

Compute Resource Allocation and Job Scheduling
& 81.97 & 0.15 \\

Code Quality and Contribution Compliance
& 86.50 & 0.98 \\

Workflow Editing and Configuration
& 86.00 & 0.80 \\

Dependency Update and CI Automation
& 87.63 & 0.33 \\

Deployment and Service Infrastructure
& \textbf{93.87} & \textbf{0.05} \\

Containerized Tool Execution and Resolution
& 85.98 & 0.53 \\

Installation and Upgrade Validation
& \textbf{76.44} & \textbf{1.26} \\

Automated Tool Version and Metadata Maintenance
& 86.43 & 0.35 \\

Documentation and Web Resource Maintenance
& 86.54 & 0.17 \\

\bottomrule
\end{tabularx}

\vspace{1mm}
\begin{minipage}{0.75\textwidth}
\footnotesize
\textit{Note.} Merge (\%) denotes the proportion of pull requests merged
within each topic. Median Days denotes the median time from pull-request
creation to merge among merged pull requests. Bold values indicate the
highest and lowest values in each column.
\end{minipage}

\end{table}

\textbf{Bivariate associations with pull-request resolution.} 

Bivariate analysis revealed substantial differences in merge outcomes across pull-request characteristics (Table~\ref{tab:pr_bivariate_merge}). Draft pull requests were rarely merged (1.17\% vs.\ 87.80\%; $\mathrm{OR}=0.002$), whereas pull requests authored by project members or owners had a higher merge rate than those from other authors (90.82\% vs.\ 80.54\%; $\mathrm{OR}=2.39$). Milestone-associated pull requests were also more likely to be merged (92.13\% vs.\ 84.40\%; $\mathrm{OR}=2.16$), and bug-labelled pull requests showed a particularly high merge rate (96.15\% vs.\ 86.22\%; $\mathrm{OR}=3.88$). In contrast, bot-authored pull requests ($\mathrm{OR}=0.23$) and externally authored pull requests ($\mathrm{OR}=0.15$) had substantially lower merge odds. Pull requests mentioning review ($\mathrm{OR}=0.45$), testing ($\mathrm{OR}=0.61$), dependencies ($\mathrm{OR}=0.57$), or security ($\mathrm{OR}=0.51$) were also associated with lower merge odds. All reported associations remained statistically significant after Benjamini--Hochberg correction.

Among the 48,376 pull requests that reached a final decision through either merging or closure without merge, several of the same characteristics showed corresponding associations with closure without integration. Draft pull requests had substantially higher odds of being closed without merge, as did bot-authored and externally authored pull requests. Review-, testing-, dependency-, and security-related pull requests also showed higher odds of closure without merge, whereas milestone-associated and bug-labelled pull requests showed lower odds. These findings indicate that contributor role, workflow state, and the technical context represented in a pull request are associated with whether proposed changes are ultimately integrated.

\begin{table}[t]
\caption{Selected bivariate associations with pull-request merging.}
\label{tab:pr_bivariate_merge}
\centering
\small
\setlength{\tabcolsep}{8pt}
\renewcommand{\arraystretch}{1.10}

\begin{tabular}{lrrr}
\toprule
\textbf{Factor} &
\textbf{Absent (\%)} &
\textbf{Present (\%)} &
\textbf{OR} \\
\midrule
Draft                & 87.80 & 1.17  & 0.002 \\
Member/owner author  & 80.54 & 90.82 & 2.39  \\
Milestone            & 84.40 & 92.13 & 2.16  \\
Bug label            & 86.22 & 96.15 & 3.88  \\
Bot author           & 88.74 & 64.49 & 0.23  \\
External author      & 87.35 & 50.95 & 0.15  \\
Mentions review      & 87.27 & 75.64 & 0.45  \\
Mentions test        & 88.90 & 83.00 & 0.61  \\
Mentions dependency  & 87.19 & 79.38 & 0.57  \\
Mentions security    & 86.64 & 76.65 & 0.51  \\
\bottomrule
\end{tabular}

\vspace{1mm}
\begin{minipage}{0.75\textwidth}
\footnotesize
\textit{Note.} Absent and Present denote pull-request merge rates when the
corresponding factor is absent or present, respectively. OR denotes the odds
ratio for merging when the factor is present versus absent. All reported
associations remained statistically significant after Benjamini--Hochberg
correction.
\end{minipage}
\end{table}

\textbf{Bivariate associations with time-to-merge.} 

Among the 42,957 merged pull requests, time-to-merge showed a related but distinct pattern (Table~\ref{tab:pr_time_merge}). Pull requests containing a body had a median time-to-merge of 0.60 days compared with 0.03 days for those without a body ($\delta=0.40$). Milestone-associated pull requests required a median of 0.91 days compared with 0.15 days ($\delta=0.33$), while pull requests targeting development branches required 1.06 days compared with 0.19 days ($\delta=0.32$). Pull requests mentioning tests (0.81 vs.\ 0.12 days; $\delta=0.30$) or containing checklists (0.93 vs.\ 0.22 days; $\delta=0.30$) also showed longer merge times.

In contrast, pull requests targeting the \texttt{main} or \texttt{master} branch were merged more quickly (0.16 vs.\ 0.80 days; $\delta=-0.25$), as were bug-labelled pull requests (0.10 vs.\ 0.30 days; $\delta=-0.14$). Requested reviewers were associated with a longer median time-to-merge (0.64 vs.\ 0.28 days), although the effect size was negligible ($\delta=0.13$). Overall, these results reinforce the distinction between integration outcome and integration speed: characteristics associated with eventual merging do not necessarily correspond to shorter time-to-merge, and several testing, coordination, and development-context signals characterize pull requests that require longer integration periods.

\begin{table}[t]
\caption{Selected factors associated with time-to-merge for pull requests.}
\label{tab:pr_time_merge}
\centering

\small
\renewcommand{\arraystretch}{1.10}
\setlength{\tabcolsep}{7pt}

\begin{tabular}{lrrr}
\toprule
\textbf{Factor} &
\multicolumn{1}{c}{\textbf{Absent}} &
\multicolumn{1}{c}{\textbf{Present}} &
\multicolumn{1}{c}{\textbf{$\delta$}} \\
&
\multicolumn{1}{c}{\textbf{Median days}} &
\multicolumn{1}{c}{\textbf{Median days}} &
\\
\midrule
Has body               & 0.03 & 0.60 & 0.40 \\
Milestone              & 0.15 & 0.91 & 0.33 \\
Targets development    & 0.19 & 1.06 & 0.32 \\
Has URL                & 0.12 & 0.81 & 0.30 \\
Checklist              & 0.22 & 0.93 & 0.30 \\
Mentions test          & 0.12 & 0.81 & 0.30 \\
Has label              & 0.12 & 0.79 & 0.29 \\
Targets main/master    & 0.80 & 0.16 & $-0.25$ \\
Mentions documentation & 0.21 & 0.76 & 0.22 \\
Code block             & 0.28 & 0.85 & 0.21 \\
Mentions review        & 0.27 & 0.83 & 0.20 \\
Requested reviewer     & 0.28 & 0.64 & 0.13 \\
Bug label              & 0.30 & 0.10 & $-0.14$ \\
\bottomrule
\end{tabular}

\vspace{1mm}
\begin{minipage}{0.75\textwidth}
\footnotesize
\textit{Note.} Time-to-merge is calculated among merged pull requests.
$\delta$ denotes Cliff's delta comparing pull requests with versus without
the corresponding factor. Positive values indicate longer time-to-merge when
the factor is present, whereas negative values indicate shorter time-to-merge.
All displayed differences remained statistically significant after
Benjamini--Hochberg correction.
\end{minipage}
\end{table}

\textbf{Multivariable models of pull-request resolution.}

The multivariable logistic models showed substantial differences in pull-request outcomes after accounting for the other included characteristics (Table~\ref{tab:pr_logistic}). In the merge model, auto-merge configuration exhibited the strongest positive association with merging ($\mathrm{OR}=154.05$), followed by milestone assignment ($\mathrm{OR}=7.52$), member/owner authorship ($\mathrm{OR}=5.06$), and contributor authorship ($\mathrm{OR}=3.74$). Pull requests generated by automation bots also had higher adjusted odds of merging ($\mathrm{OR}=1.73$). In contrast, bot authorship more generally was associated with lower odds of merging ($\mathrm{OR}=0.37$), as were pull requests targeting development branches ($\mathrm{OR}=0.23$). Draft status showed the strongest negative association with merging ($\mathrm{OR}=0.001$).

A complementary model restricted to pull requests that reached a final decision compared those closed without merging with those that were merged. Draft status showed by far the strongest positive association with closure without merge ($\mathrm{OR}=595.61$). Pull requests targeting development branches ($\mathrm{OR}=2.78$) and bot-authored pull requests ($\mathrm{OR}=2.47$) also had higher adjusted odds of being closed without merge. Conversely, auto-merge configuration ($\mathrm{OR}=0.002$), milestone assignment ($\mathrm{OR}=0.22$), member/owner authorship ($\mathrm{OR}=0.24$), contributor authorship ($\mathrm{OR}=0.34$), and automation-bot authorship ($\mathrm{OR}=0.59$) were associated with lower odds of closure without merge. Together, the two models indicate that pull-request resolution outcomes are strongly associated with workflow state, contributor role, automation, and coordination characteristics.

\begin{table}[t]
\caption{Selected adjusted associations with pull-request resolution.}
\label{tab:pr_logistic}
\centering

\small
\renewcommand{\arraystretch}{1.10}
\setlength{\tabcolsep}{9pt}

\begin{tabular}{lrr}
\toprule
\textbf{Factor} &
\multicolumn{1}{c}{\textbf{Merge OR}} &
\multicolumn{1}{c}{\textbf{Closed-unmerged OR}} \\
\midrule
Auto-merge          & 154.05 & 0.002  \\
Milestone           & 7.52   & 0.22   \\
Member/owner author & 5.06   & 0.24   \\
Contributor author  & 3.74   & 0.34   \\
Automation bot      & 1.73   & 0.59   \\
Bot author          & 0.37   & 2.47   \\
Targets development & 0.23   & 2.78   \\
Draft               & 0.001  & 595.61 \\
\bottomrule
\end{tabular}

\vspace{1mm}
\begin{minipage}{0.75\textwidth}
\footnotesize
\textit{Note.} OR denotes the adjusted odds ratio. The merge model includes
all analyzed pull requests, whereas the closed-unmerged model compares pull
requests closed without merging with merged pull requests among those reaching
a final decision. Values greater than 1 indicate higher odds of the
corresponding outcome, whereas values below 1 indicate lower odds.
\end{minipage}
\end{table}

\textbf{Cox proportional hazards model.}

To examine time-to-final-decision while retaining open pull requests as right-censored observations, we fitted a multivariable Cox proportional hazards model. Several contributor, automation, and maintenance characteristics were associated with the rate of reaching a final decision (Table~\ref{tab:pr_cox}). Pull requests mentioning auto-merge showed the strongest positive association ($HR=1.51$, 95\% CI: 1.28--1.77), followed by member/owner authorship ($HR=1.49$, 95\% CI: 1.44--1.53), automation-bot authorship ($HR=1.47$, 95\% CI: 1.25--1.73), and contributor authorship ($HR=1.29$, 95\% CI: 1.25--1.33). Pull requests describing fixes ($HR=1.15$, 95\% CI: 1.13--1.17) and containing checklists ($HR=1.13$, 95\% CI: 1.08--1.18) were also associated with faster progression toward a final decision.

In contrast, draft pull requests showed substantially slower resolution trajectories ($HR=0.29$, 95\% CI: 0.26--0.31). Bot authorship more generally was also associated with slower progression toward a final decision ($HR=0.76$, 95\% CI: 0.73--0.78), as were mentions of tool wrappers ($HR=0.81$, 95\% CI: 0.78--0.83), targeting development branches ($HR=0.81$, 95\% CI: 0.79--0.84), targeting \texttt{main}/\texttt{master} branches ($HR=0.85$, 95\% CI: 0.82--0.88), workflow mentions ($HR=0.88$, 95\% CI: 0.85--0.90), dependency mentions ($HR=0.92$, 95\% CI: 0.89--0.95), and review mentions ($HR=0.92$, 95\% CI: 0.89--0.96). These findings indicate that pull-request resolution speed is associated with contributor role, automation context, workflow state, and the type of maintenance activity represented in the pull request.

\begin{table}[t]
\caption{Selected Cox proportional hazards results for pull-request resolution.}
\label{tab:pr_cox}
\centering

\small
\renewcommand{\arraystretch}{1.10}
\setlength{\tabcolsep}{8pt}

\begin{tabular}{lcc}
\toprule
\textbf{Factor} &
\textbf{HR} &
\textbf{95\% CI} \\
\midrule
Mentions auto-merge   & 1.51 & 1.28--1.77 \\
Member/owner author   & 1.49 & 1.44--1.53 \\
Automation bot        & 1.47 & 1.25--1.73 \\
Contributor author    & 1.29 & 1.25--1.33 \\
Mentions fix          & 1.15 & 1.13--1.17 \\
Has checklist         & 1.13 & 1.08--1.18 \\
Mentions review       & 0.92 & 0.89--0.96 \\
Mentions dependency   & 0.92 & 0.89--0.95 \\
Mentions workflow     & 0.88 & 0.85--0.90 \\
Targets main/master   & 0.85 & 0.82--0.88 \\
Targets development   & 0.81 & 0.79--0.84 \\
Mentions tool wrapper & 0.81 & 0.78--0.83 \\
Bot author            & 0.76 & 0.73--0.78 \\
Draft                 & 0.29 & 0.26--0.31 \\
\bottomrule
\end{tabular}

\vspace{1mm}
\begin{minipage}{0.75\textwidth}
\footnotesize
\textit{Note.} HR denotes the hazard ratio for reaching a final pull-request
decision through either merging or closure without merge. An $HR>1$ indicates
a higher instantaneous rate of reaching a final decision, whereas an $HR<1$
indicates a lower instantaneous rate. Open pull requests were retained as
right-censored observations. Only selected statistically significant factors
are shown.
\end{minipage}
\end{table}

\textbf{Kaplan--Meier analysis.}

\begin{figure*}[t]
    \centering

    \begin{subfigure}[t]{0.32\textwidth}
        \centering
        \includegraphics[width=\linewidth]
        {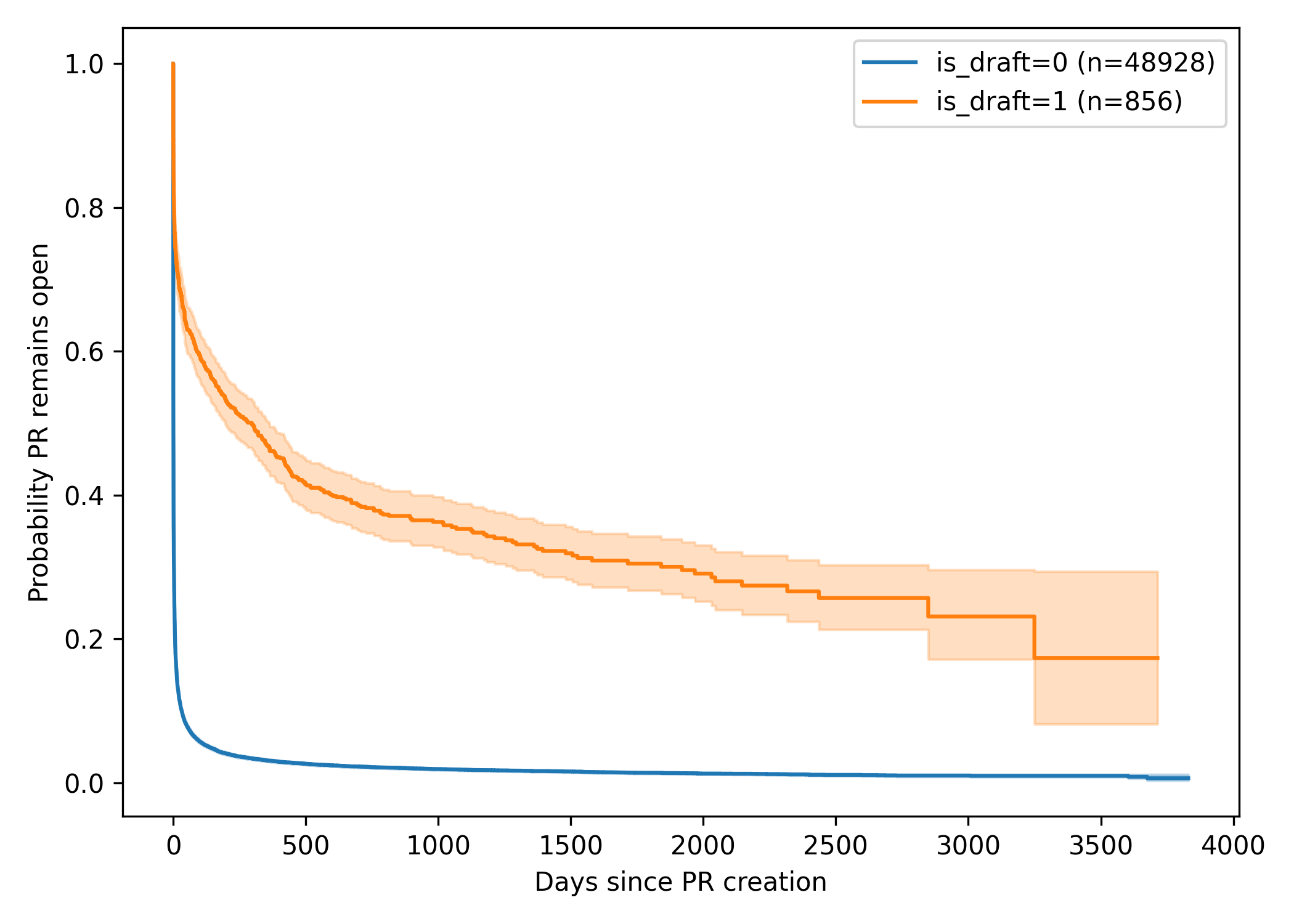}
        \caption{Draft status}
        \label{fig:pr_km_draft}
    \end{subfigure}
    \hfill
    \begin{subfigure}[t]{0.32\textwidth}
        \centering
        \includegraphics[width=\linewidth]
        {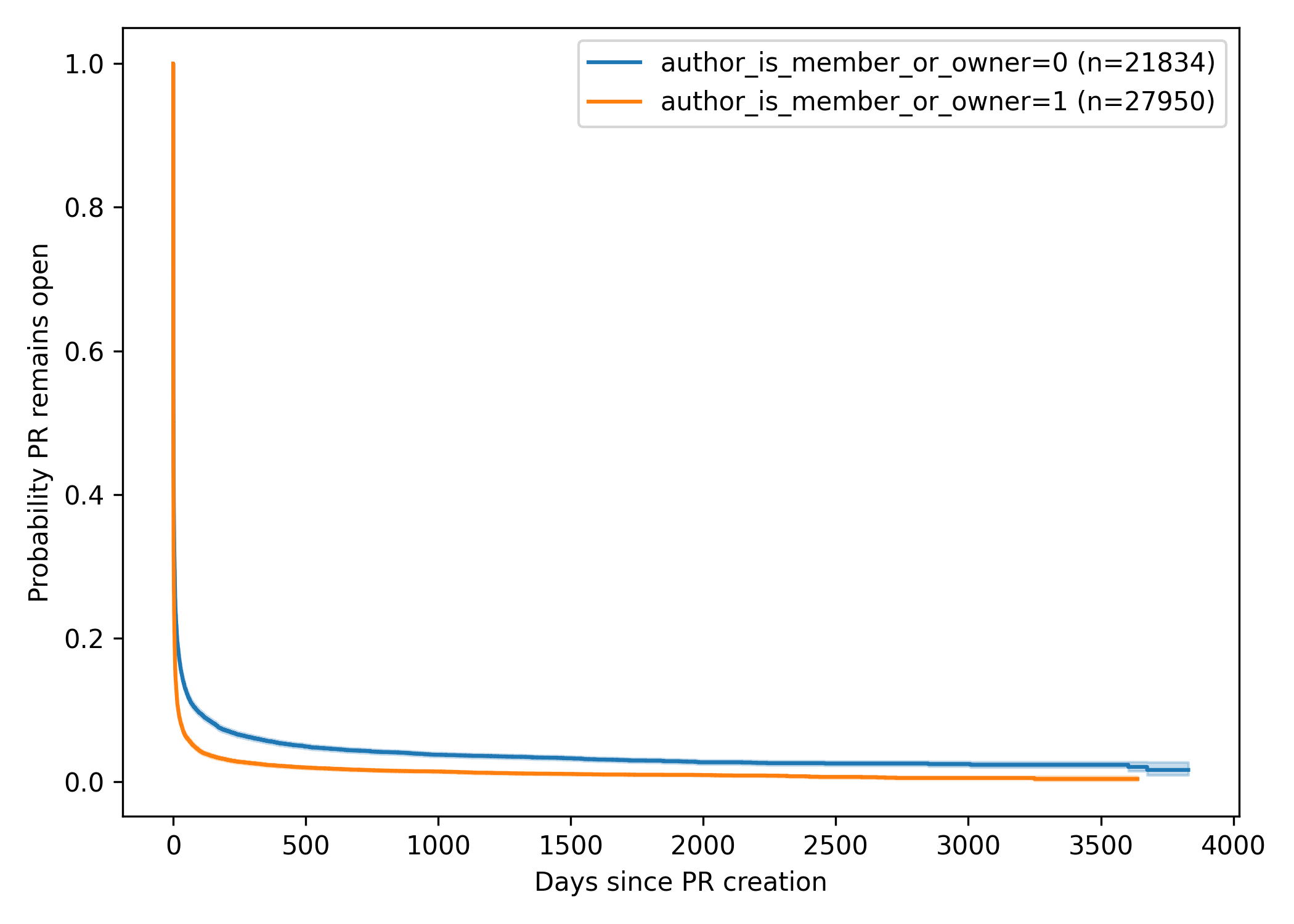}
        \caption{Member/owner authorship}
        \label{fig:pr_km_member}
    \end{subfigure}
    \hfill
    \begin{subfigure}[t]{0.32\textwidth}
        \centering
        \includegraphics[width=\linewidth]
        {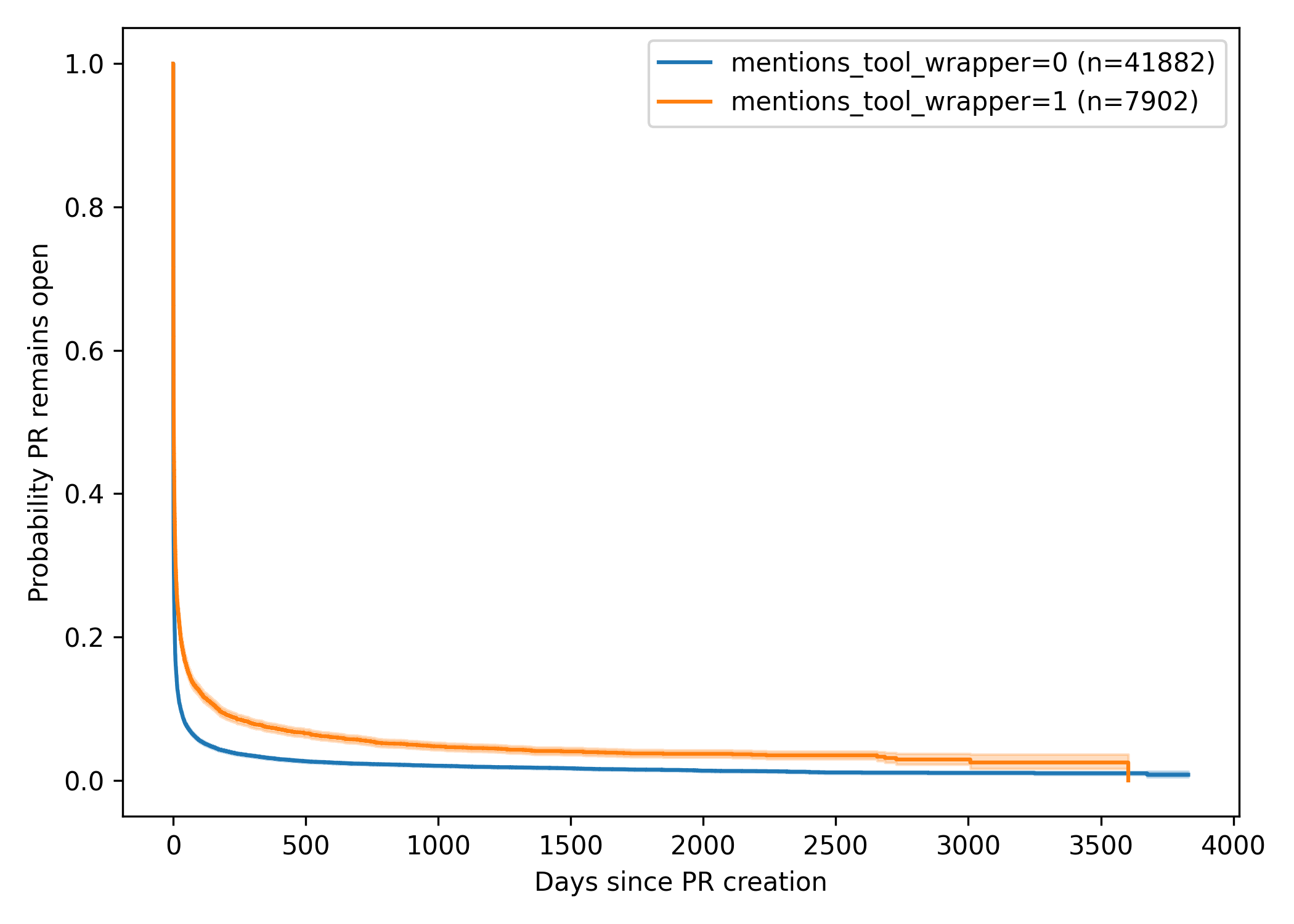}
        \caption{Tool-wrapper mention}
        \label{fig:pr_km_wrapper}
    \end{subfigure}

    \caption{Kaplan--Meier estimates of time-to-final-decision for selected
    Galaxy pull-request characteristics. The survival probability denotes
    the probability that a pull request remains unresolved. Representative
    curves are shown for factors exhibiting clear lifecycle differences;
    all selected binary characteristics were examined in the analysis.}
    \label{fig:pr_km}
\end{figure*}

We examined Kaplan--Meier curves for all selected binary pull-request
characteristics to compare their unadjusted time-to-final-decision
trajectories. For conciseness, Fig.~\ref{fig:pr_km} presents three
representative factors showing clear and contrasting patterns: draft status,
member/owner authorship, and tool-wrapper mentions. The survival probability
represents the probability that a pull request remains unresolved; therefore,
a more rapidly declining curve indicates faster progression toward a final
decision. Pull requests authored by project members or owners reached final
decisions more rapidly than those from other authors. In contrast, draft pull
requests remained unresolved substantially longer, while pull requests
mentioning tool wrappers also exhibited slower resolution trajectories.
These unadjusted patterns are consistent with the corresponding adjusted Cox
model results. All other selected binary factors were also examined and shared in the replication package \cite{alam_2026_22882299}. Open pull requests were retained as
right-censored observations.

\paragraph{\textbf{Community Forum discussion resolution.}}

Among 6,235 Community Forum discussions, 3,279 (52.59\%) had an accepted answer and 2,956 (47.41\%) did not. For 3,270 accepted-answer discussions with an available timestamp for the eventually accepted post, the median time from discussion creation to that post was 1.13 days, compared with a mean of 65.04 days. The large difference between the median and mean indicates a strongly right-skewed distribution (Fig.~6), with most eventually accepted answers appearing relatively quickly but a smaller subset occurring after much longer intervals.
\begin{figure}[t]
    \centering
    \includegraphics[width=0.70\textwidth, height=0.3\textheight]
    {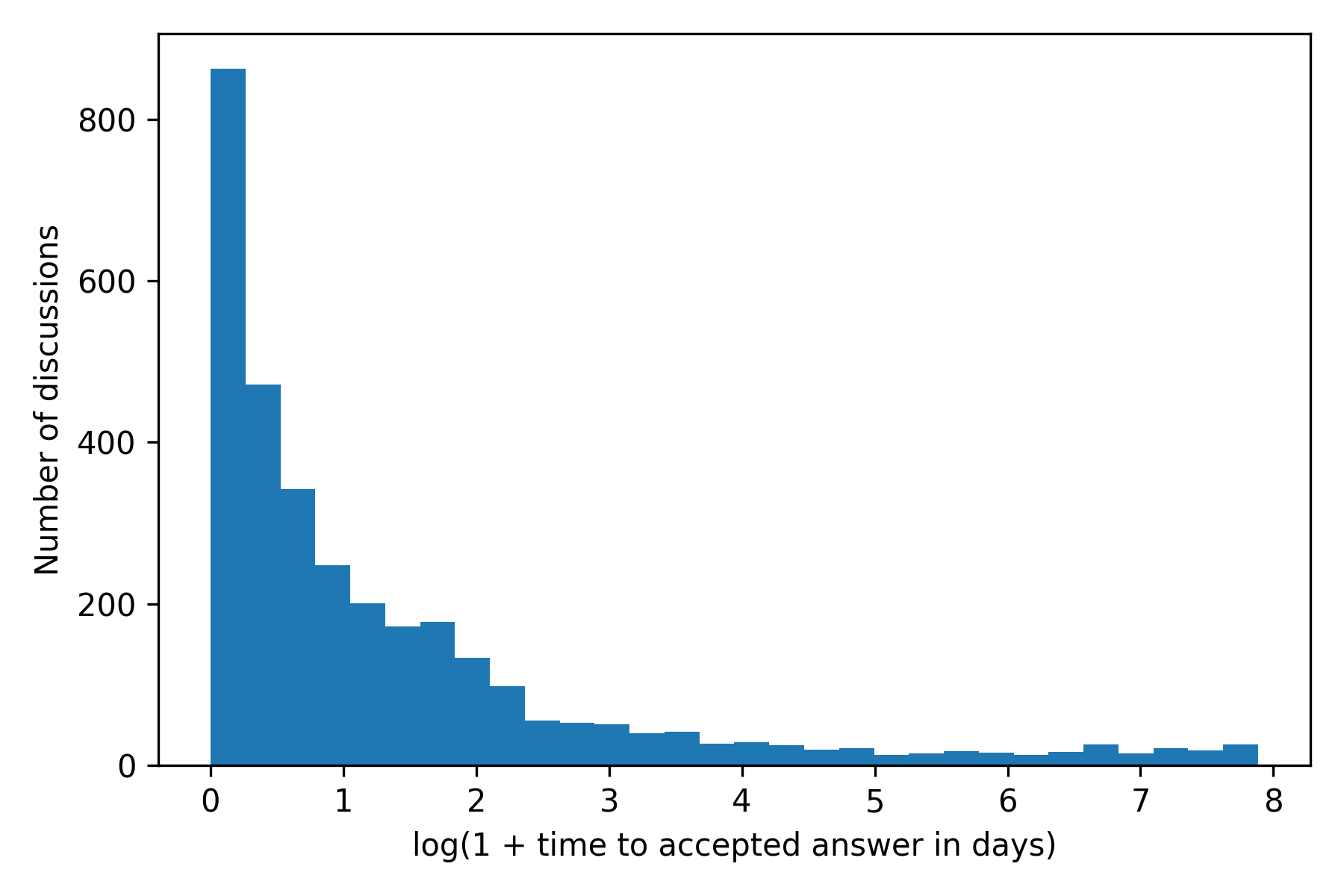}
    \caption{Log-transformed distribution of time to the creation of the accepted post}
    \label{fig:forum_accept_time}
\end{figure}

\textbf{Code blocks and resolution.}

Discussions containing code blocks had a higher accepted-answer rate than those without code blocks (58.10\% vs.\ 51.99\%). Their median time to the eventually accepted post was similar, at 1.30 days compared with 1.11 days for discussions without code blocks. Thus, code blocks were associated with a higher descriptive resolution rate, but not with clearly faster resolution.

\textbf{Technical context and resolution.}

Accepted-answer rates varied modestly across technical contexts (Table~\ref{tab:forum_technical_context}). Discussions mentioning account-related problems had the highest descriptive resolution rate (60.29\%), followed by storage (58.31\%), upload (56.85\%), version (56.43\%), and history/data concerns (56.37\%). Lower rates were observed for workflow (52.01\%) and training-related discussions (52.22\%). These differences are descriptive and are examined further using inferential and multivariable analyses.

\begin{table}[t]
\centering
\caption{Accepted-answer rates across technical contexts in Community Forum discussions.}
\label{tab:forum_technical_context}
\small
\setlength{\tabcolsep}{4pt}
\renewcommand{\arraystretch}{1.05}

\begin{tabular}{lrr}
\toprule
\textbf{Technical Context} & \textbf{N} & \textbf{Accepted (\%)} \\
\midrule
Workflow              & 546   & 52.01 \\
Training              & 766   & 52.22 \\
Reproduction evidence & 1,226 & 52.85 \\
Tool                  & 2,563 & 53.84 \\
Job execution         & 1,816 & 54.24 \\
Error                 & 2,788 & 54.70 \\
Reference genome      & 1,754 & 55.64 \\
Dependency            & 619   & 56.22 \\
History / data        & 1,641 & 56.37 \\
Version               & 1,088 & 56.43 \\
Upload                & 1,437 & 56.85 \\
Storage               & 379   & 58.31 \\
Account               & 340   & 60.29 \\
\bottomrule
\end{tabular}
\end{table}

\textbf{Resolution across forum topics.}

Resolution outcomes varied across forum topics
(Table~\ref{tab:forum_topic_resolution}). Accepted-answer rates ranged from
44.59\% for \emph{LEfSe Analysis and Visualization Errors} to 59.31\% for
\emph{Workflow Invocation and Execution Problems}, while median time to the
eventually accepted post ranged from 0.72 to 1.66 days. Mean resolution times
were substantially higher for several topics, reaching 144.45 days for
\emph{Account Authentication and Access Management}, reflecting strongly
right-skewed resolution times. Median reply counts were generally low
(0--1), whereas median views varied more widely, from 136 for
\emph{Local Galaxy Installation and Dependency Configuration} to 571 for
\emph{Kraken/Bracken Database and Taxonomic Reporting Support}. Overall,
forum topics differed not only in resolution likelihood and speed, but also
in the level of community attention they received.

\begin{table}[t]
\caption{Resolution outcomes across Galaxy Community Help Forum topics.}
\label{tab:forum_topic_resolution}
\centering

\small
\renewcommand{\arraystretch}{1.08}
\setlength{\tabcolsep}{4pt}

\begin{tabularx}{0.8\textwidth}{
@{}
>{\RaggedRight\arraybackslash}X
>{\centering\arraybackslash}p{0.06\textwidth}
>{\centering\arraybackslash}p{0.06\textwidth}
>{\centering\arraybackslash}p{0.06\textwidth}
>{\centering\arraybackslash}p{0.06\textwidth}
>{\centering\arraybackslash}p{0.06\textwidth}
@{}
}
\toprule
\textbf{Forum Topic} &
\shortstack{\textbf{Accepted}\\\textbf{(\%)}} &
\shortstack{\textbf{Median}\\\textbf{Days}} &
\shortstack{\textbf{Mean}\\\textbf{Days}} &
\shortstack{\textbf{Median}\\\textbf{Replies}} &
\shortstack{\textbf{Median}\\\textbf{Views}} \\
\midrule

LEfSe Analysis and Visualization Errors
& 44.59 & 1.11 & 48.69 & 0.0 & 399.0 \\

Kraken/Bracken Database and Taxonomic Reporting Support
& 47.95 & 1.32 & 48.33 & 1.0 & 571.0 \\

Differential Expression Input and Execution Errors
& 48.60 & 1.00 & 109.58 & 0.0 & 460.0 \\

Long-Running Tool Execution and Server Delays
& 51.58 & 1.18 & 53.79 & 0.0 & 412.0 \\

Reference Genome and Annotation Resource Availability
& 52.05 & 1.00 & 84.67 & 0.0 & 486.0 \\

File, Dataset, and Output Troubleshooting
& 52.46 & 1.15 & 66.54 & 0.0 & 422.0 \\

FTP Upload and Large-Data Transfer Problems
& 52.94 & 1.66 & 38.82 & 0.5 & 270.5 \\

Local Galaxy Installation and Dependency Configuration
& 52.94 & 1.40 & 40.74 & 0.0 & 136.0 \\

Account Authentication and Access Management
& 54.32 & 1.06 & 144.45 & 1.0 & 452.0 \\

Microbial and Taxonomic Sequence Analysis Guidance
& 56.00 & 1.10 & 44.44 & 0.0 & 422.0 \\

Storage Quota and History Cleanup Support
& 56.86 & 0.72 & 79.91 & 0.0 & 440.0 \\

Genome Assembly and Annotation Tool Failures
& 57.34 & 1.03 & 34.06 & 1.0 & 378.0 \\

Paired-End Read and FASTQ Collection Handling
& 58.33 & 0.81 & 58.70 & 0.0 & 437.5 \\

Workflow Invocation and Execution Problems
& 59.31 & 1.33 & 56.57 & 0.0 & 419.0 \\

\bottomrule
\end{tabularx}

\vspace{1mm}
\begin{minipage}{0.8\textwidth}
\footnotesize
\textit{Note.} Accepted (\%) denotes the proportion of discussions with an
accepted answer. Median and mean days denote the time from discussion creation
to the creation of the post that was eventually accepted. Replies and views
report the median values within each topic.
\end{minipage}
\end{table}

\textbf{Bivariate associations with forum resolution.}

Community engagement showed the strongest bivariate associations with
accepted-answer status. Discussions receiving likes had a substantially higher
accepted-answer rate than those without likes (69.65\% vs.\ 31.83\%;
OR=4.91), as did discussions receiving replies (69.10\% vs.\ 38.77\%;
OR=3.53) and those containing tags (58.44\% vs.\ 22.30\%; OR=4.89).
Several diagnostic and contextual characteristics were also positively
associated with accepted-answer status, including account, attachment, code
block, storage, upload, history/data, image, version, reference-genome, error,
and link-related information (Table~\ref{tab:forum_bivariate}). All
associations reported in the table remained significant after
Benjamini--Hochberg correction. Because engagement characteristics such as
replies and likes accumulate during a discussion's lifecycle, these
relationships are interpreted as associations rather than causal effects.

\begin{table}[t]
\caption{Selected bivariate associations with accepted-answer status.}
\label{tab:forum_bivariate}
\centering

\small
\renewcommand{\arraystretch}{1.10}
\setlength{\tabcolsep}{7pt}

\begin{tabularx}{0.65\textwidth}{
@{}
>{\RaggedRight\arraybackslash}X
r
r
r
@{}
}
\toprule
\textbf{Feature} &
\textbf{Absent (\%)} &
\textbf{Present (\%)} &
\textbf{OR} \\
\midrule
Likes                    & 31.83 & 69.65 & 4.91 \\
Reply                    & 38.77 & 69.10 & 3.53 \\
Tags                     & 22.30 & 58.44 & 4.89 \\
Account mention          & 52.15 & 60.29 & 1.39 \\
Attachment               & 51.71 & 58.00 & 1.29 \\
Code block               & 51.99 & 58.10 & 1.28 \\
Storage mention          & 52.22 & 58.31 & 1.28 \\
Upload mention           & 51.31 & 56.85 & 1.25 \\
History/data mention     & 51.24 & 56.37 & 1.23 \\
Image                    & 51.72 & 56.87 & 1.23 \\
Version mention          & 51.78 & 56.43 & 1.21 \\
Reference-genome mention & 51.39 & 55.64 & 1.19 \\
Error mention            & 50.88 & 54.70 & 1.17 \\
Link                     & 51.19 & 55.10 & 1.17 \\
\bottomrule
\end{tabularx}

\vspace{1mm}
\begin{minipage}{0.7\textwidth}
\footnotesize
\textit{Note.} Absent and Present denote accepted-answer rates when the
corresponding feature is absent or present, respectively. OR denotes the odds
ratio for accepted-answer status when the feature is present versus absent.
All reported associations remained statistically significant after
Benjamini--Hochberg correction ($p_{\mathrm{adj}}<0.05$).
\end{minipage}
\end{table}

\textbf{Numeric characteristics and accepted-answer status.}

Discussions with accepted answers contained more posts than those without
accepted answers (median 3 vs.\ 2; $\delta=0.44$) and received more likes
(median 1 vs.\ 0; $\delta=0.41$), both representing medium effect sizes.
Accepted discussions also had longer activity spans
($\delta=0.38$), more replies ($\delta=0.30$), and more tags
($\delta=0.28$). Differences in views, textual length, attachments,
code blocks, images, and links were statistically significant but had
negligible effect sizes.

\textbf{Bivariate associations with accepted-answer timing.}

Among the 3,270 discussions with an available accepted-answer timestamp,
several characteristics were associated with the time to the post that was
eventually accepted. Discussions receiving replies had a longer median time
than those without replies (1.52 vs.\ 0.74 days). Longer median times were
also observed for discussions mentioning errors (1.45 vs.\ 0.94 days),
versions (1.31 vs.\ 1.09 days), job execution (1.31 vs.\ 1.07 days), and
reference genomes (1.27 vs.\ 1.08 days). In contrast, discussions carrying
a history-related tag had a shorter median time (0.39 vs.\ 1.14 days).
These associations remained significant after Benjamini--Hochberg correction
($p_{\mathrm{adj}}<0.05$).

For numeric characteristics, accepted-answer time was strongly correlated
with discussion activity span ($\rho=0.76$). Smaller positive correlations
were observed for post count ($\rho=0.24$), reply count ($\rho=0.20$),
views ($\rho=0.12$), text length, and number of tags. These relationships
should be interpreted cautiously because several engagement measures
accumulate over the discussion lifecycle.

\textbf{Multivariable associations with accepted-answer status.}

The multivariable logistic regression identified several characteristics
associated with accepted-answer status (Table~\ref{tab:forum_logistic}).
Higher numbers of posts (OR=6.19, 95\% CI: 4.57--8.37) and likes
(OR=2.64, 95\% CI: 2.35--2.97) showed the strongest positive associations.
The number of tags was also positively associated with accepted-answer status
(OR=1.35, 95\% CI: 1.29--1.42), as were longer discussion activity spans
(OR=1.12, 95\% CI: 1.08--1.17) and account-related mentions
(OR=1.51, 95\% CI: 1.16--1.97). In contrast, higher reply counts
(OR=0.41, 95\% CI: 0.34--0.50) and longer discussion text
(OR=0.86, 95\% CI: 0.77--0.95) were negatively associated with
accepted-answer status after controlling for the other included variables.
All reported associations remained significant after Benjamini--Hochberg
correction. Because engagement measures such as posts, replies, likes, and
activity span accumulate during a discussion's lifecycle, these estimates
are interpreted as associations rather than causal effects.

\begin{table}[t]
\caption{Significant multivariable associations with accepted-answer status.}
\label{tab:forum_logistic}
\centering

\small
\renewcommand{\arraystretch}{1.10}
\setlength{\tabcolsep}{8pt}

\begin{tabularx}{0.65\textwidth}{
@{}
>{\RaggedRight\arraybackslash}X
r
r
r
@{}
}
\toprule
\textbf{Factor} &
\textbf{OR} &
\textbf{95\% CI} &
\textbf{$p_{\mathrm{adj}}$} \\
\midrule
Post count$^{a}$    & 6.19 & 4.57--8.37 & $<.001$ \\
Like count$^{a}$    & 2.64 & 2.35--2.97 & $<.001$ \\
Account mention     & 1.51 & 1.16--1.97 & .015 \\
Number of tags      & 1.35 & 1.29--1.42 & $<.001$ \\
Activity span$^{a}$ & 1.12 & 1.08--1.17 & $<.001$ \\
Text length$^{a}$   & 0.86 & 0.77--0.95 & .029 \\
Reply count$^{a}$   & 0.41 & 0.34--0.50 & $<.001$ \\
\bottomrule
\end{tabularx}

\vspace{1mm}
\begin{minipage}{0.65\textwidth}
\footnotesize
\textit{Note.} OR denotes odds ratio and CI denotes confidence interval.
$p_{\mathrm{adj}}$ denotes the Benjamini--Hochberg-adjusted $p$-value.
$^{a}$Variable was log-transformed before model estimation.
\end{minipage}
\end{table}

\textbf{Time-to-accepted-answer analysis.}

Kaplan--Meier analysis showed significant differences in accepted-answer
timing across several discussion characteristics. Discussions with replies
had a longer median time to the eventually accepted post than those without
replies (1.52 vs.\ 0.74 days), whereas discussions with account-related tags
(0.62 vs.\ 1.13 days) and history-related tags (0.39 vs.\ 1.14 days)
reached the eventually accepted post more quickly. Differences were also
observed for likes, tags, attachments, images, links, code blocks, and several
technical-context indicators after Benjamini--Hochberg correction. The
time-to-event analysis included 6,226 discussions, of which 3,270 experienced
the accepted-answer event; discussions without an observed event were
right-censored.

\textbf{Multivariable time-to-accepted-answer analysis.}
The Cox proportional-hazards model identified several factors associated with
the rate of reaching the eventually accepted post
(Table~\ref{tab:forum_cox}). Higher post count (HR=1.53, 95\% CI:
1.38--1.69), like count (HR=1.57, 95\% CI: 1.48--1.67), number of tags
(HR=1.16, 95\% CI: 1.13--1.18), and account-related mentions
(HR=1.18, 95\% CI: 1.03--1.36) were associated with a higher resolution
rate. In contrast, higher reply count (HR=0.88, 95\% CI: 0.82--0.94),
view count (HR=0.94, 95\% CI: 0.92--0.97), activity span
(HR=0.97, 95\% CI: 0.96--0.99), and text length
(HR=0.94, 95\% CI: 0.89--1.00) were associated with a lower resolution
rate.

\begin{table}[t]
\caption{Significant associations with time to the eventually accepted post
in the Cox proportional-hazards model.}
\label{tab:forum_cox}
\centering

\small
\renewcommand{\arraystretch}{1.10}
\setlength{\tabcolsep}{8pt}

\begin{tabularx}{0.75\textwidth}{
@{}
>{\RaggedRight\arraybackslash}X
r
r
r
@{}
}
\toprule
\textbf{Factor} &
\textbf{HR} &
\textbf{95\% CI} &
\textbf{$p$} \\
\midrule
Like count$^{a}$    & 1.57 & 1.48--1.67 & $<.001$ \\
Post count$^{a}$    & 1.53 & 1.38--1.69 & $<.001$ \\
Account mention     & 1.18 & 1.03--1.36 & .017 \\
Number of tags      & 1.16 & 1.13--1.18 & $<.001$ \\
Activity span$^{a}$ & 0.97 & 0.96--0.99 & .008 \\
View count$^{a}$    & 0.94 & 0.92--0.97 & $<.001$ \\
Text length$^{a}$   & 0.94 & 0.89--1.00 & .034 \\
Reply count$^{a}$   & 0.88 & 0.82--0.94 & $<.001$ \\
\bottomrule
\end{tabularx}

\vspace{1mm}
\begin{minipage}{0.75\textwidth}
\footnotesize
\textit{Note.} HR denotes the hazard ratio and CI denotes the confidence
interval. An HR $>1$ indicates a higher instantaneous rate of reaching the
eventually accepted post, whereas an HR $<1$ indicates a lower instantaneous
rate. $^{a}$Variable was log-transformed before model estimation.
\end{minipage}
\end{table}

Because post counts, replies, likes, views, and activity span may accumulate
over the course of a discussion, these estimates are interpreted as lifecycle
associations rather than causal or prospective predictors of resolution.

\textbf{Kaplan--Meier analysis.}

Kaplan--Meier curves further illustrated differences in accepted-answer
timing across discussion characteristics. Discussions containing tags showed
a substantially lower probability of remaining without an accepted answer
than discussions without tags. A similar pattern was observed for discussions
mentioning account-related concerns. Discussions containing code blocks also
showed a modestly faster transition toward an accepted answer. These
differences were supported by log-rank tests, although the magnitude of
separation varied across characteristics. The remaining Kaplan–Meier comparisons were examined as part of the analysis and shared in the replication
package \cite{alam_2026_22882299}.

\begin{figure*}[t]
    \centering

    \begin{subfigure}[t]{0.32\textwidth}
        \centering
        \includegraphics[width=\linewidth]
        {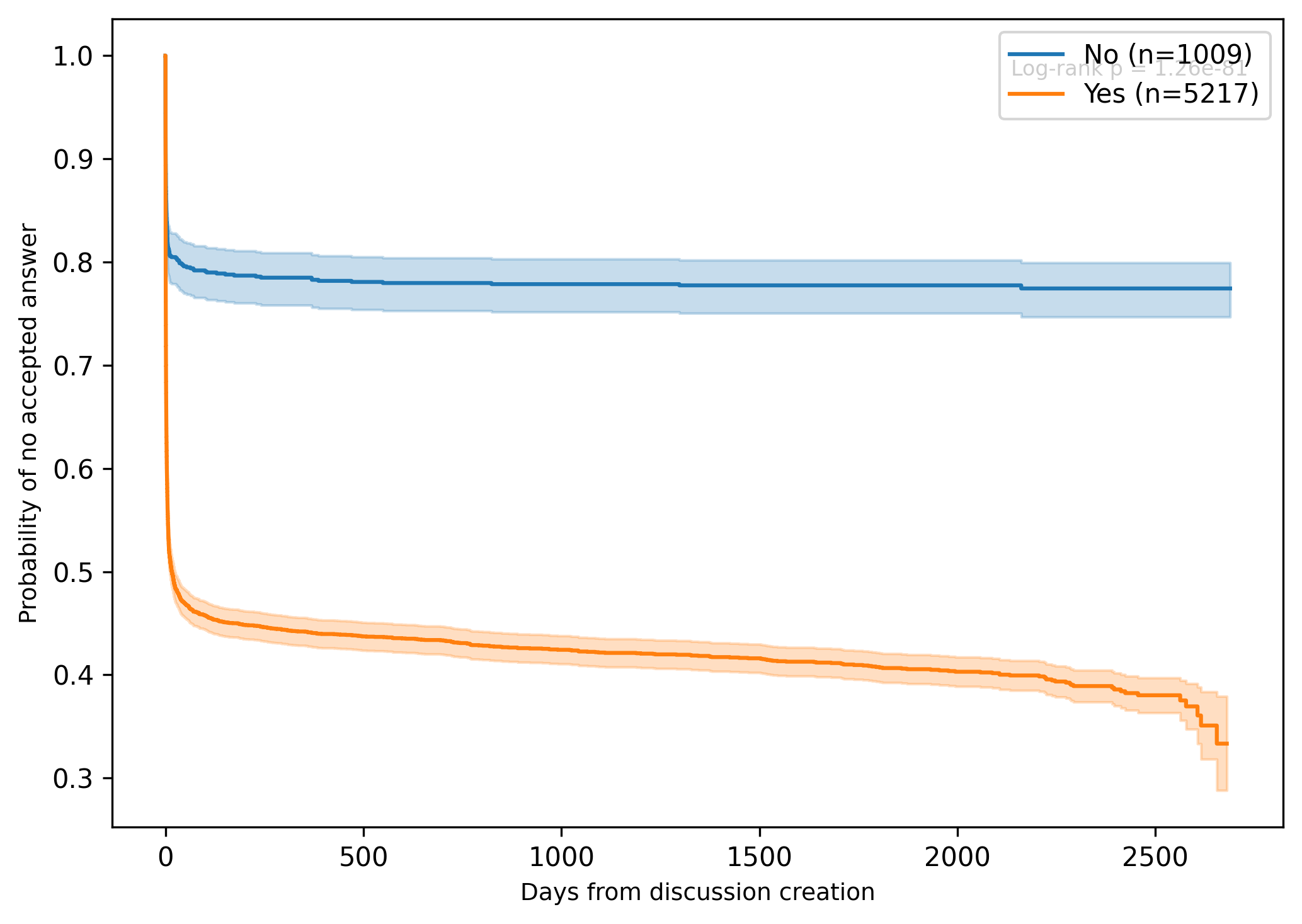}
        \caption{Presence of tags.}
        \label{fig:km_forum_tags}
    \end{subfigure}
    \hfill
    \begin{subfigure}[t]{0.32\textwidth}
        \centering
        \includegraphics[width=\linewidth]
        {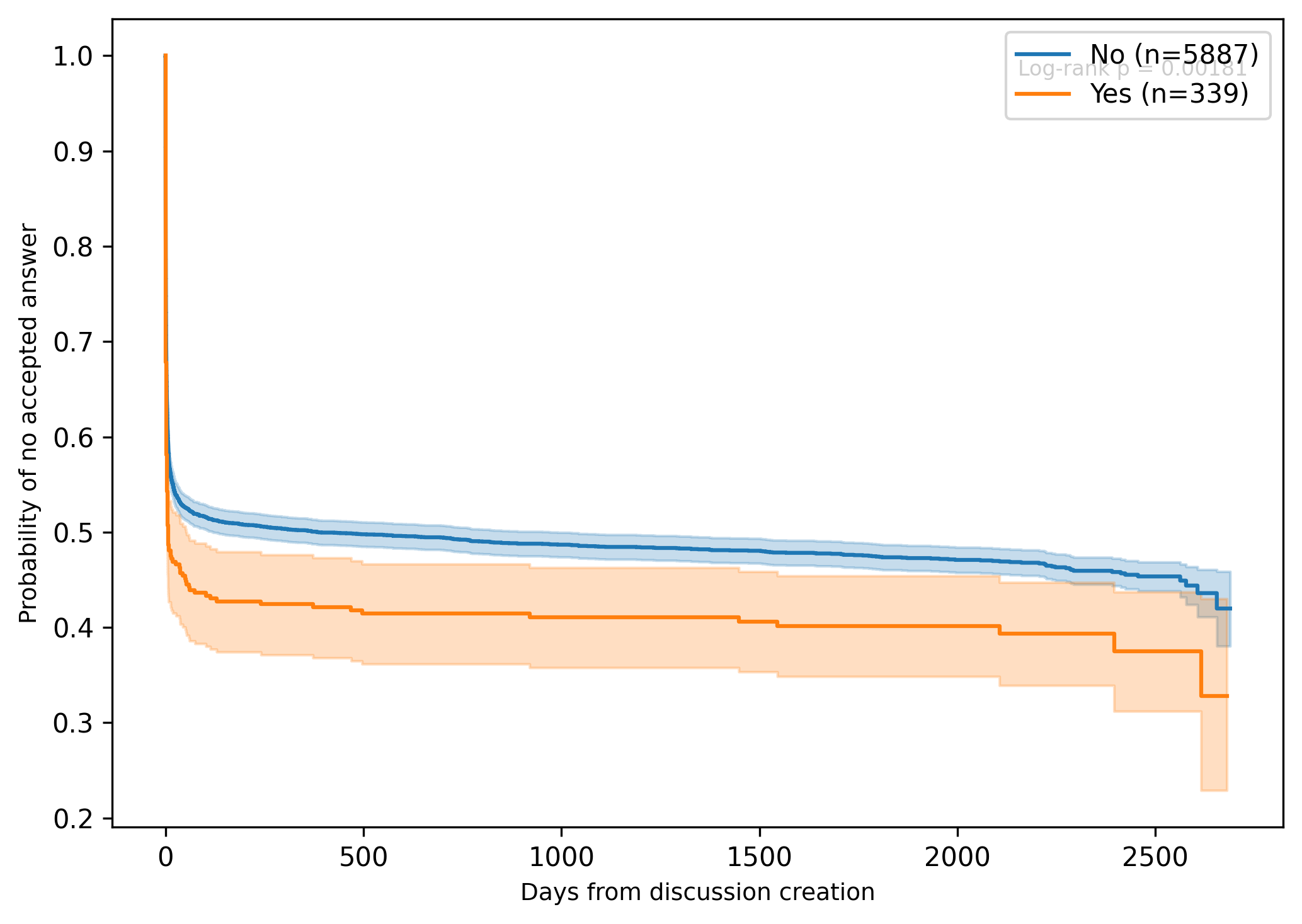}
        \caption{Account-related mention.}
        \label{fig:km_forum_account}
    \end{subfigure}
    \hfill
    \begin{subfigure}[t]{0.32\textwidth}
        \centering
        \includegraphics[width=\linewidth]
        {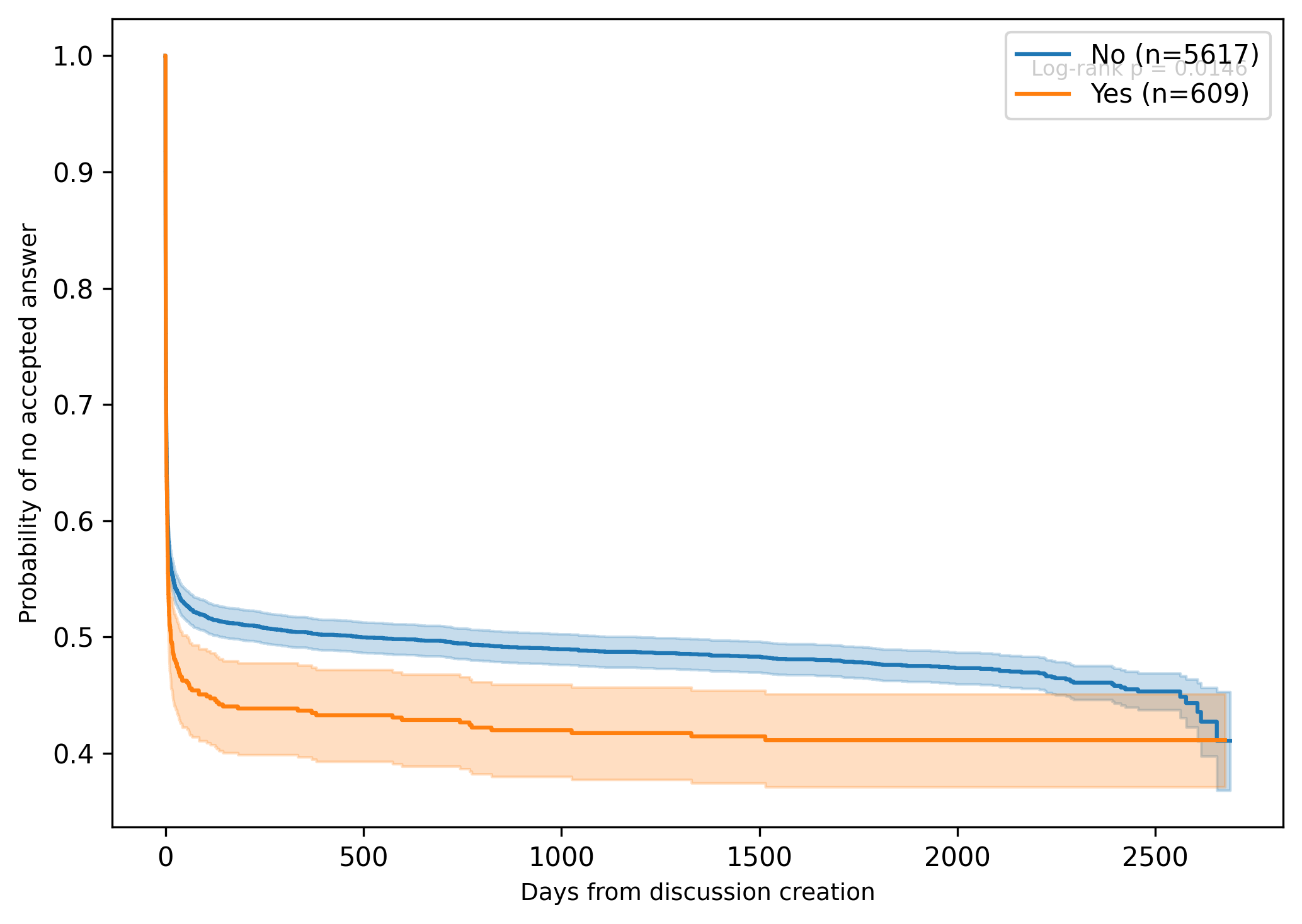}
        \caption{Presence of a code block.}
        \label{fig:km_forum_code}
    \end{subfigure}

    \caption{Kaplan--Meier estimates of time to the eventually accepted post
    for selected Galaxy Community Help Forum characteristics. Lower curves indicate
    a lower probability of remaining without an accepted answer and therefore
    faster progression toward the accepted-answer event.}
    \label{fig:forum_km_selected}
\end{figure*}

\begin{tcolorbox}[
    title=\textbf{RQ2 Summary},
    colback=gray!5,
    colframe=black!60,
    boxrule=0.5pt,
    arc=1mm,
    left=4pt,
    right=4pt,
    top=4pt,
    bottom=4pt
]
\small
Resolution outcomes and resolution speed showed distinct patterns across Galaxy's maintenance and support spaces. For issues, milestones, assignment, and diagnostic information were associated with closure and resolution trajectories. Pull-request outcomes were strongly associated with workflow state, contributor role, and automation, while forum resolution was primarily associated with engagement and organizational characteristics. Overall, the results show that understanding maintenance effectiveness requires considering both whether artifacts are resolved and how quickly resolution occurs.

\end{tcolorbox}

\subsection{RQ3: Cross-Space Traceability and Problem--Solution Connections}

\textbf{RQ3} examines how maintenance and support knowledge is connected across Galaxy's development and community-support spaces. We analyze explicit references, candidate semantic connections, and recurring technical signals across GitHub issues, pull requests, and Community Forum discussions to assess both recorded traceability and broader cross-space relatedness.

\subsubsection{Motivation}

Maintenance knowledge in a community-driven SWS can be distributed across multiple development and support spaces. A problem reported by a user may appear in a Community Forum discussion, subsequently be formalized as a GitHub issue, and later motivate an implementation in a pull request. Conversely, a repository-side change may address a recurring user-facing problem without being explicitly linked back to the corresponding support discussion. When such relationships are not explicitly recorded, problem context, implementation rationale, and reusable solutions become fragmented across the ecosystem. \textbf{RQ3} therefore investigates the extent to which Galaxy maintenance artifacts are explicitly connected, whether additional potentially related artifacts can be identified through semantic similarity and temporal proximity, and how technical concerns are distributed across repository-centered development and community-centered support spaces.

\subsubsection{Approach}

As described in Section~\ref{sec:rq3_approach}, we examined cross-space problem--solution connections using three complementary analyses. First, we quantified explicit traceability among GitHub issues, pull requests, and Community Forum discussions to determine where maintenance relationships are directly recorded. Second, we examined semantically related but explicitly unlinked artifacts to identify additional candidate connections across development and support spaces, considering their temporal ordering. Third, we compared shared technical signals across the three artifact types to determine whether similar maintenance concerns recur across spaces even when individual artifacts are not directly connected. We synthesized these results to distinguish strong, explicitly recorded traceability from weaker cross-space relatedness. Semantic matches are therefore interpreted as candidate connections rather than confirmed problem--solution relationships.

\subsubsection{Results of RQ3} The analysis reveals a strong contrast between repository-internal traceability and cross-space connectivity. Explicit relationships are concentrated within GitHub, while links between GitHub and the Community Forum are comparatively rare. At the same time, semantic matching and shared technical signals reveal additional relatedness across development and support spaces beyond what is captured through explicit references.

\paragraph{\textbf{Explicit traceability is concentrated within GitHub.}}

Across the 70,200 artifacts, we identified 16,426 distinct resolved explicit source--target relationships (Table~\ref{tab:rq3_explicit_links}). The largest category consisted of pull request-to-pull request references ($n=8{,}346$, 50.81\%), followed by pull request-to-issue ($n=5{,}036$, 30.66\%), issue-to-pull request ($n=1{,}488$, 9.06\%), and issue-to-issue references ($n=1{,}189$, 7.24\%). Together, relationships whose source and target were both GitHub artifacts accounted for 16,059 links, or 97.77\% of all resolved explicit relationships.

Pull request-to-issue relationships provide particularly direct evidence of connections between reported problems and implementation activity. Of the 5,036 such relationships, 1,605 (31.87\%) contained a closing-keyword reference such as \textit{fixes}, \textit{closes}, or \textit{resolves}. These relationships provide stronger traceability evidence than generic artifact references because the pull-request text explicitly characterizes its relationship to the referenced issue.

\begin{table}[t]
\caption{Explicit relationships identified among Galaxy maintenance and
support artifacts.}
\label{tab:rq3_explicit_links}
\centering

\small
\renewcommand{\arraystretch}{1.10}
\setlength{\tabcolsep}{9pt}

\begin{tabular}{lrr}
\toprule
\textbf{Connection Type} &
\textbf{Links} &
\textbf{(\%)} \\
\midrule
Pull request $\rightarrow$ Pull request & 8,346 & 50.81 \\
Pull request $\rightarrow$ Issue        & 5,036 & 30.66 \\
Issue $\rightarrow$ Pull request        & 1,488 & 9.06  \\
Issue $\rightarrow$ Issue               & 1,189 & 7.24  \\
GitHub $\rightarrow$ Forum              & 278   & 1.69  \\
Forum $\rightarrow$ Forum               & 73    & 0.44  \\
Forum $\rightarrow$ GitHub              & 16    & 0.10  \\
\midrule
\textbf{Total} &
\textbf{16,426} &
\textbf{100.00} \\
\bottomrule
\end{tabular}
\end{table}

\paragraph{\textbf{Explicit traceability across GitHub and the forum is sparse.}} 

Only 294 resolved explicit relationships connected GitHub artifacts with Community Forum discussions, corresponding to 1.79\% of all resolved links. Of these, 278 relationships ran from GitHub artifacts to forum discussions, whereas only 16 ran from forum discussions to GitHub issues or pull requests. Thus, although explicit relationships are strongly concentrated within GitHub, direct traceability between repository-centered development and community-centered support spaces is comparatively uncommon.

\paragraph{\textbf{Semantic matching reveals additional candidate relationships.}} After removing explicitly linked artifact pairs, semantic matching identified 4,882 candidate connections (Table~\ref{tab:rq3_semantic}). Most ($n=4{,}134$, 84.68\%) connected issues and pull requests. Within the configured temporal window, 3,258 candidate pairs placed the issue before the pull request and had a median cosine similarity of 0.361 and a median temporal gap of 63.08 days. A further 876 candidates placed the pull request before the issue, with a median similarity of 0.365 and a median gap of 10.27 days.

The remaining 748 candidates (15.32\%) bridged the Community Forum and GitHub. These included 421 forum--pull request and 327 forum--issue candidate pairs. For forum--pull request candidates, 352 placed the forum discussion first (median similarity $=0.349$; median gap $=73.43$ days), whereas 69 placed the pull request first (median similarity $=0.339$; median gap $=10.08$ days). For forum--issue candidates, 253 placed the forum discussion first (median similarity $=0.380$; median gap $=63.32$ days), whereas 74 placed the issue first (median similarity $=0.370$; median gap $=8.83$ days).

Because candidate generation allowed targets up to 180 days after but only 30 days before the source artifact, these directional counts should not be interpreted as evidence that one artifact type or platform generally precedes another. Instead, they show that semantically similar artifacts occur across development and support spaces even when no explicit traceability relationship is recorded.

\begin{table}[t]
\caption{Candidate semantic connections after removing explicitly linked
artifact pairs.}
\label{tab:rq3_semantic}
\centering

\small
\renewcommand{\arraystretch}{1.10}
\setlength{\tabcolsep}{7pt}

\begin{tabularx}{0.75\textwidth}{
@{}
>{\RaggedRight\arraybackslash}X
r
r
r
@{}
}
\toprule
\textbf{Temporal Order} &
\multicolumn{1}{c}{\textbf{Pairs}} &
\multicolumn{1}{c}{\textbf{Median Similarity}} &
\multicolumn{1}{c}{\textbf{Median Gap (days)}} \\
\midrule

Issue $\rightarrow$ Pull request
& 3,258 & 0.361 & 63.08 \\

Pull request $\rightarrow$ Issue
& 876 & 0.365 & 10.27 \\

Forum $\rightarrow$ Pull request
& 352 & 0.349 & 73.43 \\

Forum $\rightarrow$ Issue
& 253 & 0.380 & 63.32 \\

Issue $\rightarrow$ Forum
& 74 & 0.370 & 8.83 \\

Pull request $\rightarrow$ Forum
& 69 & 0.339 & 10.08 \\

\midrule
\textbf{Total}
& \textbf{4,882} & -- & -- \\
\bottomrule
\end{tabularx}

\vspace{1mm}
\begin{minipage}{0.75\textwidth}
\footnotesize
\textit{Note.} Arrows denote temporal ordering within the candidate-generation
window rather than a confirmed causal or problem--solution relationship.
Explicitly linked artifact pairs were removed before semantic matching.
\end{minipage}
\end{table}

\textbf{Manual validation of semantic candidates.}

Manual validation showed that semantic matching recovered substantively related artifacts across all three candidate families (Table~\ref{tab:rq3_manual_validation}). Among issue--pull request candidates, 18 of 119 pairs (15.1\%) were classified as exact or strongly related and 39 (32.8\%) as broadly related, resulting in 47.9\% exhibiting at least broad relatedness. The corresponding proportion was 53.8\% for forum--issue candidates, including 28 (23.5\%) exact or strongly related and 36 (30.3\%) broadly related pairs. Forum--pull request candidates showed the highest proportion of related pairs: 27 (22.7\%) were exact or strongly related and 46 (38.7\%) were broadly related, for a combined 61.3\%.

Across the validation sample, 73 pairs (20.4\%) were exact or strongly related, 121 (33.9\%) were broadly related, 58 (16.2\%) were weakly related, and 105 (29.4\%) were unrelated. Thus, semantic matching frequently surfaced substantively related artifacts that were not explicitly linked, but it also produced a non-trivial proportion of unrelated candidates. This result supports our conservative treatment of these matches as candidate semantic connections rather than confirmed traceability relationships.

\begin{table}[t]
\caption{Manual validation of candidate semantic connections.}
\label{tab:rq3_manual_validation}
\centering

\small
\renewcommand{\arraystretch}{1.10}
\setlength{\tabcolsep}{6pt}

\begin{tabularx}{0.70\textwidth}{
@{}
>{\RaggedRight\arraybackslash}X
>{\centering\arraybackslash}p{0.10\textwidth}
>{\centering\arraybackslash}p{0.10\textwidth}
>{\centering\arraybackslash}p{0.10\textwidth}
>{\centering\arraybackslash}p{0.10\textwidth}
@{}
}
\toprule
\textbf{Family} &
\textbf{Exact/Strong} &
\textbf{Broad} &
\textbf{Weak} &
\textbf{Unrelated} \\
\midrule

Issue--pull request
& 18 (15.1) & 39 (32.8) & 18 (15.1) & 44 (37.0) \\

Forum--issue
& 28 (23.5) & 36 (30.3) & 17 (14.3) & 38 (31.9) \\

Forum--pull request
& 27 (22.7) & 46 (38.7) & 23 (19.3) & 23 (19.3) \\

\bottomrule
\end{tabularx}

\vspace{1mm}
\begin{minipage}{0.70\textwidth}
\footnotesize
\textit{Note.} Values are reported as $n$ (\%) within each candidate family.
\end{minipage}
\end{table}

\textbf{Technical concerns recur across development and support spaces.}

The technical-signal analysis further shows that the three artifact types discuss overlapping technical concerns, but with substantially different emphases (Table~\ref{tab:rq3_signal_overlap}). Execution-error signals were most prevalent in forum discussions (36.42\%), compared with 22.96\% of issues and 8.18\% of pull requests. Forum discussions also showed comparatively high prevalence of history and data-management concerns (24.47\%), upload/storage/transfer concerns (21.27\%), and reference-genome or annotation concerns (13.36\%). Issues occupied a more intermediate position between user-facing support and implementation activity. History and data management (21.26\%), tool-wrapper integration (20.87\%), testing and validation (20.24\%), and infrastructure and deployment (20.05\%) all appeared frequently in issue reports.

Pull requests were most strongly characterized by implementation-oriented signals. Testing and validation appeared in 28.52\% of pull requests and training or documentation in 23.28\%, compared with 20.24\% and 18.47\% of issues and 3.88\% and 13.47\% of forum discussions, respectively. Tool-wrapper integration was also substantially more common in repository artifacts (20.87\% of issues and 14.83\% of pull requests) than in forum discussions (6.22\%).

These patterns indicate that related technical concerns are represented across the three spaces, while the form in which they appear reflects the role of each channel: the forum emphasizes user-facing failures and data handling, issues capture reported and coordinated maintenance concerns, and pull requests emphasize implementation, testing, and documentation.

\begin{table}[t]
\caption{Prevalence of technical signals across Galaxy artifact types.}
\label{tab:rq3_signal_overlap}
\centering

\small
\renewcommand{\arraystretch}{1.10}
\setlength{\tabcolsep}{7pt}

\begin{tabularx}{0.65\textwidth}{
@{}
>{\RaggedRight\arraybackslash}X
r
r
r
@{}
}
\toprule
\textbf{Technical Signal} &
\multicolumn{1}{c}{\textbf{Forum (\%)}} &
\multicolumn{1}{c}{\textbf{Issue (\%)}} &
\multicolumn{1}{c}{\textbf{Pull Request (\%)}} \\
\midrule

Execution error
& 36.42 & 22.96 & 8.18 \\

Workflow execution/invocation
& 7.94 & 13.07 & 6.84 \\

History/data management
& 24.47 & 21.26 & 15.12 \\

Tool-wrapper integration
& 6.22 & 20.87 & 14.83 \\

Testing/validation
& 3.88 & 20.24 & 28.52 \\

Dependency/package management
& 6.61 & 13.49 & 9.78 \\

Container/runtime
& 2.82 & 4.57 & 2.99 \\

Distributed execution
& 3.80 & 4.80 & 3.60 \\

Infrastructure/deployment
& 14.32 & 20.05 & 9.61 \\

Installation/configuration
& 12.69 & 17.71 & 9.67 \\

Authentication/account
& 5.50 & 4.37 & 1.93 \\

Training/documentation
& 13.47 & 18.47 & 23.28 \\

Reference genome/annotation
& 13.36 & 3.43 & 1.73 \\

Upload/storage/transfer
& 21.27 & 11.06 & 4.56 \\

\bottomrule
\end{tabularx}

\vspace{1mm}
\begin{minipage}{0.65\textwidth}
\footnotesize
\textit{Note.} Values denote the percentage of artifacts containing the
corresponding technical signal. Signals are not mutually exclusive; therefore,
a single artifact may contain multiple signals.
\end{minipage}
\end{table}

\paragraph{\textbf{Synthesis of cross-space problem--solution connections.}}

Taken together, the \textbf{RQ3} analyses reveal a clear distinction between repository-internal traceability and cross-space connectivity in the Galaxy ecosystem. Explicit relationships are strongly concentrated within GitHub, where issues and pull requests are frequently connected and implementation activity can often be traced to repository-level problem reports. In contrast, explicit relationships between GitHub and the Community Forum are rare, indicating that user-facing problems and repository-side maintenance activity are often documented in separate spaces. Nevertheless, semantic matching identified additional candidate relationships among forum discussions, issues, and pull requests after explicitly linked pairs were removed, while the technical-signal analysis showed that many of the same concerns recur across all three artifact types. These findings indicate that related maintenance knowledge spans development and support spaces even when the corresponding relationships are not explicitly recorded.

\begin{tcolorbox}[
    title=\textbf{RQ3 Summary},
    colback=gray!5,
    colframe=black!60,
    boxrule=0.5pt,
    arc=1mm,
    left=4pt,
    right=4pt,
    top=4pt,
    bottom=4pt
]
\small
The results reveal a clear difference between technical relatedness and explicit traceability in Galaxy. Explicit relationships are strongly concentrated within GitHub, accounting for 97.77\% of the 16,426 resolved relationships, whereas only 294 resolved relationships bridge GitHub and the Community Forum. Nevertheless, semantic matching identified 748 additional forum--GitHub candidate connections, and recurring technical signals show substantial overlap across the three channels. Overall, maintenance knowledge spans development and support spaces, but explicit cross-space traceability remains limited.
\end{tcolorbox}

\section{Discussion}
\label{sec:discussion}
Our findings provide an ecosystem-level perspective on how maintenance and support are organized and sustained in a large, community-driven SWS. Taken together, the three \textbf{RQs} reveal three broader insights. First, maintenance in Galaxy extends beyond conventional source-code repair to encompass workflows, scientific tools, data and reference resources, dependencies, computing infrastructure, documentation, and user-facing analytical support. Second, maintenance resolution is multidimensional: the characteristics associated with whether an artifact is ultimately resolved are not necessarily the same as those associated with how quickly resolution occurs. Third, although maintenance knowledge is strongly interconnected within repository-centered development activities, explicit connections between development and community-support spaces remain limited despite substantial technical relatedness across them. Collectively, these findings extend repository-centered views of software maintenance by showing that, in scientific workflow ecosystems, maintenance is a distributed socio-technical process spanning software evolution, scientific infrastructure, and community support. The following discussion considers the implications of these findings for understanding, studying, and supporting maintenance in Galaxy and similar scientific software ecosystems.

\subsection{Maintenance in SWSs Is an Ecosystem-Level Socio-Technical Activity}
\textbf{RQ1} shows that maintenance in Galaxy cannot be understood solely as source-code defect correction or software evolution. Across GitHub issues, pull requests, and Community Forum discussions, we observed recurring concerns involving workflow execution, data and history management, scientific tools and dependencies, distributed computing infrastructure, deployment, testing, reference resources, documentation, and user support. This breadth reflects the structure of SWSs, whose continued operation depends not only on the core platform but also on interconnected tools, workflows, software environments, data resources, computational infrastructure, and community-maintained knowledge \cite{DBLP:conf/apsec/AlamRS23, DBLP:journals/ese/AlamRRM25}. Maintenance in such systems is therefore distributed across both technical components and the communities that develop, operate, and use them.

Our findings extend prior research on scientific workflow development and maintenance. Previous studies have identified recurring challenges involving workflow execution, dependencies, data operations, documentation, scheduling, system evolution, and other aspects of scientific workflow development \cite{DBLP:journals/ese/AlamRRM25}. Research on Galaxy workflow reuse has further shown that tool upgrades, unavailable tools, incomplete workflows, dependency changes, and workflow-design problems can hinder continued workflow use \cite{DBLP:conf/apsec/AlamRS23}. Our results reinforce these observations but broaden the perspective by showing how maintenance concerns are distributed across both developer-facing and user-facing spaces within a mature scientific workflow ecosystem.

The three artifact types expose complementary layers of this maintenance process. GitHub issues primarily formalize defects, limitations, missing resources, and maintenance needs requiring developer attention, whereas pull requests capture implementation, integration, testing, automation, infrastructure changes, documentation, and other preventive or evolutionary maintenance activities. Community Forum discussions provide a different perspective by revealing how users experience these technical conditions while conducting scientific analyses. They include not only workflow and tool failures but also difficulties involving data organization, reference-genome selection, tool configuration, storage, authentication, and domain-specific analyses. This complementarity is consistent with prior software-engineering research showing that development knowledge and coordination are distributed across repositories, Q\&A systems, and other communication channels, with different platforms supporting different forms of technical interaction and knowledge sharing \cite{DBLP:journals/tse/StoreyZFSG17,DBLP:conf/socialcom/VasilescuFS13,DBLP:conf/msr/ZagalskyTGSP16}. Consequently, studies based on a single artifact type may capture only part of the maintenance process: repository data provide rich evidence of implementation and coordination, whereas community-support data expose operational, usability, and scientific-analysis difficulties that may never be formalized as repository issues.

These findings also highlight the hybrid technical and scientific character of SWS maintenance. Scientific workflows depend on evolving tools, software dependencies, data formats, reference resources, parameters, and execution environments, making their sustainability closely tied to the surrounding computational ecosystem \cite{alam2023supporting}. In this setting, a technically functioning component may still generate substantial support needs when users cannot configure its inputs, select appropriate reference resources, organize datasets correctly, or integrate it into a scientifically meaningful workflow. The boundary between software maintenance and scientific support is therefore less distinct than in conventional repository-centered views of maintenance. Our findings suggest that assessments of maintenance in SWSs may need to consider not only whether software components continue to function, but also whether the surrounding tools, data resources, infrastructure, and support knowledge enable users to reliably construct and execute scientific analyses.

More broadly, our multi-artifact findings suggest that SWS maintenance is best understood as an ecosystem-level socio-technical activity rather than as a collection of isolated software changes. Technical evolution, infrastructure operation, scientific-resource maintenance, and community support are interconnected, and changes or failures in one part of the ecosystem can generate maintenance and support needs elsewhere. For empirical software-engineering research, this perspective indicates that repository-centered evidence alone may provide an incomplete account of maintenance in scientific software. Examining development and support spaces together offers a more comprehensive view of where maintenance demands arise, how they manifest for developers and users, and how technical and scientific concerns interact across the broader ecosystem.

\subsection{Resolution Is Not a Single Maintenance Outcome}

\textbf{RQ2} shows that maintenance resolution should not be treated as a single binary outcome. Across GitHub issues, pull requests, and Community Forum discussions, characteristics associated with whether an artifact was eventually resolved were not always associated with faster resolution. For issues, milestones, assignees, and diagnostic information such as error descriptions and code blocks were associated with resolution, whereas high levels of discussion or coordination sometimes coincided with longer resolution times. This is consistent with prior research showing that bug-report quality, coordination, and other artifact characteristics influence issue resolution and bug-fixing time \cite{DBLP:journals/tse/ZimmermannPBJSW10,DBLP:conf/icse/ZhangGV13}.

A similar pattern emerged for pull requests. Contributor role, draft status, automation, and maintenance context were associated with both integration outcomes and resolution speed, reinforcing prior findings that pull-request decisions depend on a combination of technical and social factors \cite{DBLP:conf/icse/GousiosPD14,DBLP:conf/icse/GousiosZSD15,tsay2014influence,DBLP:journals/tse/ZhangYGR23}. However, characteristics such as testing, review, documentation, or dependency-related information may also indicate more complex changes requiring additional validation rather than inefficient maintenance.

Forum discussions further demonstrate that engagement should be interpreted carefully. Posts, replies, likes, and activity duration may reflect progress toward resolution, problem difficulty, or both. More generally, many coordination and engagement variables accumulate during an artifact's lifecycle and therefore should be interpreted as associations rather than causal or prospective predictors of resolution.

These findings suggest that maintenance effectiveness should be evaluated along at least two complementary dimensions: \emph{whether} artifacts are eventually resolved and \emph{how long} resolution takes. Reporting only closure, merge, or accepted-answer rates can obscure a long tail of difficult artifacts that remain unresolved for extended periods. Resolution likelihood and resolution latency should therefore be considered jointly when assessing maintenance processes in scientific software ecosystems.

\subsection{Maintenance Topics Differ in Both Resolution Likelihood and Resolution Latency}

The topic-level results show that maintenance burden is not distributed uniformly across technical concerns. Some topics exhibit high eventual resolution rates but comparatively long resolution times, while others are resolved more quickly despite lower overall resolution rates. For example, Software Dependency and Package Management issues had a relatively high closure rate but among the longest median times to closure, whereas Authentication and Runtime Issues were resolved more quickly. Pull-request topics showed similar variation, with deployment and service-infrastructure changes generally progressing faster than installation, upgrade-validation, and tool-integration work.

These differences are consistent with prior research showing that issue-resolution time and pull-request outcomes depend on the technical characteristics and complexity of the underlying work \cite{DBLP:conf/icse/ZhangGV13,DBLP:conf/icse/GousiosPD14,DBLP:journals/tse/ZhangYGR23}. In scientific workflow ecosystems, such variation may be particularly pronounced because maintenance tasks often depend on external tools, software versions, execution environments, scientific resources, and infrastructure \cite{alam2023supporting}.

These findings suggest that aggregate closure or merge rates alone can provide an incomplete picture of maintenance performance. Maintenance metrics should therefore be interpreted in relation to the type of work being performed. Topics involving dependencies, installation, tool integration, workflow execution, and infrastructure may warrant specialized testing, validation, or diagnostic support because they repeatedly exhibit distinct resolution dynamics.

\subsection{A Cross-Space Traceability Gap Separates Support from Development}

\textbf{RQ3} reveals a substantial gap between technical relatedness and explicit traceability across Galaxy's development and support spaces. Of the 16,426 resolved explicit relationships, 97.77\% occurred between GitHub artifacts, indicating strong connectivity among repository-centered maintenance activities. In contrast, only 294 relationships (1.79\%) explicitly connected GitHub artifacts with Community Forum discussions, even though forum discussions frequently addressed maintenance-relevant concerns such as workflow failures, installation problems, storage limitations, reference resources, and infrastructure issues. This imbalance suggests that problem and solution knowledge is well connected within repository-centered development but remains only sparsely linked to community-facing support. This finding is consistent with prior research showing that software-development knowledge is distributed across repositories, Q\&A platforms, and other communication channels \cite{DBLP:journals/tse/StoreyZFSG17,DBLP:conf/socialcom/VasilescuFS13,DBLP:conf/msr/ZagalskyTGSP16}. Our results extend this perspective by showing that, within a scientific workflow ecosystem, technically related maintenance knowledge can exist across development and support spaces without being explicitly connected. 

The semantic analysis provides further evidence of this fragmentation. After removing known explicit links, semantic matching identified additional candidate relationships among forum discussions, issues, and pull requests, complementing prior work on recovering missing traceability links among software artifacts \cite{DBLP:conf/sigsoft/WuZKC11,DBLP:journals/re/LudersPM23,DBLP:conf/saner/YasaOAKDUT25}. However, semantic similarity and temporal proximity indicate potential relatedness rather than confirmed problem--solution relationships. Two artifacts may concern the same component or technical problem without one directly motivating or resolving the other.

This distinction between relatedness and traceability is important for maintenance knowledge management. Explicit references preserve intentional connections that allow users and maintainers to move between problem reports, implementation activities, and support discussions. Our findings therefore suggest that the principal cross-space challenge is not necessarily the absence of relevant knowledge, but the limited mechanisms for connecting that knowledge across platforms. Strengthening bidirectional traceability between forum discussions, issues, and pull requests could help preserve user-facing problem context together with the maintenance activities and solutions that address it.

\subsection{Implications for Maintenance Tools and Community Practice}

The findings suggest several opportunities for improving Galaxy and similar
community-driven scientific software ecosystems.

First, issue and support interfaces could encourage the capture of actionable
diagnostic information. Error descriptions, code blocks, reproduction
information, versions, and URLs repeatedly appeared in the resolution
analyses. Templates that prompt users for these elements may reduce the effort
required to reconstruct the execution context of a reported problem. Such
templates should remain lightweight, however, because excessive reporting
requirements may discourage participation.

Second, maintenance tools could provide lifecycle-aware triage rather than
relying only on static labels or activity counts. For example, an issue with
many comments is not necessarily progressing efficiently, and a pull request
with extensive testing or review may simply represent a more complex change.
Triage systems should therefore distinguish between evidence of engagement,
evidence of diagnostic completeness, and evidence of prolonged unresolved
complexity.

Third, the cross-space traceability gap suggests an opportunity for automated
link recommendation. When a forum discussion resembles an existing issue or
pull request, a system could suggest the repository artifact to the user or
maintainer. Conversely, when a new issue resembles a recurring forum
discussion, maintainers could be prompted to preserve the original support
context. Such systems should present similarity as a recommendation rather
than automatically asserting equivalence, because our semantic analysis shows
that textual relatedness alone cannot confirm a problem--solution
relationship.

Fourth, resolved forum discussions represent a potentially valuable
maintenance knowledge base. Accepted answers often contain configuration
guidance, workarounds, explanations of tool behavior, or information about
known infrastructure limitations. Linking these discussions to relevant
issues, pull requests, documentation, or training materials could help convert
ephemeral support knowledge into reusable ecosystem knowledge.

Finally, the high volume of preventive, automated, and infrastructure-related
work visible in pull requests suggests that maintenance-support tools should
not focus solely on reactive bug fixing. Dependency updates, testing,
container maintenance, deployment, documentation, and tool-version
synchronization represent substantial portions of the maintenance workload.
Automation that supports these recurring tasks may reduce maintenance burden
before problems reach users.

\subsection{Actionable Recommendations}
\label{sec:recommendations}

The findings across the three \textbf{RQs} point to several practical opportunities for improving maintenance and support in Galaxy and similar community-driven scientific workflow ecosystems. \textbf{RQ1} identifies recurring maintenance areas that require sustained developer and community attention, including workflow execution, dependencies, infrastructure, tool integration, scientific resources, and user support. \textbf{RQ2} shows that resolution is multidimensional: characteristics associated with eventual resolution do not necessarily correspond to faster resolution, and different maintenance topics exhibit distinct resolution dynamics. \textbf{RQ3} further reveals that, although maintenance knowledge is strongly connected within GitHub, explicit traceability between repository-centered development and community-facing support remains limited. Based on these findings, we derive the actionable recommendations summarized in Table~\ref{tab:recommendations}.

\begin{table*}[t]
\centering
\caption{Actionable recommendations derived from the empirical findings.}
\label{tab:recommendations}
\small
\renewcommand{\arraystretch}{1.15}
\setlength{\tabcolsep}{4pt}

\begin{tabularx}{0.9\textwidth}{
p{0.15\textwidth}
p{0.35\textwidth}
X
}
\toprule
\textbf{Recommendation} &
\textbf{Empirical Basis} &
\textbf{Suggested Action} \\
\midrule

Improve diagnostic reporting &
Error descriptions, code blocks, reproduction information, and URLs were
associated with issue resolution and, for several signals, faster closure. &
Use lightweight templates requesting errors, software or tool versions,
reproduction steps, execution context, and relevant logs or links. \\\cmidrule{2-3}

Adopt lifecycle-aware triage &
Assignees and milestones were associated with favorable resolution patterns,
whereas high activity did not necessarily correspond to faster resolution. &
Prioritize artifacts using ownership, age, milestone status, diagnostic
completeness, and unresolved duration rather than activity volume alone. \\

\cmidrule{2-3}

Target difficult maintenance areas &
Dependencies, installation, upgrades, workflow execution, and tool integration
showed distinct resolution likelihoods and lifecycle durations. &
Provide targeted testing, validation, diagnostic guidance, and maintainer
support for recurring difficult areas. \\

\cmidrule{2-3}

Apply automation selectively &
Automation-related characteristics showed different associations across
contribution types and maintenance contexts. &
Automate repetitive and well-validated tasks while retaining human review
for complex or context-dependent changes. \\

\cmidrule{2-3}

Strengthen cross-space traceability &
Although 97.77\% of explicit relationships occurred within GitHub, only
294 relationships (1.79\%) connected GitHub with the Community Forum. &
Encourage bidirectional linking when forum problems lead to issues or pull
requests and when repository changes address support problems. \\

\cmidrule{2-3}

Recommend related artifacts &
Semantic matching identified additional candidate relationships after
explicitly linked artifact pairs were removed. &
Suggest potentially related issues, pull requests, and forum discussions,
with human confirmation before establishing a traceability link. \\

\cmidrule{2-3}

Reuse community-support knowledge &
Forum discussions contained recurring solutions, workarounds, configuration
guidance, and explanations of operational and scientific-analysis problems. &
Promote recurring accepted solutions into documentation, training materials,
FAQs, and troubleshooting resources. \\

\cmidrule{2-3}

Track outcome and latency jointly &
Resolution likelihood and resolution time showed different patterns across
artifact types and maintenance topics. &
Monitor both eventual resolution rates and lifecycle duration to identify
persistent maintenance bottlenecks. \\

\bottomrule
\end{tabularx}
\end{table*}

Several of these recommendations concern the information available when a maintenance or support request is created. Diagnostic information such as error descriptions, code fragments, reproduction details, versions, and relevant links can provide maintainers with useful context for understanding a reported problem. Lightweight issue and support templates could therefore encourage users to provide sufficient diagnostic evidence while avoiding excessive reporting requirements. Similarly, the differences observed across maintenance topics suggest that a uniform triage strategy may be insufficient. Dependency management, installation, tool integration, workflow execution, and infrastructure concerns may require more specialized validation or maintainer expertise than comparatively routine maintenance activities.

The resolution analyses also suggest that triage mechanisms should distinguish between \emph{activity} and \emph{progress}. An artifact with many comments, replies, reviews, or other interactions may be receiving substantial attention, but this activity can also reflect unresolved complexity. Maintenance dashboards and triage tools could therefore consider multiple lifecycle signals---such as artifact age, ownership, milestone status, diagnostic completeness, and unresolved duration---rather than using discussion volume alone as an indicator of progress. Likewise, because resolution likelihood and resolution latency capture different aspects of maintenance performance, both should be monitored when assessing maintenance health.

The cross-space findings motivate a complementary set of knowledge-management recommendations. The limited number of explicit links between GitHub and the Community Forum suggests that relevant problem context and implementation knowledge can remain separated across platforms. Encouraging bidirectional references when a forum discussion leads to an issue or pull request could preserve this context. In addition, semantic matching could support link recommendation by surfacing potentially related artifacts to maintainers or users. Such recommendations should remain human-confirmed, however, because semantic similarity indicates possible relatedness rather than a verified problem--solution relationship.

Finally, Community Forum discussions represent a potentially valuable source of reusable maintenance knowledge. Accepted answers often capture configuration guidance, workarounds, explanations of tool behavior, and solutions to recurring operational or scientific-analysis problems. Promoting recurring solutions into documentation, training resources, or troubleshooting guides could make this knowledge easier to discover beyond the original discussion. At the same time, the substantial preventive, automated, and infrastructure-oriented work visible in pull requests indicates that maintenance support should not focus exclusively on reactive defect correction. Selective automation of repetitive activities such as dependency updates, testing, validation, and configuration maintenance may help reduce recurring maintenance effort, while human review remains important for complex or context-dependent changes.

These recommendations are derived from observed associations and cross-space patterns rather than from controlled evaluations of the proposed interventions. They should therefore be interpreted as empirically grounded directions for improving maintenance practice and for future research evaluating the effectiveness of maintenance-support mechanisms in Galaxy and other scientific workflow ecosystems.

\section{Threats to Validity}
\label{sec:threats}
Threats to validity concern factors that may affect the accuracy, reliability,
or generalizability of our findings. More broadly, validity addresses
the extent to which the conclusions drawn from a study adequately reflect the
underlying phenomena being investigated~\cite{bean2007qualitative}. Such
threats may introduce bias, measurement error, or limitations in interpretation.
Recognizing these limitations is therefore important for assessing the
credibility and robustness of the results. In this study, we consider the
following potential threats to validity.

\textbf{Construct validity: }concerns the extent to which
the measures and operationalizations used in a study accurately represent the
concepts being investigated~\cite{wohlin2012experimentation}. The operationalizations used in this study may not capture all aspects of
maintenance and support. In \textbf{RQ1}, BERTopic identifies recurring lexical and
semantic patterns rather than maintenance concepts directly; consequently,
topic interpretation depends on the selected model configuration and manual
labeling. To mitigate this threat, topics were interpreted using representative
terms and artifacts and independently reviewed by two researchers with
experience in scientific workflows and software development. In \textbf{RQ2}, issue
closure, pull-request merging or final decision, and accepted answers were used
as observable proxies for resolution. These outcomes do not necessarily imply
that the underlying technical problem was completely solved. Moreover, forum
resolution time is measured to the creation of the post that was eventually
accepted, rather than the exact time at which the answer was marked as
accepted. We therefore interpret this measure as an accepted-answer lifecycle
proxy.

\textbf{Internal validity: }concerns whether the observed
relationships between study variables can be attributed to the factors being
examined rather than to alternative explanations or confounding influences
~\cite{wohlin2012experimentation}. The RQ2 analyses are observational and identify associations rather than
causal relationships. Several characteristics, including comments, replies,
likes, labels, reviews, milestones, and activity span, can accumulate during
an artifact's lifecycle and may therefore reflect both problem complexity and
resolution activity. Although multivariable models were used to account for
observed factors, unmeasured confounders such as maintainer workload,
artifact complexity, contributor expertise, or project priorities may affect
the observed relationships. We consequently avoid interpreting regression
coefficients and hazard ratios as causal or prospective effects.

\textbf{External validity: } External validity refers to the extent to which
the findings of a study can be generalized beyond the specific context,
population, or setting investigated~\cite{wohlin2012experimentation}.
This study focuses on Galaxy, a large and mature community-driven scientific
workflow ecosystem. Its technical structure, user community, and maintenance
practices may differ from those of other SWSs such as Nextflow, Snakemake, or
proprietary scientific platforms. Therefore, the observed topic distributions,
resolution patterns, and traceability characteristics should not be generalized
directly to all scientific software ecosystems. Replication across other
workflow ecosystems is needed to assess the broader applicability of the
findings.

Our analysis covers 340 active Galaxy GitHub repositories together
with discussions from the Galaxy Community Help Forum. Other communication
and collaboration channels used by the Galaxy community, such as Slack,
Discord, mailing lists, or other community platforms, were not included.
Consequently, some maintenance requests, troubleshooting activities,
coordination practices, or problem--solution interactions may remain
unobserved because they occur outside the analyzed sources. This may introduce
selection bias and limit the applicability of the findings to the complete
Galaxy communication ecosystem. Nevertheless, the selected GitHub repositories
and Community Help Forum provide structured, long-term, and publicly available
records of repository-centered development and user-facing support, enabling
systematic and reproducible empirical analysis.

\section{Conclusion}
\label{sec:conclusion}

This study provides an ecosystem-level understanding of maintenance and support in Galaxy through an analysis of 11,762 GitHub issues, 52,203 pull requests, and 6,235 Community Forum discussions. Our findings show that maintenance in a community-driven scientific workflow system extends beyond source-code evolution to include workflow execution, data management, tools and dependencies, infrastructure, testing, scientific resources, documentation, and user-facing support. Issues, pull requests, and forum discussions provide complementary perspectives on this maintenance process, capturing reported problems, implementation activities, and difficulties encountered by users in practice. Our results further show that maintenance resolution is multidimensional. Coordination, diagnostic information, contributor characteristics, automation, and engagement are associated differently with eventual resolution and resolution speed across artifact types and maintenance topics. These findings suggest that closure, merge, or accepted-answer rates alone provide an incomplete picture of maintenance performance and should be considered together with resolution time and the characteristics of the underlying maintenance work. We also identify a substantial gap between technical relatedness and explicit cross-space traceability. Although 97.77\% of the 16,426 resolved explicit relationships occur within GitHub, only 294 directly connect GitHub artifacts with Community Forum discussions. At the same time, semantic and technical similarities indicate that related problem and solution knowledge frequently spans these spaces without explicit links. This finding highlights the importance of mechanisms that preserve connections between user-facing problems, maintenance activities, and reusable solutions.

Future work could extend the analysis to additional Galaxy communication and collaboration channels, such as Slack, Discord, and mailing lists, to obtain a more complete view of how maintenance knowledge is created, transferred, and reused across the community. Longitudinal and intervention-based studies could also evaluate the effectiveness of diagnostic-support mechanisms, lifecycle-aware triage, selective automation, and human-in-the-loop approaches for recovering and validating cross-space traceability. Such work could help determine how empirically derived maintenance-support mechanisms affect resolution processes and knowledge reuse in practice.

\section{Acknowledgments}

This research is supported in part by the Natural Sciences and Engineering Research Council of Canada (NSERC) Discovery Grants program, and by the industry-stream NSERC CREATE in Software Analytics Research (SOAR). This research is also supported in-part by two Canada First Research Excellence Funds (CFREFs) grants coordinated by the Global Institute for Food Security (GIFS) and the Global Institute for Water Security (GIWS).

\section*{Ethics and Privacy Statement}

Not Applicable

\bibliographystyle{ACM-Reference-Format}
\bibliography{sample-base}

@String{Computing = "Computing" }

@String{Computer = "{IEEE} Computer" }

@String{Springer = "Springer-Verlag" }

@article{DBLP:journals/fgcs/DeelmanGST09,
  author       = {Ewa Deelman and
                  Dennis Gannon and
                  Matthew S. Shields and
                  Ian J. Taylor},
  title        = {Workflows and e-Science: An overview of workflow system features and
                  capabilities},
  journal      = {Future Gener. Comput. Syst.},
  volume       = {25},
  number       = {5},
  pages        = {528--540},
  year         = {2009},
  url          = {https://doi.org/10.1016/j.future.2008.06.012},
  doi          = {10.1016/J.FUTURE.2008.06.012},
  bibsource    = {dblp computer science bibliography, https://dblp.org}
}

@article{DBLP:journals/fgcs/SilvaFPJSD17,
  author       = {Rafael Ferreira da Silva and
                  Rosa Filgueira and
                  Ilia Pietri and
                  Ming Jiang and
                  Rizos Sakellariou and
                  Ewa Deelman},
  title        = {A characterization of workflow management systems for extreme-scale
                  applications},
  journal      = {Future Gener. Comput. Syst.},
  volume       = {75},
  pages        = {228--238},
  year         = {2017},
  url          = {https://doi.org/10.1016/j.future.2017.02.026},
  doi          = {10.1016/J.FUTURE.2017.02.026},
  bibsource    = {dblp computer science bibliography, https://dblp.org}
}

@article{DBLP:journals/fgcs/SuterCABBCCDTFGJKKLMOPP26,
  author       = {Fr{\'{e}}d{\'{e}}ric Suter and
                  Tain{\~{a}} Coleman and
                  Ilkay Altintas and
                  others},
  title        = {A terminology for scientific workflow systems},
  journal      = {Future Gener. Comput. Syst.},
  volume       = {174},
  pages        = {107974},
  year         = {2026},
  url          = {https://doi.org/10.1016/j.future.2025.107974},
  doi          = {10.1016/J.FUTURE.2025.107974},
  bibsource    = {dblp computer science bibliography, https://dblp.org}
}

@article{DBLP:journals/bioinformatics/KosterR18,
  author       = {Johannes K{\"{o}}ster and
                  Sven Rahmann},
  title        = {Snakemake - a scalable bioinformatics workflow engine},
  journal      = {Bioinform.},
  volume       = {34},
  number       = {20},
  pages        = {3600},
  year         = {2018},
  url          = {https://doi.org/10.1093/bioinformatics/bty350},
  doi          = {10.1093/BIOINFORMATICS/BTY350},
  bibsource    = {dblp computer science bibliography, https://dblp.org}
}

@article{DBLP:journals/fgcs/DeelmanVJRCMMCS15,
  author       = {Ewa Deelman and
                  Karan Vahi and
                  Gideon Juve and
                  Mats Rynge and
                  Scott Callaghan and
                  Philip Maechling and
                  Rajiv Mayani and
                  Weiwei Chen and
                  Rafael Ferreira da Silva and
                  Miron Livny and
                  R. Kent Wenger},
  title        = {Pegasus, a workflow management system for science automation},
  journal      = {Future Gener. Comput. Syst.},
  volume       = {46},
  pages        = {17--35},
  year         = {2015},
  url          = {https://doi.org/10.1016/j.future.2014.10.008},
  doi          = {10.1016/J.FUTURE.2014.10.008},
  bibsource    = {dblp computer science bibliography, https://dblp.org}
}

@article{DBLP:journals/nar/AfganNGBGSOMLSF22,
  author       = {Enis Afgan and
                  Anton Nekrutenko and
                  Bj{\"{o}}rn A. Gr{\"{u}}ning and
                  others},
  title        = {The Galaxy platform for accessible, reproducible and collaborative
                  biomedical analyses: 2022 update},
  journal      = {Nucleic Acids Res.},
  volume       = {50},
  number       = {{W1}},
  pages        = {345--351},
  year         = {2022},
  url          = {https://doi.org/10.1093/nar/gkac247},
  doi          = {10.1093/NAR/GKAC247},
  bibsource    = {dblp computer science bibliography, https://dblp.org}
}

@article{DBLP:journals/ese/KalliamvakouGBS16,
  author       = {Eirini Kalliamvakou and
                  Georgios Gousios and
                  Kelly Blincoe and
                  Leif Singer and
                  Daniel M. Germ{\'{a}}n and
                  Daniela E. Damian},
  title        = {An in-depth study of the promises and perils of mining GitHub},
  journal      = {Empir. Softw. Eng.},
  volume       = {21},
  number       = {5},
  pages        = {2035--2071},
  year         = {2016},
  url          = {https://doi.org/10.1007/s10664-015-9393-5},
  doi          = {10.1007/S10664-015-9393-5},
  bibsource    = {dblp computer science bibliography, https://dblp.org}
}

@inproceedings{DBLP:conf/icse/GousiosPD14,
  author       = {Georgios Gousios and
                  Martin Pinzger and
                  Arie van Deursen},
  editor       = {Pankaj Jalote and
                  Lionel C. Briand and
                  Andr{\'{e}} van der Hoek},
  title        = {An exploratory study of the pull-based software development model},
  booktitle    = {36th International Conference on Software Engineering, {ICSE} '14,
                  Hyderabad, India - May 31 - June fgruning2018practical07, 2014},
  pages        = {345--355},
  publisher    = {{ACM}},
  year         = {2014},
  url          = {https://doi.org/10.1145/2568225.2568260},
  doi          = {10.1145/2568225.2568260},
  bibsource    = {dblp computer science bibliography, https://dblp.org}
}

@inproceedings{DBLP:conf/indiaSE/DhasadeVC20,
  author       = {Akash Balasaheb Dhasade and
                  Akhila Sri Manasa Venigalla and
                  Sridhar Chimalakonda},
  editor       = {Sanjeev Jain and
                  Atul Gupta and
                  David Lo and
                  Diptikalyan Saha and
                  Richa Sharma},
  title        = {Towards Prioritizing GitHub Issues},
  booktitle    = {{ISEC} 2020: 13th Innovations in Software Engineering Conference,
                  Jabalpur, India, February 27-29, 2020},
  pages        = {18:1--18:5},
  publisher    = {{ACM}},
  year         = {2020},
  url          = {https://doi.org/10.1145/3385032.3385052},
  doi          = {10.1145/3385032.3385052},
  bibsource    = {dblp computer science bibliography, https://dblp.org}
}

@article{DBLP:journals/ese/HanSWDX20,
  author       = {Junxiao Han and
                  Emad Shihab and
                  Zhiyuan Wan and
                  Shuiguang Deng and
                  Xin Xia},
  title        = {What do Programmers Discuss about Deep Learning Frameworks},
  journal      = {Empir. Softw. Eng.},
  volume       = {25},
  number       = {4},
  pages        = {2694--2747},
  year         = {2020},
  url          = {https://doi.org/10.1007/s10664-020-09819-6},
  doi          = {10.1007/S10664-020-09819-6},
  bibsource    = {dblp computer science bibliography, https://dblp.org}
}

@inproceedings{DBLP:conf/msr/YangWSHKLXL23,
  author       = {Zhou Yang and
                  Chenyu Wang and
                  Jieke Shi and
                  Thong Hoang and
                  Pavneet Singh Kochhar and
                  Qinghua Lu and
                  Zhenchang Xing and
                  David Lo},
  title        = {What Do Users Ask in Open-Source {AI} Repositories? An Empirical Study
                  of GitHub Issues},
  booktitle    = {20th {IEEE/ACM} International Conference on Mining Software Repositories,
                  {MSR} 2023, Melbourne, Australia, May 15-16, 2023},
  pages        = {79--91},
  publisher    = {{IEEE}},
  year         = {2023},
  url          = {https://doi.org/10.1109/MSR59073.2023.00024},
  doi          = {10.1109/MSR59073.2023.00024},
  bibsource    = {dblp computer science bibliography, https://dblp.org}
}

@article{DBLP:journals/ese/AlamRRM25,
  author       = {Khairul Alam and
                  Banani Roy and
                  Chanchal K. Roy and
                  Kartik Mittal},
  title        = {An Empirical Investigation on the Challenges in Scientific Workflow
                  Systems Development},
  journal      = {Empir. Softw. Eng.},
  volume       = {30},
  number       = {5},
  pages        = {151},
  year         = {2025},
  url          = {https://doi.org/10.1007/s10664-025-10705-2},
  doi          = {10.1007/S10664-025-10705-2},
  bibsource    = {dblp computer science bibliography, https://dblp.org}
}

@article{DBLP:journals/corr/abs-2601-09612,
  author       = {Khairul Alam and
                  Banani Roy},
  title        = {Analyzing GitHub Issues and Pull Requests in nf-core Pipelines: Insights
                  into nf-core Pipeline Repositories},
  journal      = {CoRR},
  volume       = {abs/2601.09612},
  year         = {2026},
  url          = {https://doi.org/10.48550/arXiv.2601.09612},
  doi          = {10.48550/ARXIV.2601.09612},
  eprinttype   = {arXiv},
  eprint       = {2601.09612},
  bibsource    = {dblp computer science bibliography, https://dblp.org}
}

@inproceedings{DBLP:conf/icsm/OpenjaAK20,
  author       = {Moses Openja and
                  Bram Adams and
                  Foutse Khomh},
  title        = {Analysis of Modern Release Engineering Topics : - {A} Large-Scale
                  Study using StackOverflow -},
  booktitle    = {{IEEE} International Conference on Software Maintenance and Evolution,
                  {ICSME} 2020, Adelaide, Australia, September 28 - October 2, 2020},
  pages        = {104--114},
  publisher    = {{IEEE}},
  year         = {2020},
  url          = {https://doi.org/10.1109/ICSME46990.2020.00020},
  doi          = {10.1109/ICSME46990.2020.00020},
  bibsource    = {dblp computer science bibliography, https://dblp.org}
}

@inproceedings{DBLP:conf/wsdm/RoderBH15,
  author       = {Michael R{\"{o}}der and
                  Andreas Both and
                  Alexander Hinneburg},
  editor       = {Xueqi Cheng and
                  Hang Li and
                  Evgeniy Gabrilovich and
                  Jie Tang},
  title        = {Exploring the Space of Topic Coherence Measures},
  booktitle    = {Proceedings of the Eighth {ACM} International Conference on Web Search
                  and Data Mining, {WSDM} 2015, Shanghai, China, February 2-6, 2015},
  pages        = {399--408},
  publisher    = {{ACM}},
  year         = {2015},
  url          = {https://doi.org/10.1145/2684822.2685324},
  doi          = {10.1145/2684822.2685324},
  bibsource    = {dblp computer science bibliography, https://dblp.org}
}

@inproceedings{DBLP:conf/emnlp/ReimersG19,
  author       = {Nils Reimers and
                  Iryna Gurevych},
  editor       = {Kentaro Inui and
                  Jing Jiang and
                  Vincent Ng and
                  Xiaojun Wan},
  title        = {Sentence-BERT: Sentence Embeddings using Siamese BERT-Networks},
  booktitle    = {Proceedings of the 2019 Conference on Empirical Methods in Natural
                  Language Processing and the 9th International Joint Conference on
                  Natural Language Processing, {EMNLP-IJCNLP} 2019, Hong Kong, China,
                  November 3-7, 2019},
  pages        = {3980--3990},
  publisher    = {Association for Computational Linguistics},
  year         = {2019},
  url          = {https://doi.org/10.18653/v1/D19-1410},
  doi          = {10.18653/V1/D19-1410},
  bibsource    = {dblp computer science bibliography, https://dblp.org}
}

@inproceedings{DBLP:conf/icse/GousiosZSD15,
  author       = {Georgios Gousios and
                  Andy Zaidman and
                  Margaret{-}Anne D. Storey and
                  Arie van Deursen},
  editor       = {Antonia Bertolino and
                  Gerardo Canfora and
                  Sebastian G. Elbaum},
  title        = {Work Practices and Challenges in Pull-Based Development: The Integrator's
                  Perspective},
  booktitle    = {37th {IEEE/ACM} International Conference on Software Engineering,
                  {ICSE} 2015, Florence, Italy, May 16-24, 2015, Volume 1},
  pages        = {358--368},
  publisher    = {{IEEE} Computer Society},
  year         = {2015},
  url          = {https://doi.org/10.1109/ICSE.2015.55},
  doi          = {10.1109/ICSE.2015.55},
  bibsource    = {dblp computer science bibliography, https://dblp.org}
}

@inproceedings{DBLP:conf/msr/Abdellatif0BAS20,
  author       = {Ahmad Abdellatif and
                  Diego Costa and
                  Khaled Badran and
                  Rabe Abdalkareem and
                  Emad Shihab},
  editor       = {Sunghun Kim and
                  Georgios Gousios and
                  Sarah Nadi and
                  Joseph Hejderup},
  title        = {Challenges in Chatbot Development: {A} Study of Stack Overflow Posts},
  booktitle    = {{MSR} '20: 17th International Conference on Mining Software Repositories,
                  Seoul, Republic of Korea, 29-30 June, 2020},
  pages        = {174--185},
  publisher    = {{ACM}},
  year         = {2020},
  url          = {https://doi.org/10.1145/3379597.3387472},
  doi          = {10.1145/3379597.3387472},
  bibsource    = {dblp computer science bibliography, https://dblp.org}
}

@article{DBLP:journals/corr/abs-2205-03181,
  author       = {Mohamed Raed El aoun and
                  Heng Li and
                  Foutse Khomh and
                  Moses Openja},
  title        = {Understanding Quantum Software Engineering Challenges An Empirical
                  Study on Stack Exchange Forums and GitHub Issues},
  journal      = {CoRR},
  volume       = {abs/2205.03181},
  year         = {2022},
  url          = {https://doi.org/10.48550/arXiv.2205.03181},
  doi          = {10.48550/ARXIV.2205.03181},
  eprinttype   = {arXiv},
  eprint       = {2205.03181},
  bibsource    = {dblp computer science bibliography, https://dblp.org}
}

@article{DBLP:journals/jossw/McInnesHA17,
  author       = {Leland McInnes and
                  John Healy and
                  Steve Astels},
  title        = {hdbscan: Hierarchical density based clustering},
  journal      = {J. Open Source Softw.},
  volume       = {2},
  number       = {11},
  pages        = {205},
  year         = {2017},
  url          = {https://doi.org/10.21105/joss.00205},
  doi          = {10.21105/JOSS.00205},
  bibsource    = {dblp computer science bibliography, https://dblp.org}
}

@article{DBLP:journals/ese/HataNBKT22,
  author       = {Hideaki Hata and
                  Nicole Novielli and
                  Sebastian Baltes and
                  Raula Gaikovina Kula and
                  Christoph Treude},
  title        = {GitHub Discussions: An exploratory study of early adoption},
  journal      = {Empir. Softw. Eng.},
  volume       = {27},
  number       = {1},
  pages        = {3},
  year         = {2022},
  url          = {https://doi.org/10.1007/s10664-021-10058-6},
  doi          = {10.1007/S10664-021-10058-6},
  bibsource    = {dblp computer science bibliography, https://dblp.org}
}

@inproceedings{DBLP:conf/icse-chase/HellmanCUCG22,
  author       = {Jazlyn Hellman and
                  Jiahao Chen and
                  Md. Sami Uddin and
                  Jinghui Cheng and
                  Jin L. C. Guo},
  title        = {Characterizing User Behaviors in Open-Source Software User Forums:
                  An Empirical Study},
  booktitle    = {15th {IEEE/ACM} International Workshop on Cooperative and Human Aspects
                  of Software Engineering , CHASE@ICSE 2022, Pittsburgh, PA, USA, May
                  21-22, 2022},
  pages        = {46--55},
  publisher    = {{IEEE}},
  year         = {2022},
  url          = {https://doi.org/10.1145/3528579.3529178},
  doi          = {10.1145/3528579.3529178},
  bibsource    = {dblp computer science bibliography, https://dblp.org}
}

@inproceedings{DBLP:conf/msr/AlamR26,
  author       = {Khairul Alam and
                  Banani Roy},
  title        = {Analyzing GitHub Issues and Pull Requests in nf-core Pipelines: Insights
                  into nf-core Pipeline Repositories},
  booktitle    = {Proceedings of the 23rd International Conference on Mining Software
                  Repositories, {MSR} 2026, Rio de Janeiro, Brazil, April 13-14, 2026},
  pages        = {508--519},
  publisher    = {{ACM}},
  year         = {2026},
  url          = {https://doi.org/10.1145/3793302.3793376},
  doi          = {10.1145/3793302.3793376},
  bibsource    = {dblp computer science bibliography, https://dblp.org}
}

@inproceedings{DBLP:conf/icse/AnvikHM06,
  author       = {John Anvik and
                  Lyndon Hiew and
                  Gail C. Murphy},
  editor       = {Leon J. Osterweil and
                  H. Dieter Rombach and
                  Mary Lou Soffa},
  title        = {Who should fix this bug?},
  booktitle    = {28th International Conference on Software Engineering {(ICSE} 2006),
                  Shanghai, China, May 20-28, 2006},
  pages        = {361--370},
  publisher    = {{ACM}},
  year         = {2006},
  url          = {https://doi.org/10.1145/1134285.1134336},
  doi          = {10.1145/1134285.1134336},
  bibsource    = {dblp computer science bibliography, https://dblp.org}
}

@article{DBLP:journals/tse/ZimmermannPBJSW10,
  author       = {Thomas Zimmermann and
                  Rahul Premraj and
                  Nicolas Bettenburg and
                  Sascha Just and
                  Adrian Schr{\"{o}}ter and
                  Cathrin Weiss},
  title        = {What Makes a Good Bug Report?},
  journal      = {{IEEE} Trans. Software Eng.},
  volume       = {36},
  number       = {5},
  pages        = {618--643},
  year         = {2010},
  url          = {https://doi.org/10.1109/TSE.2010.63},
  doi          = {10.1109/TSE.2010.63},
  bibsource    = {dblp computer science bibliography, https://dblp.org}
}

@inproceedings{DBLP:conf/icse/ZhangGV13,
  author       = {Hongyu Zhang and
                  Liang Gong and
                  Steven Versteeg},
  editor       = {David Notkin and
                  Betty H. C. Cheng and
                  Klaus Pohl},
  title        = {Predicting bug-fixing time: an empirical study of commercial software
                  projects},
  booktitle    = {35th International Conference on Software Engineering, {ICSE} '13,
                  San Francisco, CA, USA, May 18-26, 2013},
  pages        = {1042--1051},
  publisher    = {{IEEE} Computer Society},
  year         = {2013},
  url          = {https://doi.org/10.1109/ICSE.2013.6606654},
  doi          = {10.1109/ICSE.2013.6606654},
  bibsource    = {dblp computer science bibliography, https://dblp.org}
}

@article{DBLP:journals/tse/ZhangYGR23,
  author       = {Xunhui Zhang and
                  Yue Yu and
                  Georgios Gousios and
                  Ayushi Rastogi},
  title        = {Pull Request Decisions Explained: An Empirical Overview},
  journal      = {{IEEE} Trans. Software Eng.},
  volume       = {49},
  number       = {2},
  pages        = {849--871},
  year         = {2023},
  url          = {https://doi.org/10.1109/TSE.2022.3165056},
  doi          = {10.1109/TSE.2022.3165056},
  bibsource    = {dblp computer science bibliography, https://dblp.org}
}

@article{DBLP:journals/ese/BaruaTH14,
  author       = {Anton Barua and
                  Stephen W. Thomas and
                  Ahmed E. Hassan},
  title        = {What are developers talking about? An analysis of topics and trends
                  in Stack Overflow},
  journal      = {Empir. Softw. Eng.},
  volume       = {19},
  number       = {3},
  pages        = {619--654},
  year         = {2014},
  url          = {https://doi.org/10.1007/s10664-012-9231-y},
  doi          = {10.1007/S10664-012-9231-Y},
  bibsource    = {dblp computer science bibliography, https://dblp.org}
}

@inproceedings{DBLP:conf/socialcom/VasilescuFS13,
  author       = {Bogdan Vasilescu and
                  Vladimir Filkov and
                  Alexander Serebrenik},
  title        = {StackOverflow and GitHub: Associations between Software Development
                  and Crowdsourced Knowledge},
  booktitle    = {International Conference on Social Computing, SocialCom 2013, SocialCom/PASSAT/BigData/EconCom/BioMedCom
                  2013, Washington, DC, USA, 8-14 September, 2013},
  pages        = {188--195},
  publisher    = {{IEEE} Computer Society},
  year         = {2013},
  url          = {https://doi.org/10.1109/SocialCom.2013.35},
  doi          = {10.1109/SOCIALCOM.2013.35},
  bibsource    = {dblp computer science bibliography, https://dblp.org}
}

@article{DBLP:journals/ese/SilvaGG21,
  author       = {Camila Mariane C. Silva and
                  Matthias Galster and
                  Fabian Gilson},
  title        = {Topic modeling in software engineering research},
  journal      = {Empir. Softw. Eng.},
  volume       = {26},
  number       = {6},
  pages        = {120},
  year         = {2021},
  url          = {https://doi.org/10.1007/s10664-021-10026-0},
  doi          = {10.1007/S10664-021-10026-0},
  bibsource    = {dblp computer science bibliography, https://dblp.org}
}

@article{DBLP:journals/infsof/AgrawalFM18,
  author       = {Amritanshu Agrawal and
                  Wei Fu and
                  Tim Menzies},
  title        = {What is wrong with topic modeling? And how to fix it using search-based
                  software engineering},
  journal      = {Inf. Softw. Technol.},
  volume       = {98},
  pages        = {74--88},
  year         = {2018},
  url          = {https://doi.org/10.1016/j.infsof.2018.02.005},
  doi          = {10.1016/J.INFSOF.2018.02.005},
  bibsource    = {dblp computer science bibliography, https://dblp.org}
}

@inproceedings{DBLP:conf/msr/Treude019,
  author       = {Christoph Treude and
                  Markus Wagner},
  editor       = {Margaret{-}Anne D. Storey and
                  Bram Adams and
                  Sonia Haiduc},
  title        = {Predicting good configurations for GitHub and stack overflow topic
                  models},
  booktitle    = {Proceedings of the 16th International Conference on Mining Software
                  Repositories, {MSR} 2019, 26-27 May 2019, Montreal, Canada},
  pages        = {84--95},
  publisher    = {{IEEE} / {ACM}},
  year         = {2019},
  url          = {https://doi.org/10.1109/MSR.2019.00022},
  doi          = {10.1109/MSR.2019.00022},
  bibsource    = {dblp computer science bibliography, https://dblp.org}
}

@inproceedings{DBLP:conf/msr/ZagalskyTGSP16,
  author       = {Alexey Zagalsky and
                  Carlos G{\'{o}}mez Teshima and
                  Daniel M. Germ{\'{a}}n and
                  Margaret{-}Anne D. Storey and
                  Germ{\'{a}}n Poo{-}Caama{\~{n}}o},
  editor       = {Miryung Kim and
                  Romain Robbes and
                  Christian Bird},
  title        = {How the {R} community creates and curates knowledge: a comparative
                  study of stack overflow and mailing lists},
  booktitle    = {Proceedings of the 13th International Conference on Mining Software
                  Repositories, {MSR} 2016, Austin, TX, USA, May 14-22, 2016},
  pages        = {441--451},
  publisher    = {{ACM}},
  year         = {2016},
  url          = {https://doi.org/10.1145/2901739.2901772},
  doi          = {10.1145/2901739.2901772},
  bibsource    = {dblp computer science bibliography, https://dblp.org}
}

@article{DBLP:journals/tse/StoreyZFSG17,
  author       = {Margaret{-}Anne D. Storey and
                  Alexey Zagalsky and
                  Fernando Marques Figueira Filho and
                  Leif Singer and
                  Daniel M. Germ{\'{a}}n},
  title        = {How Social and Communication Channels Shape and Challenge a Participatory
                  Culture in Software Development},
  journal      = {{IEEE} Trans. Software Eng.},
  volume       = {43},
  number       = {2},
  pages        = {185--204},
  year         = {2017},
  url          = {https://doi.org/10.1109/TSE.2016.2584053},
  doi          = {10.1109/TSE.2016.2584053},
  bibsource    = {dblp computer science bibliography, https://dblp.org}
}

@inproceedings{DBLP:conf/icsm/aounLKO21,
  author       = {Mohamed Raed El Aoun and
                  Heng Li and
                  Foutse Khomh and
                  Moses Openja},
  title        = {Understanding Quantum Software Engineering Challenges An Empirical
                  Study on Stack Exchange Forums and GitHub Issues},
  booktitle    = {{IEEE} International Conference on Software Maintenance and Evolution,
                  {ICSME} 2021, Luxembourg, September 27 - October 1, 2021},
  pages        = {343--354},
  publisher    = {{IEEE}},
  year         = {2021},
  url          = {https://doi.org/10.1109/ICSME52107.2021.00037},
  doi          = {10.1109/ICSME52107.2021.00037},
  bibsource    = {dblp computer science bibliography, https://dblp.org}
}

@inproceedings{DBLP:conf/nips/Song0QLL20,
  author       = {Kaitao Song and
                  Xu Tan and
                  Tao Qin and
                  Jianfeng Lu and
                  Tie{-}Yan Liu},
  editor       = {Hugo Larochelle and
                  Marc'Aurelio Ranzato and
                  Raia Hadsell and
                  Maria{-}Florina Balcan and
                  Hsuan{-}Tien Lin},
  title        = {MPNet: Masked and Permuted Pre-training for Language Understanding},
  booktitle    = {Advances in Neural Information Processing Systems 33: Annual Conference
                  on Neural Information Processing Systems 2020, NeurIPS 2020, December
                  6-12, 2020, virtual},
  year         = {2020},
  url          = {https://proceedings.neurips.cc/paper/2020/hash/c3a690be93aa602ee2dc0ccab5b7b67e-Abstract.html},
  bibsource    = {dblp computer science bibliography, https://dblp.org}
}

@article{DBLP:journals/jcst/YangLXWS16,
  author       = {Xinli Yang and
                  David Lo and
                  Xin Xia and
                  Zhiyuan Wan and
                  Jian{-}Ling Sun},
  title        = {What Security Questions Do Developers Ask? {A} Large-Scale Study of
                  Stack Overflow Posts},
  journal      = {J. Comput. Sci. Technol.},
  volume       = {31},
  number       = {5},
  pages        = {910--924},
  year         = {2016},
  url          = {https://doi.org/10.1007/s11390-016-1672-0},
  doi          = {10.1007/S11390-016-1672-0},
  bibsource    = {dblp computer science bibliography, https://dblp.org}
}

@inproceedings{DBLP:conf/msr/ScocciaMA21,
  author       = {Gian Luca Scoccia and
                  Patrizio Migliarini and
                  Marco Autili},
  title        = {Challenges in Developing Desktop Web Apps: a Study of Stack Overflow
                  and GitHub},
  booktitle    = {18th {IEEE/ACM} International Conference on Mining Software Repositories,
                  {MSR} 2021, Madrid, Spain, May 17-19, 2021},
  pages        = {271--282},
  publisher    = {{IEEE}},
  year         = {2021},
  url          = {https://doi.org/10.1109/MSR52588.2021.00039},
  doi          = {10.1109/MSR52588.2021.00039},
  bibsource    = {dblp computer science bibliography, https://dblp.org}
}

@inproceedings{DBLP:conf/apsec/AlamRS23,
  author       = {Khairul Alam and
                  Banani Roy and
                  Alexander Serebrenik},
  title        = {Reusability Challenges of Scientific Workflows: {A} Case Study for
                  Galaxy},
  booktitle    = {30th Asia-Pacific Software Engineering Conference, {APSEC} 2023, Seoul,
                  Republic of Korea, December 4-7, 2023},
  pages        = {289--298},
  publisher    = {{IEEE}},
  year         = {2023},
  url          = {https://doi.org/10.1109/APSEC60848.2023.00039},
  doi          = {10.1109/APSEC60848.2023.00039},
  bibsource    = {dblp computer science bibliography, https://dblp.org}
}

@article{DBLP:journals/fgcs/BoulakiaBCCFGHL17,
  author       = {Sarah Cohen Boulakia and
                  Khalid Belhajjame and
                  Olivier Collin and
                  J{\'{e}}r{\^{o}}me Chopard and
                  Christine Froidevaux and
                  Alban Gaignard and
                  Konrad Hinsen and
                  Pierre Larmande and
                  Yvan Le Bras and
                  Fr{\'{e}}d{\'{e}}ric Lemoine and
                  Fabien Mareuil and
                  Herv{\'{e}} M{\'{e}}nager and
                  Christophe Pradal and
                  Christophe Blanchet},
  title        = {Scientific workflows for computational reproducibility in the life
                  sciences: Status, challenges and opportunities},
  journal      = {Future Gener. Comput. Syst.},
  volume       = {75},
  pages        = {284--298},
  year         = {2017},
  url          = {https://doi.org/10.1016/j.future.2017.01.012},
  doi          = {10.1016/J.FUTURE.2017.01.012},
  bibsource    = {dblp computer science bibliography, https://dblp.org}
}

@inproceedings{DBLP:conf/saner/AlamR26,
  author       = {Khairul Alam and
                  Banani Roy},
  title        = {What Drives Issue Resolution Speed? An Empirical Study of Scientific
                  Workflow Systems on GitHub},
  booktitle    = {{IEEE} International Conference on Software Analysis, Evolution and
                  Reengineering, {SANER} 2026, Limassol, Cyprus, March 17-20, 2026},
  pages        = {1--6},
  publisher    = {{IEEE}},
  year         = {2026},
  url          = {https://doi.org/10.1109/SANER67736.2026.00116},
  doi          = {10.1109/SANER67736.2026.00116},
  bibsource    = {dblp computer science bibliography, https://dblp.org}
}

@article{DBLP:journals/corr/abs-2203-05794,
  author       = {Maarten Grootendorst},
  title        = {BERTopic: Neural topic modeling with a class-based {TF-IDF} procedure},
  journal      = {CoRR},
  volume       = {abs/2203.05794},
  year         = {2022},
  url          = {https://doi.org/10.48550/arXiv.2203.05794},
  doi          = {10.48550/ARXIV.2203.05794},
  eprinttype   = {arXiv},
  eprint       = {2203.05794},
  bibsource    = {dblp computer science bibliography, https://dblp.org}
}

@article{DBLP:journals/jmlr/BleiNJ03,
  author       = {David M. Blei and
                  Andrew Y. Ng and
                  Michael I. Jordan},
  title        = {Latent Dirichlet Allocation},
  journal      = {J. Mach. Learn. Res.},
  volume       = {3},
  pages        = {993--1022},
  year         = {2003},
  url          = {https://jmlr.org/papers/v3/blei03a.html},
  bibsource    = {dblp computer science bibliography, https://dblp.org}
}

@article{DBLP:journals/corr/abs-1802-03426,
  author       = {Leland McInnes and
                  John Healy},
  title        = {{UMAP:} Uniform Manifold Approximation and Projection for Dimension
                  Reduction},
  journal      = {CoRR},
  volume       = {abs/1802.03426},
  year         = {2018},
  url          = {http://arxiv.org/abs/1802.03426},
  eprinttype   = {arXiv},
  eprint       = {1802.03426},
  bibsource    = {dblp computer science bibliography, https://dblp.org}
}

@article{DBLP:journals/nar/GruningFYWEEHBV17,
  author       = {Bj{\"{o}}rn A. Gr{\"{u}}ning and
                  J{\"{o}}rg Fallmann and
                  Dilmurat Yusuf and
                  Sebastian Will and
                  Anika Erxleben and
                  Florian Eggenhofer and
                  Torsten Houwaart and
                  B{\'{e}}r{\'{e}}nice Batut and
                  Pavankumar Videm and
                  Andrea Bagnacani and
                  Markus Wolfien and
                  Steffen Lott and
                  Youri Hoogstrate and
                  Wolfgang R. Hess and
                  Olaf Wolkenhauer and
                  Steve Hoffmann and
                  Altuna Akalin and
                  Uwe Ohler and
                  Peter F. Stadler and
                  Rolf Backofen},
  title        = {The {RNA} workbench: best practices for {RNA} and high-throughput
                  sequencing bioinformatics in Galaxy},
  journal      = {Nucleic Acids Res.},
  volume       = {45},
  number       = {Webserver-Issue},
  pages        = {W560--W566},
  year         = {2017},
  url          = {https://doi.org/10.1093/nar/gkx409},
  doi          = {10.1093/NAR/GKX409},
  bibsource    = {dblp computer science bibliography, https://dblp.org}
}

@article{DBLP:journals/tsc/LinLFCPLFH09,
  author       = {Cui Lin and
                  Shiyong Lu and
                  Xubo Fei and
                  Artem Chebotko and
                  Darshan Pai and
                  Zhaoqiang Lai and
                  Farshad Fotouhi and
                  Jing Hua},
  title        = {A Reference Architecture for Scientific Workflow Management Systems
                  and the {VIEW} {SOA} Solution},
  journal      = {{IEEE} Trans. Serv. Comput.},
  volume       = {2},
  number       = {1},
  pages        = {79--92},
  year         = {2009},
  url          = {https://doi.org/10.1109/TSC.2009.4},
  doi          = {10.1109/TSC.2009.4},
  bibsource    = {dblp computer science bibliography, https://dblp.org}
}

@article{DBLP:journals/concurrency/LudascherABHJJLTZ06,
  author       = {Bertram Lud{\"{a}}scher and
                  Ilkay Altintas and
                  Chad Berkley and
                  Dan Higgins and
                  Efrat Jaeger and
                  Matthew B. Jones and
                  Edward A. Lee and
                  Jing Tao and
                  Yang Zhao},
  title        = {Scientific workflow management and the Kepler system},
  journal      = {Concurr. Comput. Pract. Exp.},
  volume       = {18},
  number       = {10},
  pages        = {1039--1065},
  year         = {2006},
  url          = {https://doi.org/10.1002/cpe.994},
  doi          = {10.1002/CPE.994},
  bibsource    = {dblp computer science bibliography, https://dblp.org}
}

@article{DBLP:journals/bib/Leipzig17,
  author       = {Jeremy Leipzig},
  title        = {A review of bioinformatic pipeline frameworks},
  journal      = {Briefings Bioinform.},
  volume       = {18},
  number       = {3},
  pages        = {530--536},
  year         = {2017},
  url          = {https://doi.org/10.1093/bib/bbw020},
  doi          = {10.1093/BIB/BBW020},
  bibsource    = {dblp computer science bibliography, https://dblp.org}
}

@article{DBLP:journals/datasci/LamprechtGKMAPA20,
  author       = {Anna{-}Lena Lamprecht and
                  Leyla J. Garc{\'{\i}}a and
                  Mateusz Kuzak and
                  Carlos Martinez{-}Ortiz and
                  Ricardo Arcila and
                  Eva Mart{\'{\i}}n del Pico and
                  Victoria Dominguez Del Angel and
                  Stephanie van de Sandt and
                  Jon C. Ison and
                  Paula Andrea Mart{\'{\i}}nez and
                  Peter McQuilton and
                  Alfonso Valencia and
                  Jennifer L. Harrow and
                  Fotis E. Psomopoulos and
                  Josep Llu{\'{\i}}s Gelp{\'{\i}} and
                  Neil P. Chue Hong and
                  Carole A. Goble and
                  Salvador Capella{-}Guti{\'{e}}rrez},
  title        = {Towards {FAIR} principles for research software},
  journal      = {Data Sci.},
  volume       = {3},
  number       = {1},
  pages        = {37--59},
  year         = {2020},
  url          = {https://doi.org/10.3233/ds-190026},
  doi          = {10.3233/DS-190026},
  bibsource    = {dblp computer science bibliography, https://dblp.org}
}

@article{DBLP:journals/access/AlsharaSSS23,
  author       = {Zakarea Alshara and
                  Hamzeh Eyal Salman and
                  Anas Shatnawi and
                  Abdelhak{-}Djamel Seriai},
  title        = {ML-Augmented Automation for Recovering Links Between Pull-Requests
                  and Issues on GitHub},
  journal      = {{IEEE} Access},
  volume       = {11},
  pages        = {5596--5608},
  year         = {2023},
  url          = {https://doi.org/10.1109/ACCESS.2023.3236392},
  doi          = {10.1109/ACCESS.2023.3236392},
  bibsource    = {dblp computer science bibliography, https://dblp.org}
}

@inproceedings{DBLP:conf/sigsoft/WuZKC11,
  author       = {Rongxin Wu and
                  Hongyu Zhang and
                  Sunghun Kim and
                  Shing{-}Chi Cheung},
  editor       = {Tibor Gyim{\'{o}}thy and
                  Andreas Zeller},
  title        = {ReLink: recovering links between bugs and changes},
  booktitle    = {SIGSOFT/FSE'11 19th {ACM} {SIGSOFT} Symposium on the Foundations of
                  Software Engineering {(FSE-19)} and ESEC'11: 13th European Software
                  Engineering Conference (ESEC-13), Szeged, Hungary, September 5-9,
                  2011},
  pages        = {15--25},
  publisher    = {{ACM}},
  year         = {2011},
  url          = {https://doi.org/10.1145/2025113.2025120},
  doi          = {10.1145/2025113.2025120},
  bibsource    = {dblp computer science bibliography, https://dblp.org}
}

@inproceedings{DBLP:conf/saner/YasaOAKDUT25,
  author       = {Ayberk Yasa and
                  Cemhan Kaan {\"{O}}zaltan and
                  G{\"{o}}rkem Ayten and
                  Fatih Kaplama and
                  {\"{O}}mercan Devran and
                  Baykal Mehmet U{\c{c}}ar and
                  Eray T{\"{u}}z{\"{u}}n},
  title        = {Evaluating ReLink for Traceability Link Recovery in Practice},
  booktitle    = {{IEEE} International Conference on Software Analysis, Evolution and
                  Reengineering, {SANER} 2025, Montreal, QC, Canada, March 4-7, 2025},
  pages        = {80--90},
  publisher    = {{IEEE}},
  year         = {2025},
  url          = {https://doi.org/10.1109/SANER64311.2025.00016},
  doi          = {10.1109/SANER64311.2025.00016},
  bibsource    = {dblp computer science bibliography, https://dblp.org}
}

@article{DBLP:journals/jss/RuanCPZ19,
  author       = {Hang Ruan and
                  Bihuan Chen and
                  Xin Peng and
                  Wenyun Zhao},
  title        = {DeepLink: Recovering issue-commit links based on deep learning},
  journal      = {J. Syst. Softw.},
  volume       = {158},
  year         = {2019},
  url          = {https://doi.org/10.1016/j.jss.2019.110406},
  doi          = {10.1016/J.JSS.2019.110406},
  bibsource    = {dblp computer science bibliography, https://dblp.org}
}

@inproceedings{DBLP:conf/icse/TreudeBS11,
  author       = {Christoph Treude and
                  Ohad Barzilay and
                  Margaret{-}Anne D. Storey},
  editor       = {Richard N. Taylor and
                  Harald C. Gall and
                  Nenad Medvidovic},
  title        = {How do programmers ask and answer questions on the web?},
  booktitle    = {Proceedings of the 33rd International Conference on Software Engineering,
                  {ICSE} 2011, Waikiki, Honolulu , HI, USA, May 21-28, 2011},
  pages        = {804--807},
  publisher    = {{ACM}},
  year         = {2011},
  url          = {https://doi.org/10.1145/1985793.1985907},
  doi          = {10.1145/1985793.1985907},
  bibsource    = {dblp computer science bibliography, https://dblp.org}
}

@article{DBLP:journals/re/LudersPM23,
  author       = {Clara Marie L{\"{u}}ders and
                  Tim Pietz and
                  Walid Maalej},
  title        = {On understanding and predicting issue links},
  journal      = {Requir. Eng.},
  volume       = {28},
  number       = {4},
  pages        = {541--565},
  year         = {2023},
  url          = {https://doi.org/10.1007/s00766-023-00406-x},
  doi          = {10.1007/S00766-023-00406-X},
  bibsource    = {dblp computer science bibliography, https://dblp.org}
}

@inproceedings{DBLP:conf/icse/BacchelliLR10,
  author       = {Alberto Bacchelli and
                  Michele Lanza and
                  Romain Robbes},
  editor       = {Jeff Kramer and
                  Judith Bishop and
                  Premkumar T. Devanbu and
                  Sebasti{\'{a}}n Uchitel},
  title        = {Linking e-mails and source code artifacts},
  booktitle    = {Proceedings of the 32nd {ACM/IEEE} International Conference on Software
                  Engineering - Volume 1, {ICSE} 2010, Cape Town, South Africa, 1-8
                  May 2010},
  pages        = {375--384},
  publisher    = {{ACM}},
  year         = {2010},
  url          = {https://doi.org/10.1145/1806799.1806855},
  doi          = {10.1145/1806799.1806855},
  bibsource    = {dblp computer science bibliography, https://dblp.org}
}

@inproceedings{DBLP:conf/cscw/BreuPSZ10,
  author       = {Silvia Breu and
                  Rahul Premraj and
                  Jonathan Sillito and
                  Thomas Zimmermann},
  editor       = {Kori Inkpen and
                  Carl Gutwin and
                  John C. Tang},
  title        = {Information needs in bug reports: improving cooperation between developers
                  and users},
  booktitle    = {Proceedings of the 2010 {ACM} Conference on Computer Supported Cooperative
                  Work, {CSCW} 2010, Savannah, Georgia, USA, February 6-10, 2010},
  pages        = {301--310},
  publisher    = {{ACM}},
  year         = {2010},
  url          = {https://doi.org/10.1145/1718918.1718973},
  doi          = {10.1145/1718918.1718973},
  bibsource    = {dblp computer science bibliography, https://dblp.org}
}

@article{barker2022introducing,
  title={Introducing the FAIR Principles for research software},
  author={Barker, Michelle and Chue Hong, Neil P and Katz, Daniel S and Lamprecht, Anna-Lena and Martinez-Ortiz, Carlos and Psomopoulos, Fotis and Harrow, Jennifer and Castro, Leyla Jael and Gruenpeter, Morane and Martinez, Paula Andrea and others},
  journal={Scientific data},
  volume={9},
  number={1},
  pages={622},
  year={2022},
  publisher={Nature Publishing Group UK London}
}

@article{blankenberg2014dissemination,
  title={Dissemination of scientific software with Galaxy ToolShed},
  author={Blankenberg, Daniel and Von Kuster, Gregory and Bouvier, Emil and Baker, Dannon and Afgan, Enis and Stoler, Nicholas and Galaxy Team and Taylor, James and Nekrutenko, Anton},
  journal={Genome biology},
  volume={15},
  number={2},
  pages={403},
  year={2014},
  publisher={Springer}
}

@article{deelman2009workflows,
  title={Workflows and e-Science: An overview of workflow system features and capabilities},
  author={Deelman, Ewa and Gannon, Dennis and Shields, Matthew and Taylor, Ian},
  journal={Future generation computer systems},
  volume={25},
  number={5},
  pages={528--540},
  year={2009},
  publisher={Elsevier}
}

@article{giardine2005galaxy,
  title={Galaxy: a platform for interactive large-scale genome analysis},
  author={Giardine, Belinda and Riemer, Cathy and Hardison, Ross C and Burhans, Richard and Elnitski, Laura and Shah, Prachi and Zhang, Yi and Blankenberg, Daniel and Albert, Istvan and Taylor, James and others},
  journal={Genome research},
  volume={15},
  number={10},
  pages={1451--1455},
  year={2005},
  publisher={Cold Spring Harbor Laboratory Press}
}

@article{talia2013workflow,
  title={Workflow systems for science: Concepts and tools},
  author={Talia, Domenico},
  journal={International Scholarly Research Notices},
  volume={2013},
  number={1},
  pages={404525},
  year={2013},
  publisher={Wiley Online Library}
}

@article{bean2007qualitative,
  title={Qualitative research design: An interactive approach},
  author={Bean, Cynthia J},
  journal={Organizational Research Methods},
  volume={10},
  number={2},
  pages={393},
  year={2007},
  publisher={SAGE PUBLICATIONS, INC.}
}

@article{wohlin2012experimentation,
  title={Experimentation in software engineering},
  author={Wohlin, Claes and Runeson, Per and H{\"o}st, Martin and Ohlsson, Magnus C and Regnell, Bj{\"o}rn and Wessl{\'e}n, Anders},
  year={2012},
  journal={Springer Science \& Business Media}
}

@article{batut2018asaim,
  title={ASaiM: a Galaxy-based framework to analyze microbiota data},
  author={Batut, B{\'e}r{\'e}nice and Gravouil, Kevin and Defois, Clemence and Hiltemann, Saskia and Brug{\`e}re, Jean-Fran{\c{c}}ois and Peyretaillade, Eric and Peyret, Pierre},
  journal={GigaScience},
  volume={7},
  number={6},
  pages={giy057},
  year={2018},
  publisher={Oxford University Press}
}

@article{gruning2018practical,
  title={Practical computational reproducibility in the life sciences},
  author={Gr{\"u}ning, Bj{\"o}rn and Chilton, John and K{\"o}ster, Johannes and Dale, Ryan and Soranzo, Nicola and Van Den Beek, Marius and Goecks, Jeremy and Backofen, Rolf and Nekrutenko, Anton and Taylor, James},
  journal={Cell systems},
  volume={6},
  number={6},
  pages={631--635},
  year={2018},
  publisher={Elsevier}
}

@article{goecks2010galaxy,
  title={Galaxy: a comprehensive approach for supporting accessible, reproducible, and transparent computational research in the life sciences},
  author={Goecks, Jeremy and Nekrutenko, Anton and Taylor, James and Galaxy Team team@ galaxyproject. org},
  journal={Genome biology},
  volume={11},
  number={8},
  pages={R86},
  year={2010},
  publisher={Springer}
}

@article{di2017nextflow,
  title={Nextflow enables reproducible computational workflows},
  author={Di Tommaso, Paolo and Chatzou, Maria and Floden, Evan W and Barja, Pablo Prieto and Palumbo, Emilio and Notredame, Cedric},
  journal={Nature biotechnology},
  volume={35},
  number={4},
  pages={316--319},
  year={2017},
  publisher={Nature Publishing Group US New York}
}

@book{vasiliev2020natural,
  title={Natural language processing with Python and spaCy: A practical introduction},
  author={Vasiliev, Yuli},
  year={2020},
  publisher={No Starch Press}
}

@article{kalbfleisch2023fifty,
  title={Fifty years of the Cox model},
  author={Kalbfleisch, John D and Schaubel, Douglas E},
  journal={Annual Review of Statistics and Its Application},
  volume={10},
  number={1},
  pages={1--23},
  year={2023},
  publisher={Annual Reviews}
}

@article{d2021methods,
  title={Methods to Analyse Time-to-Event Data: The Kaplan-Meier Survival Curve},
  author={D’Arrigo, Graziella and Leonardis, Daniela and Abd ElHafeez, Samar and Fusaro, Maria and Tripepi, Giovanni and Roumeliotis, Stefanos},
  journal={Oxidative medicine and cellular longevity},
  volume={2021},
  number={1},
  pages={2290120},
  year={2021},
  publisher={Wiley Online Library}
}

@book{hardeniya2016natural,
  title={Natural language processing: python and NLTK},
  author={Hardeniya, Nitin and Perkins, Jacob and Chopra, Deepti and Joshi, Nisheeth and Mathur, Iti},
  year={2016},
  publisher={Packt Publishing Ltd}
}

@article{galaxy2024galaxy,
  title={The Galaxy platform for accessible, reproducible, and collaborative data analyses: 2024 update},
  author={The Galaxy Community},
  journal={Nucleic acids research},
  volume={52},
  number={W1},
  pages={W83--W94},
  year={2024},
  publisher={Oxford University Press}
}

@article{galaxy2026galaxy,
  title={Galaxy for accessible, reproducible, and collaborative data analyses: 2026 update},
  author={The Galaxy Community},
  journal={Nucleic Acids Research},
  pages={gkag469},
  year={2026},
  publisher={Oxford University Press}
}

@inproceedings{tsay2014influence,
  title={Influence of social and technical factors for evaluating contribution in GitHub},
  author={Tsay, Jason and Dabbish, Laura and Herbsleb, James},
  booktitle={Proceedings of the 36th international conference on Software engineering},
  pages={356--366},
  year={2014}
}

@article{benjamini1995controlling,
  title={Controlling the false discovery rate: a practical and powerful approach to multiple testing},
  author={Benjamini, Yoav and Hochberg, Yosef},
  journal={Journal of the Royal statistical society: series B (Methodological)},
  volume={57},
  number={1},
  pages={289--300},
  year={1995},
  publisher={Wiley Online Library}
}

@article{freeman2007analysis,
  title={The analysis of categorical data: Fisher’s exact test},
  author={Freeman, Jenny V and Campbell, Michael J},
  journal={Scope},
  volume={16},
  number={2},
  pages={11--12},
  year={2007}
}

@article{cohen1960coefficient,
  title={A coefficient of agreement for nominal scales},
  author={Cohen, Jacob},
  journal={Educational and psychological measurement},
  volume={20},
  number={1},
  pages={37--46},
  year={1960},
  publisher={Sage Publications Sage CA: Thousand Oaks, CA}
}

@article{landis1977measurement,
  title={The measurement of observer agreement for categorical data},
  author={Landis, J Richard and Koch, Gary G},
  journal={biometrics},
  pages={159--174},
  year={1977},
  publisher={JSTOR}
}

@misc{huggingface_models,
  title        = {Hugging Face Models},
  author       = {Hugging Face Community},
  year         = {2026},
  howpublished = {\url{https://huggingface.co/models}},
  note         = {Accessed: 2026-07-07}
}

@misc{github-issue-tracker,
  author       = {Preston-Werner, Tom},
  title        = {GitHub Issue Tracker!},
  howpublished = {GitHub Blog},
  year         = {2009},
  month        = {April},
  day          = {15},
  note         = {Updated: January 4, 2019; Accessed: 2026-04-09},
  url          = {https://github.blog/news-insights/github-issue-tracker/}
}

@misc{pr-tracker-action,
  author       = {{GitHub Marketplace}},
  title        = {Pull Request Tracker – GitHub Action},
  year         = {2025},
  howpublished = {GitHub Marketplace page},
  url          = {https://github.com/marketplace/actions/pull-request-tracker},
  note         = {Accessed: 2026-04-09}
}

@misc{sentence_transformers_huggingface,
  title        = {Sentence Transformers — Hugging Face Hub},
  author       = {{Sentence Transformers community}},
  year         = {2025},
  howpublished = {\url{https://huggingface.co/sentence-transformers}},
  note         = {Accessed: 2025-10-07}
}

@misc{galaxycommunityhelp,
  author       = {{Galaxy Community}},
  title        = {Galaxy Community Help},
  howpublished = {\url{https://help.galaxyproject.org/}},
  note         = {Accessed: July 12, 2026}
}

@misc{galaxyIWC2026,
  author       = {{Intergalactic Workflow Commission}},
  title        = {Galaxy Workflow Library},
  year         = {2026},
  howpublished = {\url{https://iwc.galaxyproject.org/}},
  note         = {Accessed: 2026-08-01}
}

@misc{galaxyTraining2026,
  author       = {{Galaxy Training Network}},
  title        = {Galaxy Training Network},
  year         = {2026},
  howpublished = {\url{https://training.galaxyproject.org/}},
  note         = {Accessed: 2026-08-01}
}

@phdthesis{alam2023supporting,
  title={Supporting complex workflows for data-intensive discovery reliably and efficiently},
  author={Alam, Khairul and others},
  year={2023},
  school={University of Saskatchewan}
}

@dataset{alam_2026_22882299,
  author       = {Alam, Khairul},
  title        = {Understanding Maintenance and Support in a
                   Community-Driven Scientific Workflow Ecosystem: A
                   Cross-Space Study of Galaxy
                  },
  month        = sep,
  year         = 2026,
  publisher    = {Zenodo},
  doi          = {10.5281/zenodo.22882299},
  url          = {https://doi.org/10.5281/zenodo.22882299}
}

\appendix

\end{document}